\documentclass[usenatbib]{mn2e}    

\usepackage[caption=false]{subfig}
\usepackage[final]{graphicx}
\usepackage{epstopdf}
\usepackage{mathptmx}
\usepackage{float}
\usepackage{fixltx2e}
\usepackage{natbib}
\usepackage{amsmath}
\usepackage{amssymb}
\usepackage[usenames,dvipsnames]{color}
\usepackage{appendix}
\usepackage[mathscr]{euscript}

\newcommand{\mvir}{M_{\text{200}}}
\newcommand{\rvir}{r_{200}}
\newcommand{\msun}{\mathrm{M}_\odot}
\newcommand{\mstar}{M_\text{star}}

\newcommand{\hi}{\textsc{\mbox{H\hspace{1pt}i}}}

\newcommand{\eagle}{EAGLE}

\title[Satellite metallicities]{The origin of the enhanced metallicity of satellite galaxies}
\author[Y.~Bah\'{e} et al.]{\parbox[t]{\textwidth}{
Yannick~M.~Bah\'{e}\thanks{ybahe@mpa-garching.mpg.de}$^{1}$, Joop Schaye$^{2}$, Robert A.~Crain$^{3}$, Ian G.~McCarthy$^{3}$, \makebox{Richard G.~Bower$^{4}$}, Tom Theuns$^{4}$, Sean L.~McGee$^{5}$, and James W.~Trayford$^{4}$}
\vspace*{12pt} \\
$^1$ Max-Planck-Institut f\"{u}r Astrophysik, Karl-Schwarzschild Str. 1, 85748 Garching, Germany\\
$^2$ Leiden Observatory, Leiden University, PO Box 9513, 2300 RA Leiden, The Netherlands\\
$^3$ Astrophysics Research Institute, Liverpool John Moores University, 146 Brownlow Hill, Liverpool, L3 5RF, UK\\
$^4$ Institute for Computational Cosmology, Department of Physics, University of Durham, South Road, Durham DH1 3LE, UK\\
$^5$ School of Physics and Astronomy, University of Birmingham, Edgbaston, Birmingham, B15 2TT, UK\\
}

\begin{document}
\label{firstpage}
\maketitle

\begin{abstract}
Observations of galaxies in the local Universe have shown that both the ionized gas and the stars of satellites are more metal-rich than of equally massive centrals. To gain insight into the connection between this metallicity enhancement and other differences between centrals and satellites, such as their star formation rates, gas content, and growth history, we study the metallicities of $>$3600 galaxies with $\mstar > 10^{10}\, \msun$ in the cosmological hydrodynamical \eagle{} 100 Mpc `Reference' simulation, including $\sim$1500 in the vicinity of galaxy groups and clusters ($\mvir \geq 10^{13} \msun$). The simulation predicts excess gas and stellar metallicities in satellites consistent with observations, except for stellar metallicities at $\mstar \lesssim 10^{10.2} \msun$ where the predicted excess is smaller than observed. The exact magnitude of the effect depends on galaxy selection, aperture, and on whether the metallicity is weighted by stellar mass or luminosity. The stellar metallicity excess in clusters is also sensitive to the efficiency scaling of star formation feedback. We identify stripping of low-metallicity gas from the galaxy outskirts, as well as suppression of metal-poor inflows towards the galaxy centre, as key drivers of the enhancement of gas metallicity. Stellar metallicities in satellites are higher than in the field as a direct consequence of the more metal-rich star forming gas, whereas stripping of stars and suppressed stellar mass growth, as well as differences in accreted vs.~in-situ star formation between satellites and the field, are of secondary importance.

\end{abstract}



\defcitealias{Pasquali_et_al_2010}{P10}
\defcitealias{Pasquali_et_al_2012}{P12}

\section{Introduction}
\label{sec:introduction}

The internal properties of galaxies in dense environments are known to differ systematically from isolated galaxies, for example their colour (e.g.~\citealt{Peng_et_al_2010}), star formation rate (e.g.~\citealt{Kauffmann_et_al_2004, Wetzel_et_al_2012}), morphology (\citealt{Dressler_1980}) and atomic hydrogen content (e.g.~\citealt{Fabello_et_al_2012, Hess_Wilcots_2013}). Processes associated with galaxies becoming satellites have emerged as the primary driver of these trends \citep{Peng_et_al_2012}, with satellites in more massive haloes generally exhibiting greater differences from centrals. However, a detailed understanding of the physics responsible for the differences between centrals and satellite galaxies has so far proved elusive, although a large number of mechanisms have been proposed that could play a role: ram pressure stripping of galactic gas in the cold (\citealt{Gunn_Gott_1972}) or hot phase (\citealt{Larson_et_al_1980}), tidal forces (e.g.~\citealt{Moore_et_al_1996}), or galaxy--galaxy `harrassment' (\citealt{Moore_et_al_1996, Moore_et_al_1998}).

A promising way to make progress from the observational side is to better constrain the evolutionary history of satellite galaxies. Because the long timescales of galaxy evolution preclude direct observations of changes in individual galaxies, this requires recourse to indirect methods such as comparing galaxy populations at different cosmic epochs or analysing tracers that encode a record of a galaxy's history. One example is the ages of individual stars, knowledge of which allows the star formation history of a galaxy to be reconstructed \citep{Weisz_et_al_2014, Weisz_et_al_2015}. However, this method is limited to galaxies in the immediate vicinity of the Milky Way due to its requirement for high spatial resolution. An alternative tracer, which is observable to much larger distances, is the elemental composition or `metallicity' of a galaxy: this reflects both the star formation history (because stars synthesize new heavy elements), as well as gas inflows that supply fresh, metal-poor gas \citep{White_Rees_1978} and outflows, which remove metal-enriched material from the galaxy (e.g.~\citealt{Larson_1974, Dekel_Silk_1986}). Metallicities can typically be measured for two particular components of a galaxy: its ionized gas, where individual elements such as oxygen and hydrogen lead to prominent emission lines (e.g.~\citealt{Brinchmann_et_al_2004, Tremonti_et_al_2004}), and from absorption lines in stellar atmospheres \citep{Gallazzi_et_al_2005}.

Over the last decades, observations have shown that metallicity correlates with other galaxy properties. Early reports of an increased metallicity in more massive galaxies by e.g.~\citet{Lequeux_et_al_1979} were confirmed by analyses of the Sloan Digital Sky Survey (SDSS): \citet{Tremonti_et_al_2004} showed that the gas-phase metallicity of star forming galaxies in SDSS increases strongly with the stellar mass, and interpreted this as evidence for the efficiency of outflows in removing metals from lower-mass galaxies, while \citet{Gallazzi_et_al_2005} reached a similar conclusion from an analysis of stellar metallicities in SDSS. \citet{Lara-Lopez_et_al_2010} and \citet{Mannucci_et_al_2010} demonstrated an additional (inverse) dependence of metallicity on the star formation rate of galaxies, which has since been studied by many other authors (e.g.~\citealt{Andrews_Martini_2013, Lara-Lopez_et_al_2013}; see also \citealt{Bothwell_et_al_2013}) and interpreted as the effect of metal-poor gas inflows boosting star formation and diluting metallicity at the same time (see also \citealt{Ellison_et_al_2008a, Finlator_Dave_2008, Zhang_et_al_2009}).

In addition, mounting evidence indicates that metallicity is also affected by a galaxy's external environment at fixed stellar mass. \citet{Cooper_et_al_2008} demonstrated that (gas) metallicity is enhanced in dense environments, while \citet{Ellison_et_al_2008b} found that the opposite is true for galaxies in close pairs. Making use of the SDSS group catalogue of \citet{Yang_et_al_2007}, which splits galaxies into centrals and satellites, \citet[hereafter P10]{Pasquali_et_al_2010} found that satellite galaxies have higher stellar metallicity, as well as older stellar ages, than centrals of the same stellar mass, and that this difference increases towards lower stellar mass and higher host halo mass. These authors suggested stripping of stars, and the resulting reduction in stellar mass at constant metallicity, as an explanation for the stellar metallicity excess in satellites. In a similar way, \citet[hereafter P12]{Pasquali_et_al_2012} demonstrated the existence of a metallicity excess in the ionised gas of star-forming satellites relative to centrals. 

Although simple chemical evolution models can give some insight into the physical origin of these metallicity relations (e.g.~\citealt{Garnett_2002, Tremonti_et_al_2004, Peng_Maiolino_2014, Lu_et_al_2015}), a robust interpretation requires recourse to more sophisticated calculations. \citetalias{Pasquali_et_al_2010} compared their observational results to predictions from the semi-analytic galaxy formation model of \citet{Wang_et_al_2008}, and found that the model could reproduce the age difference between centrals and satellites as a consequence of star formation quenching after a galaxy becomes a satellite, which typically happens earlier in more massive haloes. However, they found that the \citet{Wang_et_al_2008} model predicts stellar metallicities in satellites that are nearly equivalent to those of centrals, in contrast to their observations. \citetalias{Pasquali_et_al_2010} concluded that this failure might point to an oversimplified treatment of environmental processes such as tidal stripping of stars in the model.

Cosmological hydrodynamical simulations are potentially a more powerful tool to understand the physics behind the elevated metallicities in satellites, because they self-consistently model the formation of galaxies and their environment, including the baryonic component, without explicitly distinguishing between centrals and satellites. Coupled with increasingly realistic `sub-grid' physics prescriptions to describe unresolved processes like radiative cooling, star formation, and feedback, such simulations have now evolved to the point where the modelled galaxy populations resemble observations in several key properties such as their stellar mass, star formation rate, and metallicity \citep{Vogelsberger_et_al_2014, Schaye_et_al_2015}. In a recent study, \citet{Genel_2016} used the Illustris simulation \citep{Vogelsberger_et_al_2014} to gain insight into the elevated gas-phase metallicities in satellite galaxies (see also \citealt{Dave_et_al_2011, DeRossi_et_al_2015}, who reported excess metallicity in satellites compared to centrals in earlier simulations). The Illustris simulation was found to qualitatively reproduce the observational result of \citetalias{Pasquali_et_al_2012}, the elevated metallicity in satellites being driven by differences in the radial distribution of star-forming gas as well as different star formation histories of satellites \citep{Genel_2016}.

In this paper, we perform an analysis of the \eagle{} simulation (\citealt{Schaye_et_al_2015, Crain_et_al_2015}) to gain further insight into the nature of satellite metallicities. Our aim is twofold: on the one hand, we want to test whether \eagle{} -- which differs from Illustris in several key aspects including the hydrodynamics scheme and implementation of feedback from star formation -- is able to reproduce the observed metallicity differences between satellites and centrals. This is an important test of the model, and also serves to establish whether the agreement with observations in terms of gas-phase metallicity reported by \citet{Genel_2016} is primarily a consequence of the specific model used for Illustris, or rather a more generic success of modern cosmological simulations. Secondly, we will use the detailed particle information and evolutionary history of the simulated galaxies from \eagle{} to study the origin of this metallicity enhancement.

While \eagle{} has been calibrated to match the masses and sizes of observed present-day galaxies, the metallicities were not explicitly constrained, and can hence be regarded as a prediction of the simulation. This is in contrast to Illustris, where the metallicity of outflowing gas is reduced by means of an adjustable parameter in order to match the normalisation of the observed mass-metallicity relation \citep{Vogelsberger_et_al_2013}. As shown by \citet{Schaye_et_al_2015}, the observed mass--metallicity relation for both star forming gas and stars is nevertheless broadly reproduced for massive ($\mstar > 10^{10}\, \msun$) galaxies in the largest-volume \eagle{} simulation, while at lower masses, the predicted metallicities are systematically too high. This discrepancy is eased in higher-resolution \eagle{} simulations -- in which the gas metallicities are consistent with observations for $\mstar \gtrsim 10^{8.5} \msun$, although stellar metallicities are still somewhat higher than observed \citep{Schaye_et_al_2015} -- but because these are computationally much more challenging, they were restricted to a relatively small box with side length of 25 comoving Mpc, and hence lack the massive haloes whose satellites we wish to study. For this reason, we here mostly restrict our analysis to the study of satellites with $\mstar > 10^{10} \msun$, for which the offset between different resolution runs is $\lesssim 0.15$ dex.

The remainder of this paper is structured as follows. In \S \ref{sec:eagle}, we briefly review the relevant characteristics of the \eagle{} simulation and describe our galaxy selection and method for tracing galaxies between different snapshots. Predictions for the gas-phase and stellar metallicities of satellite galaxies are presented and compared to both observations and alternative theoretical models in \S \ref{sec:trends}. \S \ref{sec:origin_zgas} illuminates the nature of differences in the gas-phase metallicity, highlighting gas stripping and suppressed gas inflows as the two dominant mechanisms responsible. We then investigate the action of indirect effects such as stellar mass stripping on stellar metallicities in \S \ref{sec:origin_zstar}, and demonstrate a direct connection between the excess in gas-phase and stellar metallicities in \eagle{}. Our results are summarized and discussed in \S \ref{sec:summary}. 

Throughout the paper, we use a flat $\Lambda$CDM cosmology with parameters as determined by \citet{Planck_2014} (Hubble parameter $h \equiv $ H$_{0}/(100\,{\rm km}\,{\rm s}^{-1}{\rm Mpc}^{-1}) = 0.6777$, dark energy density parameter $\Omega_\Lambda = 0.693$ (dark energy equation of state parameter $w=-1$), matter density parameter $\Omega_{\rm M} = 0.307$, and baryon density parameter $\Omega_{\rm b} = 0.04825$). The solar metallicity and oxygen abundance are assumed to be Z$_\odot$ = 0.012 \citep{Allende-Prieto_et_al_2001} and 12+log(O/H) = 8.69 \citep{Asplund_et_al_2009}, respectively. Unless specified otherwise, all masses and distances are given in physical units. In our plots, dark shaded regions denote $1\sigma$ uncertainties calculated as explained in Section \ref{sec:zgas}, while light shaded bands (where shown) indicate galaxy-to-galaxy scatter (central 50 per cent, i.e.~stretching from the 25th to the 75th percentile), unless explicitly stated otherwise.


\section{The \eagle{} simulations}
\label{sec:eagle}

\subsection{Simulation characteristics}
The ``Evolution and Assembly of GaLaxies and their Environments'' (\eagle) project
consists of a suite of cosmological hydrodynamical simulations of varying size, resolution and sub-grid physics models. For a detailed description, the interested reader is referred to \citet{Schaye_et_al_2015} and \citet{Crain_et_al_2015}; here we only give a concise summary of those aspects that are directly relevant to our work.

The analysis presented in this paper is based mainly on the largest `Reference' \eagle{} simulation (Ref-L100N1504 in the terminology of \citealt{Schaye_et_al_2015}, although for brevity we will usually refer to it here simply as `Ref-L100'), which fills a cubic volume of side length 100 comoving Mpc (`cMpc') with $N = 1504^3$ dark matter particles ($m_\text{DM} = 9.70 \times 10^6\, \msun$) and an initially equal number of gas particles ($m_\text{gas} = 1.81 \times 10^6\, \msun$). The simulation was started at $z = 127$ from cosmological initial conditions \citep{Jenkins_2013}, and evolved to $z = 0$ using a modified version of the \textsc{gadget-3} code \citep{Springel_2005}. These changes include a number of hydrodynamics updates collectively referred to as ``Anarchy'' (Dalla Vecchia, in prep.; see also \citealt{Hopkins_2013}, Appendix A of \citealt{Schaye_et_al_2015}, and \citealt{Schaller_et_al_2015}) which mitigate many of the shortcomings of `traditional' SPH codes, such as the treatment of surface discontinuities (e.g.~\citealt{Agertz_et_al_2007, Mitchell_et_al_2009}).

The Plummer-equivalent gravitational softening length is 0.7 proper kpc (`pkpc') at redshifts $z < 2.8$, and 2.66 comoving kpc (`ckpc'), i.e.~1/25 of the mean inter-particle separation, at earlier times. The simulation is therefore capable of marginally resolving the Jeans scale of gas with density and temperature characteristic of the warm, diffuse ISM\footnote{But see the discussion in \citet{Hu_et_al_2016} concerning the definition of mass resolution in SPH simulations.}, but the same is not true for cold molecular gas. A temperature floor $T_\text{eos} (\rho)$  is therefore imposed on gas with $n_\text{H} > 0.1$ cm$^{-3}$, in the form of a polytropic equation of state $P \propto \rho^\gamma$ with index $\gamma = 4/3$ and normalised to $T_\text{eos}$ = 8 000 K at $n_\text{H} = 10^{-1} \text{cm}^{-3}$ (see \citealt{Schaye_DallaVecchia_2008} for further details). In addition, gas at densities $n_\text{H} \geq 10^{-5}$ cm$^{-3}$ is prevented from cooling below 8 000 K. 

The \eagle{} code includes significantly improved sub-grid physics prescriptions, described in detail in section 4 of \citet{Schaye_et_al_2015}. These include element-by-element radiative gas cooling \citep{Wiersma_et_al_2009a} in the presence of the Cosmic Microwave Background (CMB) and an evolving \citet{Haardt_Madau_2001} UV/X-ray background, reionization of hydrogen at $z = 11.5$ and helium at $z \approx 3.5$ \citep{Wiersma_et_al_2009b}, star formation implemented as a pressure law \citep{Schaye_DallaVecchia_2008} with a metallicity-dependent density threshold of $$n_H^*(Z) = 10^{-1} \text{cm}^{-3} \left(\frac{Z}{0.002} \right)^{-0.64}$$ limited to a maximum of 10 cm$^{-3}$ (following \citealt{Schaye_2004}) and adopting a universal \citet{Chabrier_2003} stellar initial mass function (IMF) with minimum and maximum stellar masses of 0.1 and 100 $\msun$, respectively, as well as energy feedback from star formation \citep{DallaVecchia_Schaye_2012} and accreting supermassive black holes (AGN feedback; \citealt{Rosas-Guevara_et_al_2015, Schaye_et_al_2015}) in thermal form. 

Three aspects in the implementation of energy feedback from star formation merit explicit mention here, in light of the potential of feedback-driven outflows to influence galaxy metallicities (see e.g.~\citealt{Oppenheimer_Dave_2008}). Firstly, because the feedback efficiency cannot be predicted from first principles, its efficiency was calibrated to reproduce the $z\approx 0$ galaxy stellar mass function and sizes (see \citealt{Crain_et_al_2015} for an in-depth discussion of this issue). Secondly, the feedback parameterisation depends only on local gas quantities, in contrast to e.g.~the widely-used practice of scaling the parameters with the (global) velocity dispersion of a galaxy's dark matter halo (e.g.~\citealt{Okamoto_et_al_2005, Oppenheimer_Dave_2006, Vogelsberger_et_al_2013, Puchwein_Springel_2013}). Finally, star formation feedback in \eagle{} is made efficient not by temporarily disabling hydrodynamic forces or cooling for affected particles (e.g.~\citealt{Springel_Hernquist_2003, Stinson_et_al_2006, Vogelsberger_et_al_2013}), but instead by stochastically heating a fraction of particles by a temperature increment of $\Delta T = 10^{7.5}\, \text{K}$ \citep{DallaVecchia_Schaye_2012}. 

Enrichment of gas is modelled on an element-by-element basis following \citet{Wiersma_et_al_2009b}. This model includes contributions from AGB stars, type Ia and II supernovae, and explicitly tracks the metallicity of the nine elements that \citet{Wiersma_et_al_2009a} found to dominate the radiative cooling rate (H, He, C, N, O, Ne, Mg, Si, and Fe)\footnote{In addition, Ca and S are tracked assuming a fixed mass ratio relative to Si of 0.094 and 0.605, respectively (see \citealt{Wiersma_et_al_2009b}).}, as well as the total metal content of SPH particles. When a gas particle is converted into a star particle, it inherits the element abundances of its parent, which thereafter remain constant.

For better consistency with the underlying SPH formalism, the metallicity used to calculate e.g.~gas cooling rates is calculated as the ratio of the SPH-smoothed metal (or individual element) mass density and the SPH smoothed total gas density (as described by \citealt{Okamoto_et_al_2005} and \citealt{Tornatore_et_al_2007}). \citet{Wiersma_et_al_2009b} discuss how the fact that this `smoothed metallicity' of an SPH particle is influenced by the metallicity of its neighbour particles also suppresses numerical fluctuations in metallicity arising from the inherent lack of metal mixing in SPH simulations without requiring the implementation of uncertain additional physics such as diffusion. The results presented in the remainder of this paper are generally based upon these smoothed metallicities, except where explicitly stated otherwise.

In post-processing, \citet{Trayford_et_al_2015, Trayford_et_al_2016} calculated the amount of stellar light emitted in the \eagle{} simulation, with a stellar population synthesis approach based on the \citet{Bruzual_Charlot_2003} simple stellar population models. Note that, although \citet{Trayford_et_al_2015} include a prescription for dust extinction in their model, the luminosities used in this work do not take this effect into account. Because only a small part of our analysis is based on stellar luminosities, this is not expected to have a significant impact on our results. 

\subsection{Galaxy selection}
\subsubsection{Selection of galaxies and haloes at $z \approx 0$}
From the (100 cMpc)$^3$ \eagle{} Reference simulation, Ref-L100, we select galaxies from the snapshot at $z=0.1$, which approximately coincides with the median redshift of the SDSS derived galaxy samples used by \citetalias{Pasquali_et_al_2010} and \citetalias{Pasquali_et_al_2012}\footnote{We have verified that our results are qualitatively unchanged when the analysis is performed at $z=0$ instead.}. Galaxies are selected as 
self-bound subhaloes within a friends-of-friends (FOF) halo -- identified using the \textsc{subfind} algorithm (\citealt{Dolag_et_al_2009}; see also \citealt{Springel_et_al_2001b}) --- with a stellar mass of $\mstar \geq 10^{9}\, \msun$; as discussed above, we mostly restrict ourselves to the subset of these with $\mstar > 10^{10}\, \msun$, but will occasionally also extend our analysis to $10^9\, \msun \leq \mstar < 10^{10}\, \msun$. Stellar masses are computed throughout this paper as the total mass of all gravitationally bound star particles within a spherical aperture of 30 pkpc, centered on the particle for which the gravitational potential is minimum. Although stars beyond this radius have been shown to contribute non-negligibly to the total stellar mass of very massive galaxies (e.g.~\citealt{DSouza_et_al_2015}), \citet{Schaye_et_al_2015} show that a spherical 30 pkpc aperture roughly mimics the Petrosian radius often used by optical surveys such as the SDSS. For consistency, galaxy star formation rates are also computed within the same aperture.

The observational work of \citetalias{Pasquali_et_al_2010} and \citetalias{Pasquali_et_al_2012} has shown that differences between central and satellite galaxies are greatest for satellites in the most massive haloes. We therefore focus here on haloes at the mass scale of galaxy groups and (small) clusters, $10^{13} \msun \leq \mvir \lesssim 10^{14.5} \msun$, where $\mvir$ is the total mass within a spherical aperture of radius $r_{200}$ that is centered on the potential minimum of the halo and within which the mean density equals 200 times the critical density of the Universe, $\rho_\text{crit}$. In less massive haloes, the number of satellite galaxies with $\mstar \geq 10^{10} \msun$ becomes small, and their mass approaches that of the most massive galaxy in the halo (i.e.~the central), which makes the distinction between central and satellite less meaningful than in more massive systems. Clusters more massive than $\sim$10$^{14.5}\, \msun$, on the other hand, are too rare to be found in a (100 cMpc)$^3$ simulation such as \eagle{}. In total, the simulation contains 154 haloes in this mass range at $z=0.1$, nine of which can be classified as galaxy clusters ($\mvir \geq 10^{14}\, \msun$). For simplicity, we will refer to all these haloes as `groups', except where we are specifically distinguishing between systems above and below a threshold of $\mvir = 10^{14}\, \msun$.

In this paper, we follow the standard terminology of referring as the `central' galaxy to that living in the most massive subhalo in a FOF halo, which typically also sits at the minimum of its gravitational potential well (e.g.~\citealt{Yang_et_al_2005}). The galaxies hosted by all other subhaloes are `satellite' galaxies. It is unclear, however, to what extent this classification is physically meaningful (see e.g.~\citealt{Bahe_et_al_2013}) or agrees with observational central/satellite classifications, which are inevitably based on the distribution of galaxies alone, instead of the underlying dark matter structure (e.g.~\citealt{Yang_et_al_2005}). We therefore also collect \emph{all} galaxies located within $\leq 5 r_{200}$ from the centre of a group halo into a set of `group galaxies'. This enables us to investigate trends with halo-centric distance, noting that mounting evidence from observations (e.g.~\citealt{Lu_et_al_2012, Wetzel_et_al_2012}) and theory (e.g.~\citealt{Bahe_et_al_2013}) indicates that galaxies are affected by the group/cluster environment significantly beyond the virial radius. For a clear distinction, we then select as `field' galaxies all those centrals that are not located within $5 r_{200}$ of any of our group/cluster haloes, but the much larger number of centrals in the field than near groups/clusters means that virtually identical results are obtained when comparing to all centrals instead (as was done, for example, by \citetalias{Pasquali_et_al_2010} and \citetalias{Pasquali_et_al_2012}).

\begin{figure*}
  \centering
    \includegraphics[width=2.1\columnwidth]{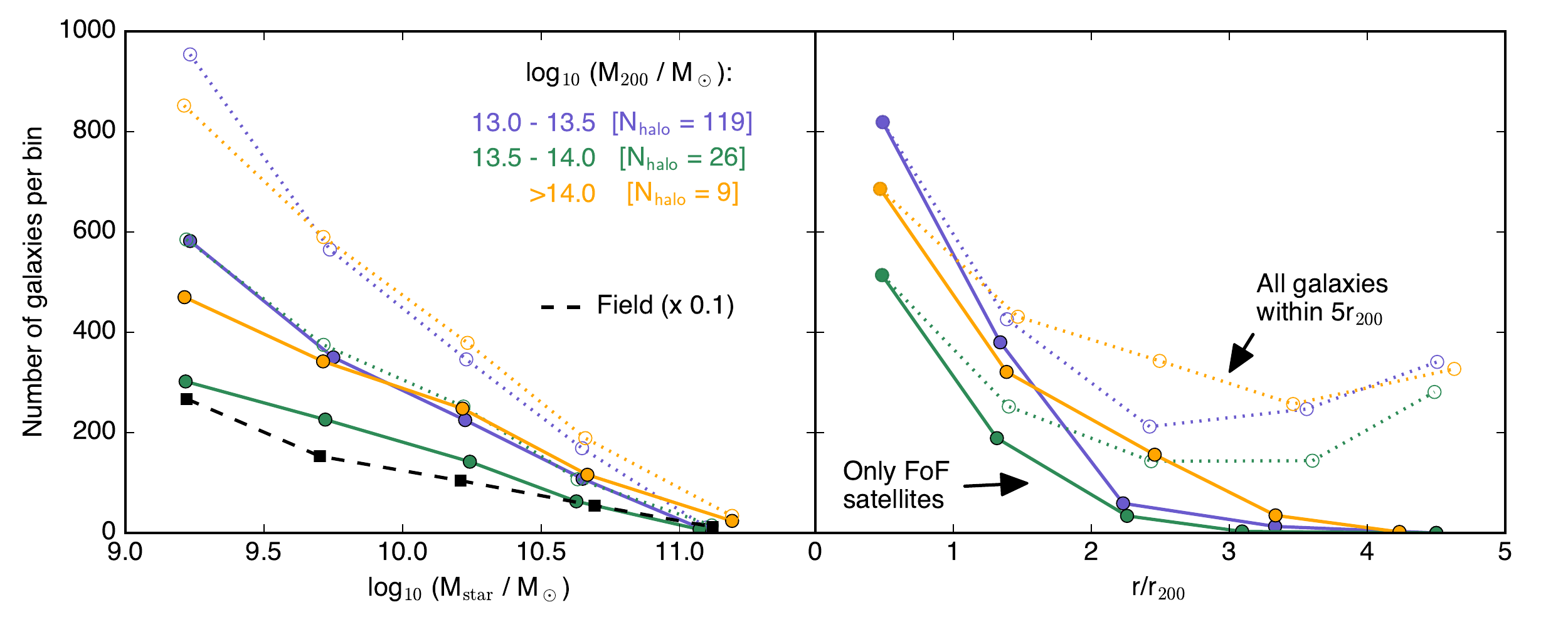}
       \caption{The number of group/cluster galaxies in the 100 cMpc \eagle{} Reference simulation as a function of host mass (differently coloured lines) and, respectively, stellar mass (\textbf{left-hand panel}), or their distance from the host centre (\textbf{right-hand panel}). Dotted lines include all galaxies within $5 r_{200}$ from the host centre, whereas the solid lines shows only those that are identified as part of the host friends-of-friends (FoF) group. In the left-hand panel, the corresponding number of field galaxies is shown as a black dashed line, reduced by a factor of 10 to fit onto the same axis. The number of haloes in each bin is given in the top-right corner of the left-hand panel.}
    \label{fig:sats.numbers}
  \end{figure*}

In Fig.~\ref{fig:sats.numbers}, the number of group and field galaxies in the \eagle{} Ref-L100 simulation is shown as a function of stellar mass (left panel), and of the distance from the group centre in units of $r_{200}$ (right). In both cases, dotted lines represent all group galaxies, while the corresponding trends for only those galaxies that are part of the group's FOF halo (the~`satellites') are shown as solid lines. The latter account for roughly half of all group galaxies, but show, as expected, a stronger concentration towards smaller halo-centric radii\footnote{The small population of FOF satellites at large $r$ is caused by extremely elongated FOF groups.} ($r \lesssim 2\,\rvir$). Fig.~\ref{fig:sats.numbers} confirms that the Ref-L100 simulation contains enough group galaxies to study trends in their metallicity and compare to the field: even in the least densely populated halo mass bin, $13.5 \leq \log_{10} (\mvir / \msun) < 14.0$, there are 740 satellites, 212 of which have $\mstar \geq 10^{10} \msun$. Note also that the number of `field' galaxies vastly outnumbers that of group galaxies, in all stellar mass bins. 

\subsubsection{Galaxy tracing}
\label{sec:tracing}
To understand the mechanisms that drive environmental metallicity trends at $z \approx 0$, it will be necessary to trace galaxies across cosmic time by identifying the progenitors in earlier snapshots. For this purpose, we employ a tracing algorithm similar to that described by \citet{Bahe_McCarthy_2015}. In brief, for every pair of adjacent snapshots (i.e. those following each other in time) we identify all subhaloes that share a significant number of dark matter particles ($N \geq 20$), and then select the subhaloes linked by the largest numbers of particles as each other's progenitor and descendent, respectively. In doing so, we take into account that any one subhalo in one snapshot may share particles with more than one subhalo in the other, and that subhaloes may temporarily evade identification by the \textsc{subfind} algorithm. For a more detailed description, the interested reader is referred to appendix A of \citet{Bahe_McCarthy_2015} where the algorithm is described in detail.


\section{Satellite metallicities at redshift z $\approx$ 0}
\label{sec:trends}
In this section, we present the relations between stellar mass and, respectively, the oxygen abundance of star-forming gas and stellar metallicity (\S \ref{sec:sdss}) predicted by the \eagle{} Ref-L100 simulation for field and satellite galaxies in different mass haloes. In both cases, we will compare these to observational data derived from SDSS spectra. In \S\ref{sec:radpos}, we investigate the effect of galaxy position within their parent halo. These results are then compared to other models, both within and outside of the \eagle{} suite (\S \ref{sec:modelcomp}).

\subsection{Comparison to observations from the SDSS}
\label{sec:sdss}

\subsubsection{Metallicity of star-forming gas}
\label{sec:zgas}
In observations, gas-phase metallicities are typically derived spectroscopically from nebular emission lines (see e.g.~\citealt{Brinchmann_et_al_2004, Tremonti_et_al_2004, Zahid_et_al_2014}). Because oxygen has traditionally been used as the `canonical' metal for this purpose, the metallicity is typically expressed in terms of the quantity 12+log(O/H), where `O' and `H' are the number densities of oxygen and hydrogen, respectively. Based on the metallicity determinations of \citet{Tremonti_et_al_2004}, and the SDSS galaxy group catalogue of \citet{Yang_et_al_2007}, \citetalias{Pasquali_et_al_2012} studied the relation between gas-phase metallicity and stellar mass in a sample of $\sim$84 000 star-forming galaxies in the SDSS, split into centrals ($\sim$70 000) and satellites ($\sim$14 000). They found that the metallicity of satellites is systematically enhanced compared to centrals of the same stellar mass, an effect that is stronger for satellites of lower stellar mass and those inhabiting more massive haloes. Note that this result is robust, at least to first order, against systematic uncertainties in the overall calibration of observational gas-phase metallicity measurements from emission lines (see e.g.~\citealt{Kennicutt_et_al_2003,Kewley_Ellison_2008}) because it only relies on the determination of \emph{relative} metallicity differences.

However, the \eagle{} simulations have neither the resolution nor the sub-grid physics to model individual star forming regions. Instead, we calculate galaxy-averaged values of 12+log(O/H) directly from the smoothed abundances of oxygen and hydrogen, weighted by the star formation rate (SFR) of individual particles to mimic the larger contribution to observed metallicity measurements from more active star forming regions whose emission lines are stronger. 

This strategy implies that our metallicity measurement ignores all particles with a density below the star formation threshold of \eagle{} (see \S\ref{sec:eagle}). Note that, because this threshold is itself a (physically motivated) function of metallicity \citep{Schaye_2004}, the metallicity measurement might therefore be subject to biases, but we have tested this by instead computing metallicities for particles above a fixed \emph{density} threshold ($n_\text{H} \geq 0.01$ cm$^{-3}$) and obtained similar results.

It is also important to keep in mind that a determination of gas-phase metallicities from nebular emission lines is only possible for star-forming galaxies. The sample selection of \citet{Tremonti_et_al_2004}, and hence also of \citetalias{Pasquali_et_al_2012}, is based on spectral features, especially the strength of the H$\beta$ line, and the [N II]/H$\alpha$ vs.~[O III]/H$\beta$ line ratios to exclude active galactic nuclei (see e.g.~\citealt{Baldwin_et_al_1981}). In the absence of mock spectra to reproduce this selection exactly for the \eagle{} galaxies, we select star forming galaxies based solely on their specific star formation rate (sSFR $\equiv$ SFR/$\mstar$) within an aperture of 30 pkpc. Our default threshold of sSFR $> 10^{-11}$ yr$^{-1}$ is motivated by the observed bimodality of the sSFR distribution in the local Universe, with a minimum at approximately this value (e.g.~\citealt{Wetzel_et_al_2012}). To explore the sensitivity of our results to the adopted threshold, and for improved consistency with the observational analysis of \citetalias{Pasquali_et_al_2012}, we also consider an alternative, stricter cut at sSFR = 10$^{-10.5}$ yr$^{-1}$, which may correspond more closely to the sample selection of that study (see their figure 13).

\begin{figure}
  \centering
    \includegraphics[width=\columnwidth]{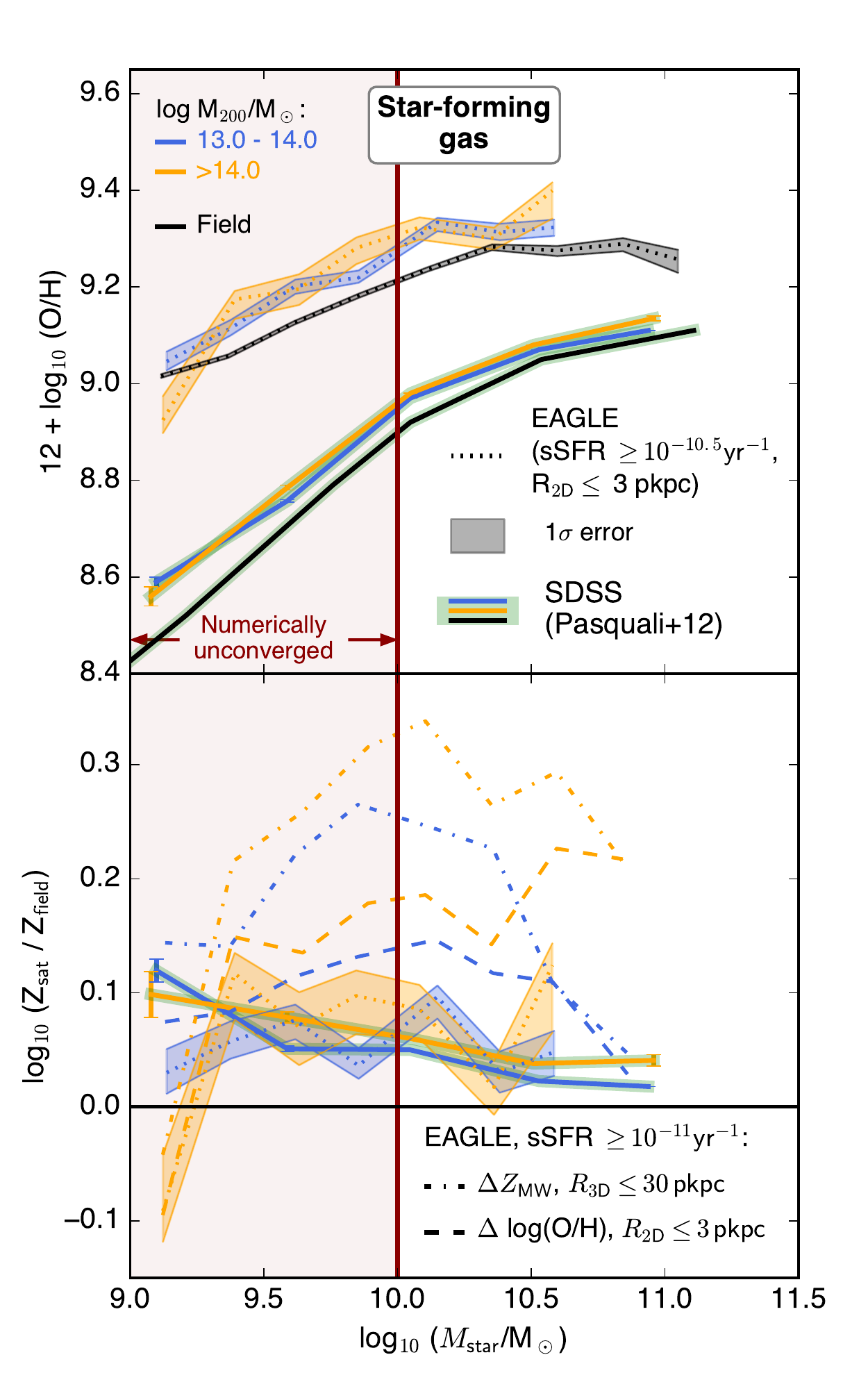}    
       \caption{\textbf{Top panel:} Gas-phase oxygen abundance in star-forming galaxies in the field (black) and satellites (blue/orange). \eagle{} galaxies with sSFR $\geq 10^{-10.5} \mathrm{yr}^{-1}$ are shown as shaded bands whose width indicates the statistical $1\sigma$ uncertainty on the median trend (central dotted line). Lines are drawn only for bins containing at least ten galaxies. Observational data from \citetalias{Pasquali_et_al_2012} are shown as thin solid lines on green background. The red shaded region on the left is potentially affected by numerical resolution in \eagle{}. \textbf{Bottom:} Logarithmic metallicity ratio between satellite and field galaxies. In addition to the data plotted in the top panel (dotted lines), \eagle{} predictions are shown for galaxies with sSFR $ \geq 10^{-11}\, \mathrm{yr}^{-1}$ (dashed), and additionally for the \emph{total} metallicity difference, defined as mass-weighted mean within a 3D aperture of 30 pkpc, (dash-dot lines). With the `strict' sSFR threshold (dotted lines), which corresponds approximately to the selection of \citetalias{Pasquali_et_al_2012}, the environmental predictions of \eagle{} agree with the SDSS data.}
    \label{fig:gasz}
  \end{figure}

In the top panel of Fig.~\ref{fig:gasz}, we present the relation between stellar mass and oxygen abundance 12+log(O/H) of star forming gas in \eagle{}, adopting the stricter threshold of sSFR $> 10^{-10.5}$ yr$^{-1}$ (dotted lines). The black line represents field galaxies, whereas satellites are shown with blue and gold lines, the former representing those in the halo mass interval $\mvir = 10^{13}$ -- $10^{14}\, \msun$ and the latter those in more massive haloes (i.e.~clusters). The width of the dark shaded bands indicates the statistical $1\sigma$ uncertainty on the median oxygen abundance (central line), i.e.~it extends from $f_\text{low}$ to $f_\text{high}$ where $f = $ 12+log(O/H) and $f_\text{low (high)} = \tilde{f} + (P_\text{15.9 (84.1)}-\tilde{f}) / \sqrt{N}$; $\tilde{f}$ here denotes the median and $P_n$ the $n$th percentile of the distribution in a bin with $N$ galaxies. The galaxy-to-galaxy scatter is indicated by the light shaded band which extends from the 25th to the 75th percentile; for clarity this is omitted for the cluster satellite bin (gold). For approximate consistency with the SDSS observations, we only calculate the contribution from gas particles that are part of the galaxy's subhalo and lie within a (2D) radial aperture (projected in the simulation $xy$-plane) of 3 pkpc, which is centered on the potential minimum of the galaxy subhalo. This corresponds approximately to the extent of the SDSS fibers at the median redshift of the galaxies considered by \citetalias{Pasquali_et_al_2012}.

For ease of comparison, we also reproduce the data from \citetalias{Pasquali_et_al_2012} in Fig.~\ref{fig:gasz}, with thin solid lines in the same colours as for \eagle{} but underlined in green. Statistical $1\sigma$ uncertainties are here shown with error bars; we note that these are calculated as $1\sigma$ error on the mean gas-phase metallicity (weighted by 1/$V_\text{max}$, where $V_\text{max}$ denotes the comoving volume within which the galaxy would have been included in the sample), propagating errors in individual measurements. 

The absolute oxygen abundances of star-forming galaxies within $R_\text{2D} \leq 3$ pkpc predicted by \eagle{} (dotted lines) are higher than what is inferred from SDSS (solid lines), by up to $\sim$0.5 dex. As discussed by \citet{Schaye_et_al_2015}, there are significant systematic uncertainties in the observational measurements, related to the calibration of strong-line indices (e.g.~\citealt{Kewley_Ellison_2008}), condensation onto dust grains (e.g.~\citealt{Mattsson_Andersen_2012}), and determination of stellar masses \citep{Conroy_et_al_2009},  and also on the simulation side due to uncertain nucleosynthetic yields (e.g.~\citealt{Wiersma_et_al_2009b}) in addition to our rather simplistic match to the SDSS fiber size and sample selection. We therefore caution against over-interpreting this discrepancy. At a qualitative level, \eagle{} reproduces the observational results of higher gas-phase oxygen abundance in more massive galaxies (as already shown by \citealt{Schaye_et_al_2015})\footnote{\citet{Schaye_et_al_2015} did not impose an aperture of $R_\text{2D} \leq 3$ pkpc in their analysis, so that the absolute values of 12+log$_{10}$ (O/H) for \eagle{} galaxies shown in their figure 13 are slightly lower than those plotted here, by $\lesssim$0.2 dex.}.

Satellite galaxies in \eagle{} are, overall, more metal-rich than equally massive field galaxies (comparing the blue/yellow and black dotted lines), which qualitatively agrees with the observations of \citetalias{Pasquali_et_al_2012}. For satellites with very low mass, $\mstar \approx 10^9 \msun$, the simulation predicts satellite metallicities that are not significantly different from the field, which is in conflict with observations. However, we reiterate that predictions for galaxies with $\mstar < 10^{10} \msun$ (the area shaded red in Fig.~\ref{fig:gasz}) are possibly affected by numerical resolution, which may at least partly account for the discrepancy in the relative difference between field and satellites.

The relatively small number of galaxies ($N = 59$ in the cluster bin with $\mstar > 10^{10} \msun$) precludes a meaningful statement on the impact of halo mass. Within the uncertainties, there is no significant difference between group and cluster satellites, whereas observationally, a slightly enhanced excess is seen in the latter. 

In order to more clearly highlight the environmental impact on galaxy metallicity, we plot in the bottom panel of Fig.~\ref{fig:gasz} the \emph{logarithmic ratio} between the median metallicity of the satellite and field galaxy populations; lines have the same meaning as in the top panel. $1\sigma$ errors are calculated by adding the uncertainties on the field and satellite populations in quadrature; in practice, the latter dominates this combined uncertainty. This plot removes the impact of the different mass--metallicity relations in the field between \eagle{} and SDSS, and allows a direct quantitative comparison of the environmental effect alone: the simulation prediction is in good agreement with observations down to $\mstar \approx 10^{9.5} \msun$. Importantly, this comparison is also more robust to the above-mentioned large systematic uncertainties in the calibration of observational metallicity indicators.

It is important to keep in mind, however, that our galaxy selection (sSFR $> 10^{-10.5}$ yr$^{-1}$) is at best a crude match to that of \citetalias{Pasquali_et_al_2012}: we have made no attempt to reject galaxies harbouring active nuclei (AGN), and furthermore the median sSFR of our galaxies is still systematically lower than theirs, by $\gtrsim 0.2$ dex\footnote{Although we note that the SDSS sSFR estimates have recently been revised downwards by this amount \citep{Chang_et_al_2015}.}. To estimate the impact of such selection differences, we also show the satellite metallicity excess obtained from our fiducial, physically motivated sSFR threshold of $10^{-11}$ yr$^{-1}$, as dashed lines. The impact of this change is substantial: it increases the environmental excess to $\sim$0.1 dex in groups and $\sim$0.2 dex in clusters, several times larger than that obtained with our only moderately stricter sSFR threshold of $10^{-10.5}$ yr$^{-1}$ (dotted lines). At least within \eagle{}, the environmental gas-phase metallicity excess is evidently sensitive to galaxy selection, implying that the apparently good agreement between \eagle{} and SDSS may be subject to significant systematic uncertainty.   

As a final test, we also explore the impact of relaxing the relatively small aperture that was matched to the SDSS fiber size, the definition of metallicity as the abundance of oxygen alone, and the weighting between different gas particles according to their star formation rates. Instead, we compute the mass-weighted mean of the \emph{total} metal abundance of star forming gas particles within a 3D radius of 30 pkpc. This result, which arguably represents a more `physical' measure of the star forming gas metallicity, is shown in the bottom panel of Fig.~\ref{fig:gasz} as dash-dot lines and shows yet stronger environmental impact, of up to 0.34 dex in cluster galaxies of $\mstar \approx 10^{10} \msun$. From more detailed tests varying the aperture, metallicity definition, and weighting scheme separately (not shown), we conclude that the largest effect arises from the difference in aperture. 

We conclude from this analysis that \eagle{} predicts an approximately realistic environmental effect on satellite gas metallicities, and that the `true' effect, integrated over an entire galaxy, is significantly greater than what is deduced from observations of the innermost galaxy region alone.

\subsubsection{Stellar metallicity}
\label{sec:zstar}

As an alternative to the determination of gas-phase oxygen abundances from emission lines, metallicities can also be measured for the stellar component of galaxies through modelling of their absorption lines; in contrast to gas metallicity such a measurement is possible for both star forming and passive galaxies. Using this technique, \citet{Gallazzi_et_al_2005} derived the stellar metallicities and ages of almost 200 000 galaxies from the SDSS DR2, and demonstrated that a subset of $\sim$44 000 of these have spectra of sufficiently high signal-to-noise (i.e.~$S/N \geq 20$) to allow a meaningful determination of these quantities (with uncertainties $\leq 0.3$ dex). Similar to the positive correlation between gas-phase oxygen abundance and stellar mass reported by \citet{Tremonti_et_al_2004}, these authors demonstrated an increase in stellar metallicity with increasing stellar mass. By combining these data with the group catalogue of \citet{Yang_et_al_2007}, \citetalias{Pasquali_et_al_2010} found an additional dependence of stellar metallicity on environment, in the sense that stars in satellites are metal-richer than those in field galaxies of the same stellar mass, qualitatively similar to the enhancement in the gas-phase metallicity of star forming galaxies discussed above \citepalias{Pasquali_et_al_2012}. 

\begin{figure}
  \centering
    \includegraphics[width=\columnwidth]{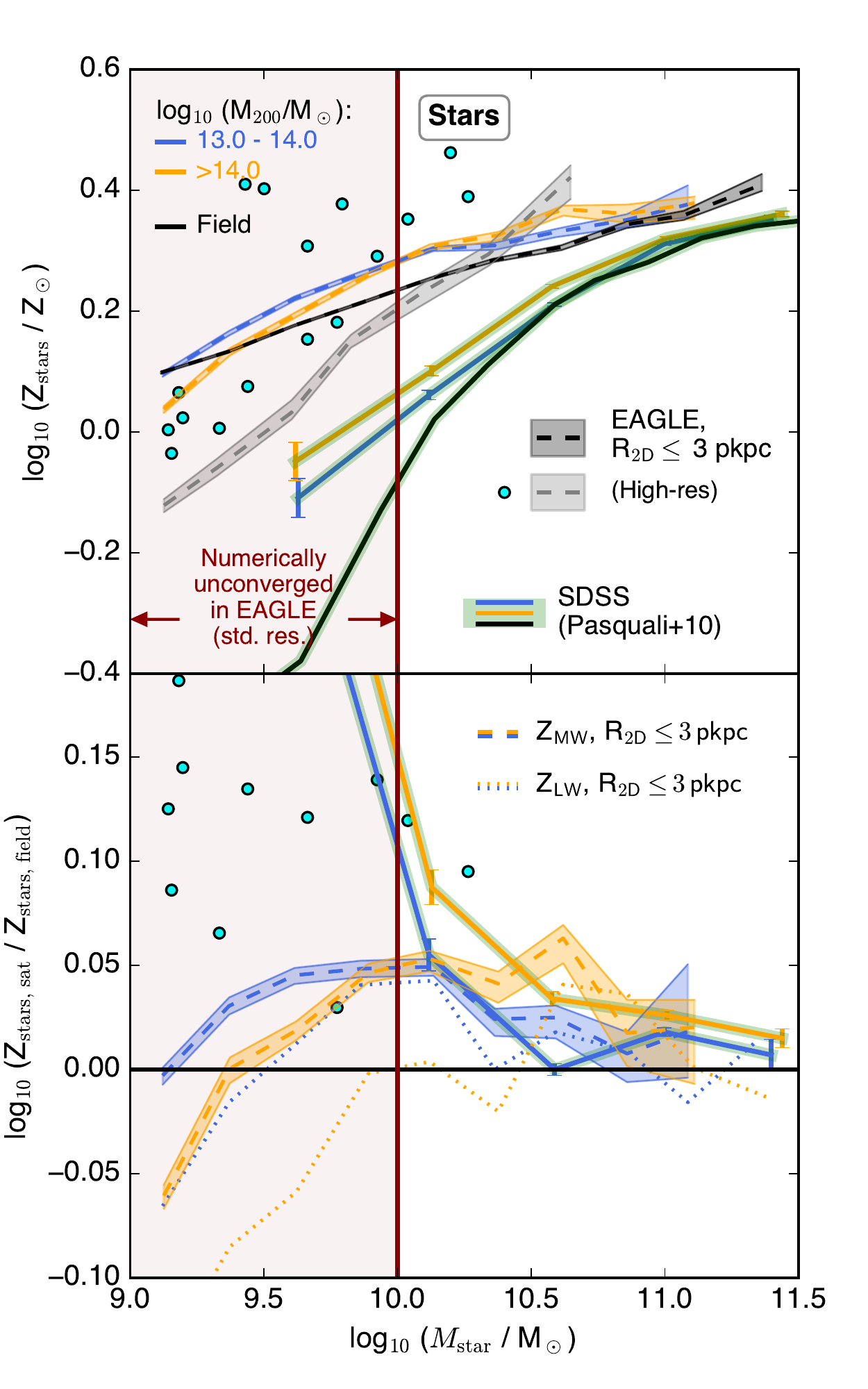}    
       \caption{Stellar metallicities in field galaxies (black) and satellites (orange/blue), in analogy to Fig.~\ref{fig:gasz}. Solid lines underlined in green represent the observational measurements of \citetalias{Pasquali_et_al_2010}, adjusted to a solar metallicity of $Z_\odot = 0.012$ (see text). Also shown are predictions from the 25 cMpc \eagle{} high resolution run (Recal-L025N0752), as cyan circles (satellites) and grey band (field). The \textbf{top} panel shows absolute metallicities, whereas the logarithmic ratio between satellite and field galaxies is shown in the \textbf{bottom}. The bottom panel also contains \eagle{} predictions for the difference in light-weighted stellar metallicity (dotted lines); see text for details. As in the case of gas-phase oxygen abundance, the \eagle{} simulations qualitatively reproduce the observed enhancement of stellar metallicities in satellite galaxies, albeit not perfectly.}
    \label{fig:stars.total-zmet}
  \end{figure}

In Fig.~\ref{fig:stars.total-zmet}, we compare \eagle{} to the observational data of \citetalias{Pasquali_et_al_2010}. The layout is analogous to Fig.~\ref{fig:gasz} and shows the stellar metallicity of \eagle{} galaxies within $R_\text{2D} = 3\, \text{pkpc}$ as shaded bands (their width again indicating the 1$\sigma$ uncertainty on the median, shown as dashed lines), and those measured from SDSS observations as thin solid lines in corresponding colours, underlined in green. Note that the latter have been adjusted to a solar metallicity of $Z_\odot = 0.012$ \citep{Allende-Prieto_et_al_2001} by multiplying with a correction factor of $0.02/0.012$, i.e.~a (logarithmic) increase of 0.22 dex. The top panel shows metallicities relative to solar, while the bottom panel shows the logarithmic ratio between satellite and field galaxies of similar stellar mass. In contrast to Fig.~\ref{fig:gasz}, we here include all simulated galaxies\footnote{The observational sample selection of \citet{Gallazzi_et_al_2005} is based on spectral S/N $> 20$, but they have shown their results are robust to relaxing this criterion. We have therefore made no attempt to reproduce their sample selection with parameters predicted by the simulation.}, and compute stellar metallicity as the mean mass-weighted total metallicity of the selected star particles (belonging to the subhalo of the galaxy and within $R_\text{2D} \leq 3$ pkpc). 

The comparison yields a qualitatively similar result to that in Fig.~\ref{fig:gasz} for the case of gas-phase oxygen abundance: in general, \eagle{} reproduces the observed excess in metallicity for satellite galaxies compared to equally-massive field galaxies, an effect that is more pronounced for satellites orbiting in more massive haloes (gold). Also reproduced is the increase of stellar metallicity with stellar mass, as already shown by \citet{Schaye_et_al_2015}, albeit with a slope that is too shallow at $\mstar \lesssim 10^{10.5} \msun$ and a normalisation that is slightly too high (by $\sim$0.05 dex at the high-$\mstar$ end).

As with gas metallicity, we explore the impact of weighting variations on the environmental stellar metallicity excess in the bottom panel of Fig.~\ref{fig:stars.total-zmet}. Our fiducial approach, mass-weighting the metallicity of individual star particles (dashed lines) is contrasted with the result using the same aperture, but using r-band light-weighted metallicities (generated using stellar population synthesis (SPS) based on \citealt{Bruzual_Charlot_2003} models; see \citealt{Trayford_et_al_2015}), shown as dotted lines. The impact of this change is non-negligible: using light-, rather than mass-weighted metallicities, the difference between field and cluster satellites is close to zero at $\mstar > 10^{10} \msun$, and \emph{negative} for lower masses; in group satellites the difference is less pronounced, but again weighting by r-band light yields a somewhat smaller environmental difference. Weighting by g- and i-band luminosity instead (not shown), yields qualitatively similar results, with a slightly stronger difference between mass- and light-weighted metallicities with g-band (by $\sim$0.02 dex), and a slightly smaller one in the i-band.

We have also tested for an influence of aperture, by comparing to metallicities averaged within $R_\text{3D} \leq 30$ pkpc (not shown). In contrast to what we found for gas metallicity above, this change only has a small influence on the environmental stellar metallicity excess in \eagle{}, of $< 0.01$ dex at $\mstar \geq 10^{10}\, \msun$.

At face value, the light-weighted metallicity average corresponds more closely to the \citet{Gallazzi_et_al_2005} and \citet{Pasquali_et_al_2010} analysis, since in the real Universe, intrinsically brighter stars contribute more strongly to the integrated spectrum. While the discrepancy between \eagle{} and the observations therefore likely implies a shortcoming on the modelling side, it is less clear at which point exactly the failure occurs: on the one hand, it could be that the environmental metallicity difference is genuinely too small, and only a fortuitous coincidence results in mass-weighted simulation results approximately corresponding to (light-weighted) observational data. On the other hand, it is also conceivable that the observed metallicities are actually reproduced, but the emitted light is not, for example because of shortcomings in the simulated passive galaxy fraction (since the galaxy light is typically dominated by the youngest stars), or the relatively simplistic SPS post-processing that ignores, for example, the influence of dust reddening. As with the impact of galaxy selection on gas-phase metallicity differences, we therefore caution that a quantitative comparison of the simulated and observed stellar metallicity excess in satellites is subject to significant systematic uncertainty (see also \citealt{Guidi_et_al_2015}). However, given the qualitative agreement -- if the difference between mass- and light-weighted metallicity excess is similar in SDSS than in \eagle{}, the observations should \emph{underestimate} the effect of environment in a mass-weighted sense -- it is still meaningful to investigate in more detail the origin of the environmental effect in the simulation, which we will return to in Section \ref{sec:origin_zstar}.

For less massive galaxies ($\mstar \lesssim 10^{10} \msun$), \citetalias{Pasquali_et_al_2010} find a rapidly increasing offset between centrals and satellites, which is driven primarily by a steepening of the mass--metallicity relation in the field. This effect is not reproduced by the \eagle{} Ref-L100 simulation, where stellar metallicities at $\mstar \approx 10^9 \msun$ are consistent with the field in the case of groups (green), or even slightly below it in the case of clusters (red, by $\sim$0.1 dex). As mentioned above, limited numerical resolution may be of significance here (as in Fig.~\ref{fig:gasz}, we conservatively consider the regime shaded in red, $\mstar < 10^{10} \msun$, as unconverged). In principle, more robust predictions can therefore be made from another simulation in the \eagle{} suite, whose mass resolution is a factor of eight better than in Ref-L100. However, computational constraints have limited this simulation (Recal-L025N0752 in the terminology of \citealt{Schaye_et_al_2015}) to a box size of only (25 cMpc)$^3$, i.e.~a factor of 4$^3$ = 64 smaller than the Ref-L100 run. As a result, Recal-L025N0752 contains only one halo on the scale of galaxy groups, with $\mvir \approx 10^{13.2}\, \msun$ and 16 satellite galaxies with $\mstar > 10^{9} \,\msun$. While any conclusion from such a small sample is necessarily only tentative, we nevertheless plot these high-resolution satellites in Fig.~\ref{fig:stars.total-zmet}, as cyan circles; the corresponding field trend is shown in the top panel in grey.

In the higher resolution simulation, the stellar metallicity of satellites is enhanced by $\gtrsim 0.05$ dex even at $\mstar = 10^9 \msun$, with the most extreme satellite having a metallicity that is almost a factor of 3 (0.5 dex) higher than the typical level in the field at its mass; although the small number of satellites precludes robust statistical analyses, the typical enhancement at $\mstar \approx 10^9 \msun$ is around 0.15 dex. While this is higher than in the standard resolution run Ref-L100, it still falls significantly short of the difference found in the SDSS ($\sim$0.3 dex at $\mstar = 10^{9.5} \msun$). Furthermore, the top panel clearly shows that the most extreme offsets are caused by satellites with anomalously high absolute metallicities, whereas in the data of \citetalias{Pasquali_et_al_2010}, it is a rapidly dropping metallicity in centrals that drives the growing discrepancy towards lower mass. In \eagle{}, on the other hand, the slope of the high-resolution field mass--metallicity relation is approximately constant between $\mstar = 10^9$ and $10^{10.5} \msun$ and, although steeper than that of Ref-L100, it is still not quite as steep as observed. We therefore conclude that the stellar metallicities of low-mass satellites constitute a marginally significant tension between \eagle{} and SDSS, a point to which we will return in Section \ref{sec:modelcomp}.

\subsection{Influence of galaxy position within haloes}
\label{sec:radpos}
So far, we have distinguished between satellite galaxies only by the mass of the halo in which they reside. Previous studies have shown that a second parameter which influences the property of satellite galaxies is their position within the halo (e.g.~\citealt{de_Lucia_et_al_2012, Petropoulou_et_al_2012,Wetzel_et_al_2012,Hess_Wilcots_2013}), in the sense that galaxies nearer the halo centre differ more strongly from the field population than those residing at the halo periphery. This is commonly attributed to the general anticorrelation between time since infall and radial position due to dynamical friction, so that galaxies at the smallest radii will typically have been accreted earliest and thus have been affected most by the group/cluster environment \citep{de_Lucia_et_al_2012}. A second contribution is the increasing strength of external influences such as tidal forces or ram pressure acting on galaxies at progressively smaller distances from the group centre. 

\begin{figure}
  \centering
    \includegraphics[width=\columnwidth]{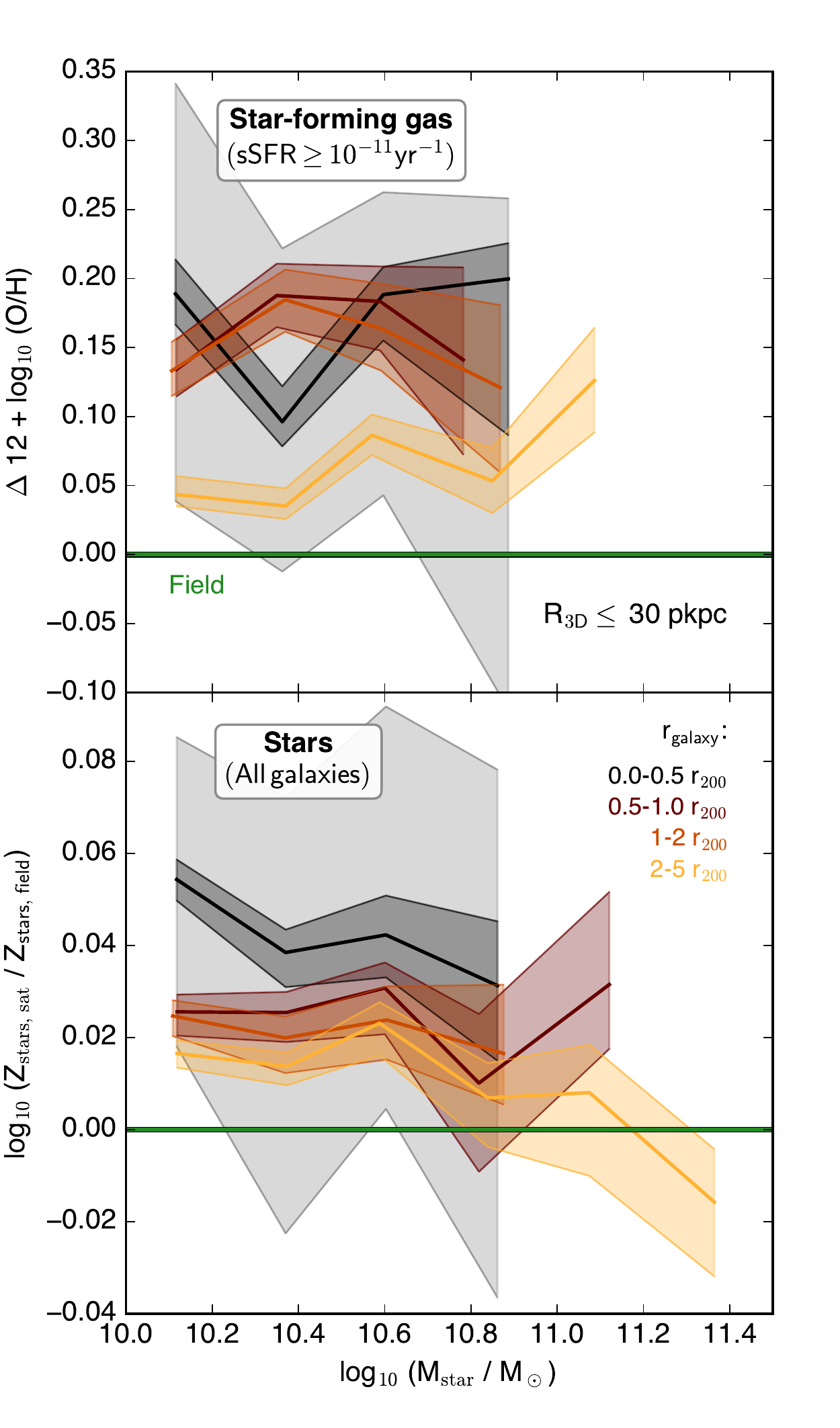}    
       \caption{The satellite metallicity excess of star forming gas (\textbf{top}) and stars (\textbf{bottom}) for galaxies at varying distance from a group or cluster. The solid green line marks the zero level, i.e. the location of field galaxies. No strong radial trend exists for gas metallicity at $r < 2 r_{200}$ (top), whereas stellar metallicity (bottom) is enhanced more strongly for galaxies near the group/cluster centre (black). In both cases, metallicities are enhanced compared to the field even in the far outskirts, at $r > 2 r_{200}$.}
           \label{fig:rrel}
  \end{figure}

In Fig.~\ref{fig:rrel} we explore the impact of halo-centric radius on galaxy metallicity, focusing on oxygen abundance in the star-forming gas phase in the top panel, and stellar metallicity in the bottom. In both cases, metallicities are normalised to the field value at a given stellar mass, and galaxies are now split into four bins according to their distance from the halo centre in units of the halo radius $r_{200}$ as indicated in the top-left corner of the bottom panel; those which are closest to the centre ($r < 0.5 r_{200}$) are shown in black, and galaxies in the far outskirts ($2 \leq r/r_{200} < 5$) in yellow. Note that we here include \emph{all} galaxies in the respective radial ranges, irrespective of whether they are identified as belonging to the FOF halo itself or not\footnote{We have tested that, when only satellite galaxies are considered instead, the radial variation is nearly insignificant out to 5 $r_{200}$. This is likely a consequence of most far-out satellites being members of massive substructures that are linked to the main halo by the FOF algorithm.}, and compute metallicities within a 3D aperture of 30 pkpc radius, as we are not comparing directly to SDSS data. 

Perhaps surprisingly, the predicted effect of halo-centric radius on metallicity is rather small. The oxygen abundance of star-forming gas is significantly higher than in the field (by $\gtrsim 0.1$ dex) even at $r > 2 r_{200}$ (yellow), and is essentially constant at smaller radii ($r < 2 r_{200}$). Stellar metallicities (bottom) exhibit similar behaviour with approximate consistency between the three bins at $r > 0.5\, \rvir$, but a somewhat higher excess of up to 0.06 dex in the innermost bin ($r < 0.5 \rvir$, black). These predictions complement existing evidence for a far-reaching zone of influence around galaxy groups and clusters, both from observations (e.g.~\citealt{Balogh_et_al_1999, von_der_Linden_et_al_2010, Lu_et_al_2012, Wetzel_et_al_2012}) and theory (e.g.~\citealt{Bahe_et_al_2013, Bahe_McCarthy_2015}).

\subsection{Sensitivity to modelling details}
\label{sec:modelcomp}
Our analysis in Section \ref{sec:sdss} above has shown that a robust comparison of the \eagle{} predictions to observational data from the SDSS is subject to non-negligible uncertainties, in particular due to galaxy selection and aperture in the case of star forming gas, and the weighting scheme in the case of stellar metallicities. It is therefore instructive to also compare our results from the \eagle{} Reference simulations to predictions from other recent theoretical models to assess their sensitivity to modelling and parameterisation details, before investigating in more detail their physical origin. We first test different simulations from the \eagle{} suite that vary the AGN and star formation feedback (\S \ref{sec:subgrid}), and then compare to predictions from other simulations (\S \ref{sec:simcomp}).

\subsubsection{\eagle{} subgrid variations}
\label{sec:subgrid}
Besides the ``Reference'' (Ref) model realised in a 100 cMpc box, the \eagle{} simulation suite also includes a range of simulations in which individual features of the galaxy formation model have been varied, as described in detail by \citet{Crain_et_al_2015}. Most of these variation runs were realised only in a (25 cMpc)$^3$ volume and therefore contain only a few satellite galaxies with $\mstar \geq 10^{10}\, \msun$. However, a subset of them was also run in a (50 cMpc)$^3$ volume, which allows for a more meaningful analysis of satellite properties (typically $\gtrsim 100$ satellites with $\mstar \geq 10^{10}\, \msun$). The particle mass of these variation runs is the same as in Ref-L100 ($m_\text{gas} = 1.81 \times 10^6\, \msun$).

Apart from a run with the (fiducial) Ref model, the (50 cMpc)$^{3}$ simulations include three models (`FBConst', `FBZ', and `FB$\sigma$') that vary the scaling of the star formation feedback efficiency. Specifically, what is varied is the fraction of the energy budget available for feedback, $f_\text{th}$, where $f_\text{th} = 1$ corresponds to to the energy available from Type-II supernovae ($10^{51}$ ergs each) resulting from a Chabrier IMF. FBConst uses a constant value of $f_\text{th} = 1$, whereas in FBZ and FB$\sigma$, $f_\text{th}$ is a smoothly varying function of metallicity and local dark matter velocity dispersion, respectively. For further details, the interested reader is referred to \citet{Crain_et_al_2015}. Although, like Ref, all these models match the observed $z=0.1$ galaxy stellar mass function, they consistently produce galaxies that are too compact for $\mstar \gtrsim 10^9 \msun$ \citep{Crain_et_al_2015}. In addition, several runs have varied the parameterisation of AGN feedback, including one model (`NoAGN') that disables it entirely, and one (`AGNdT9') in which AGN heat gas by a temperature increment of $10^9$ K, as opposed to $10^{8.5}$ K in Ref.\footnote{As shown by \citet{Schaye_et_al_2015}, this difference between AGNdT9 and Ref has a significant impact on the gas content of galaxy groups and clusters.}

\begin{figure}
  \centering
    \includegraphics[width=\columnwidth]{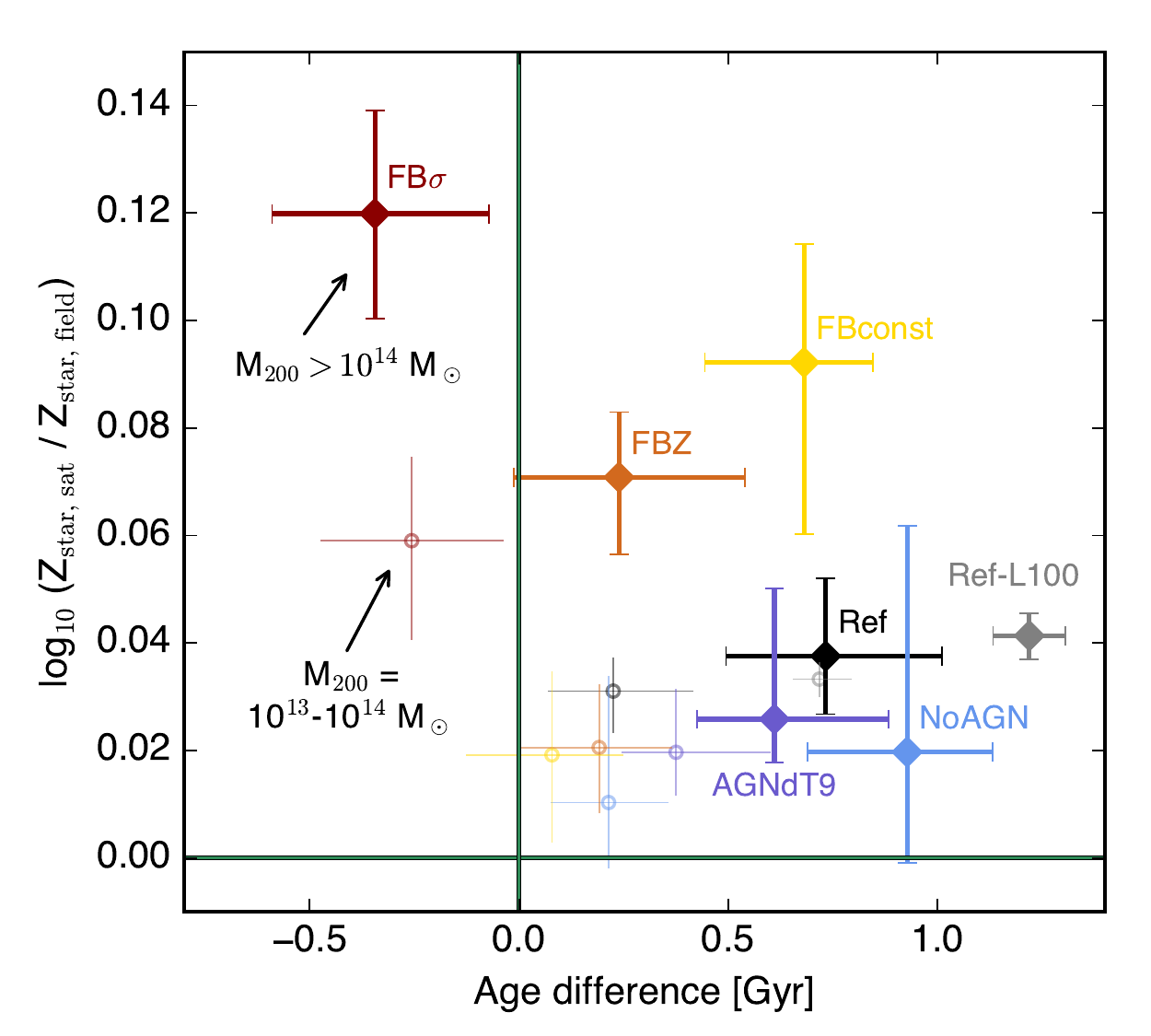}    
       \caption{The difference between satellite and field galaxies in \eagle{} (50 cMpc)$^3$ subgrid variation runs, in terms of mean stellar age (x-axis) and stellar metallicity (y-axis); shown are median values with statistical $1\sigma$ uncertainties indicated by error bars. Large diamonds with thick error bars represent satellites in (one) cluster of $\mvir \approx 10^{14} \msun$, while small open circles and thin error bars denote the prediction for group haloes ($\mvir = 10^{13}$--10$^{14} \msun$). For comparison, the two grey symbols show the corresponding values from the larger (100 cMpc)$^3$ Ref simulation discussed in the rest of this paper.}
           \label{fig:subgridcomp}
\end{figure}

In Fig.~\ref{fig:subgridcomp}, we compare the difference between satellite and field galaxies predicted by these variation runs, in terms of stellar age and stellar metallicity, plotted on the $x$- and $y$-axes, respectively. The motivation for analysing the former is that the metallicity of star forming gas, and hence the stars formed therein, is expected to increase with cosmic time, so that a lower stellar age is expected to correlate with higher metallicity, and vice versa. We do not show the corresponding difference in the metallicity of star forming gas, because -- within the even larger statistical uncertainties arising from the additional restriction that satellites must be star forming -- none of the models we have tested predict gas metallicity differences that deviate significantly from the Ref model at $z = 0.1$.

Given the limited volume of the 50 cMpc variation runs, we bin together all galaxies with $\mstar \geq 10^{10}\, \msun$, and only distinguish two bins in halo mass, $\mvir = 10^{13}$--$10^{14}\, \msun$ (groups) and $\mvir \geq 10^{14}\, \msun$ (clusters; this bin contains only one object with mass just above $10^{14}\, \msun$). In Fig.~\ref{fig:subgridcomp}, the `group' bin is shown as small open circles with thin error bars, whereas the cluster bin is represented by large filled diamonds and thick error bars. Different colours represent different models: Ref is shown in black, the star formation feedback variation runs in shades of yellow/red, and the AGN feedback variation runs in shades of blue. For comparison, we also show the prediction from the Ref-L100  simulation, in grey; the metallicity excesses of the two Ref runs are consistent with each other, while the age excess is significantly smaller in the 50 Mpc simulation, both on a group and cluster scale.

In the two AGN variation runs (blue/purple), both the metallicity and age excess are consistent with the prediction from Ref\footnote{At low significance, the NoAGN model (blue) predicts a smaller metallicity excess than Ref in groups, potentially indicating an importance of AGN feedback on this mass scale}, indicating that AGN feedback is not a significant driver of the environmental differences. However, the star formation feedback variation runs (yellow, red, and orange) \emph{all} predict a stellar metallicity excess on a cluster scale that is larger than in Ref, in particular for the FB$\sigma$ model (+0.08 dex), in which the feedback strength is varied not with the density and metallicity of the ambient gas as in Ref, but the velocity dispersion of the local DM particles. The satellites in FB$\sigma$ are also significantly younger (relative to the field) than in Ref (by 1 Gyr), and even younger than field galaxies in the same simulation (by 0.3 Gyr), which plausibly explains this metallicity offset. The reason might be that the DM velocity dispersion is in part reflecting that of the cluster halo, not the galaxy subhalo, leading to very inefficient feedback \citep{Crain_et_al_2015} that allows star formation in satellites to continue to later times than in the field population.

At a smaller magnitude, the FBZ model (orange, in which the feedback strength is varied with local gas metallicity as in Ref, but not with density, rendering the feedback numerically ineffective in dense regions; \citealt{Crain_et_al_2015}) also predicts younger ages and higher metallicities, but the third variation run (FBconst, yellow) predicts a higher metallicity excess at the same age difference as in Ref. A further investigation would be beyond the scope of this paper, but it seems clear already that the stellar metallicity of satellite galaxies is a potentially powerful diagnostic of feedback scaling prescriptions.

\subsubsection{Galaxy formation models other than \eagle{}}
\label{sec:simcomp}
A complementary test is offered by comparisons to two simulations that do not form part of the \eagle{} suite, and whose modelling techniques vary more significantly than the subgrid variation runs discussed above. The first of these is the Illustris simulation \citep{Vogelsberger_et_al_2014, Nelson_et_al_2015}, and the second the latest version of the Munich semi-analytic galaxy formation model (SAM) introduced by \citet[H15]{Henriques_et_al_2015}. We briefly review their key differences with respect to \eagle{}, before comparing their predictions on the metallicity of satellite galaxies.

Like \eagle{}, Illustris is a cosmological hydrodynamical simulation, with comparable volume ($\sim$100$^3$ cMpc$^3$) and resolution (gravitational softening length $\sim$1 pkpc). One key difference is the hydrodynamics scheme: \eagle{} uses an improved version of the SPH method (Dalla Vecchia in prep.; \citealt{Schaye_et_al_2015}) whereas Illustris is based on the moving mesh code \textsc{Arepo} \citep{Springel_2010a}. A second distinguishing feature is the implementation of energy feedback from star formation. In \eagle{}, a small number of particles is heated to a high temperature \citep{DallaVecchia_Schaye_2012}, with efficiency dependent on the local gas density and metallicity, and without hydrodynamical decoupling or disabled cooling of heated particles. The Illustris model implements feedback in a kinetic way, with wind velocity and mass loading scaled to the local DM velocity dispersion; hydrodynamical forces are temporarily disabled to allow winds to escape from the dense star forming regions \citep{Springel_Hernquist_2003, Stinson_et_al_2006, Vogelsberger_et_al_2013}. In addition, the Illustris model includes an adjustable metal loading factor that specifies the metallicity of winds in relation to the ambient ISM; as discussed by \citet{Vogelsberger_et_al_2013}, this parameter is a key factor behind the relatively good match to the observed mass-metallicity relation. 

In contrast, the H15 SAM is based on the DM-only Millennium Simulation \citep{Springel_et_al_2005}, and takes into account baryonic processes such as gas cooling, star formation, feedback, and chemical enrichment by means of analytic formulae whose free parameters are calibrated with an MCMC technique to reproduce observational data including the abundance and passive fraction of galaxies from $z=3$ to $z=0$ (see also \citealt{Henriques_et_al_2013}). One key advantage of the SAM approach is its reduced computational cost, which allows the simulation of much larger galaxy samples, and hence smaller statistical uncertainties, than what is currently feasible with fully hydrodynamical simulations such as \eagle{} or Illustris: the Millennium Simulation covers a volume of (500 $h^{-1}$ Mpc)$^3$ and includes almost 60,000 groups and clusters with $\mvir \geq 10^{13} \msun$ at $z=0.1$, compared to 154 in \eagle{} Ref-L100.

\begin{figure}
  \centering
    \includegraphics[width=\columnwidth]{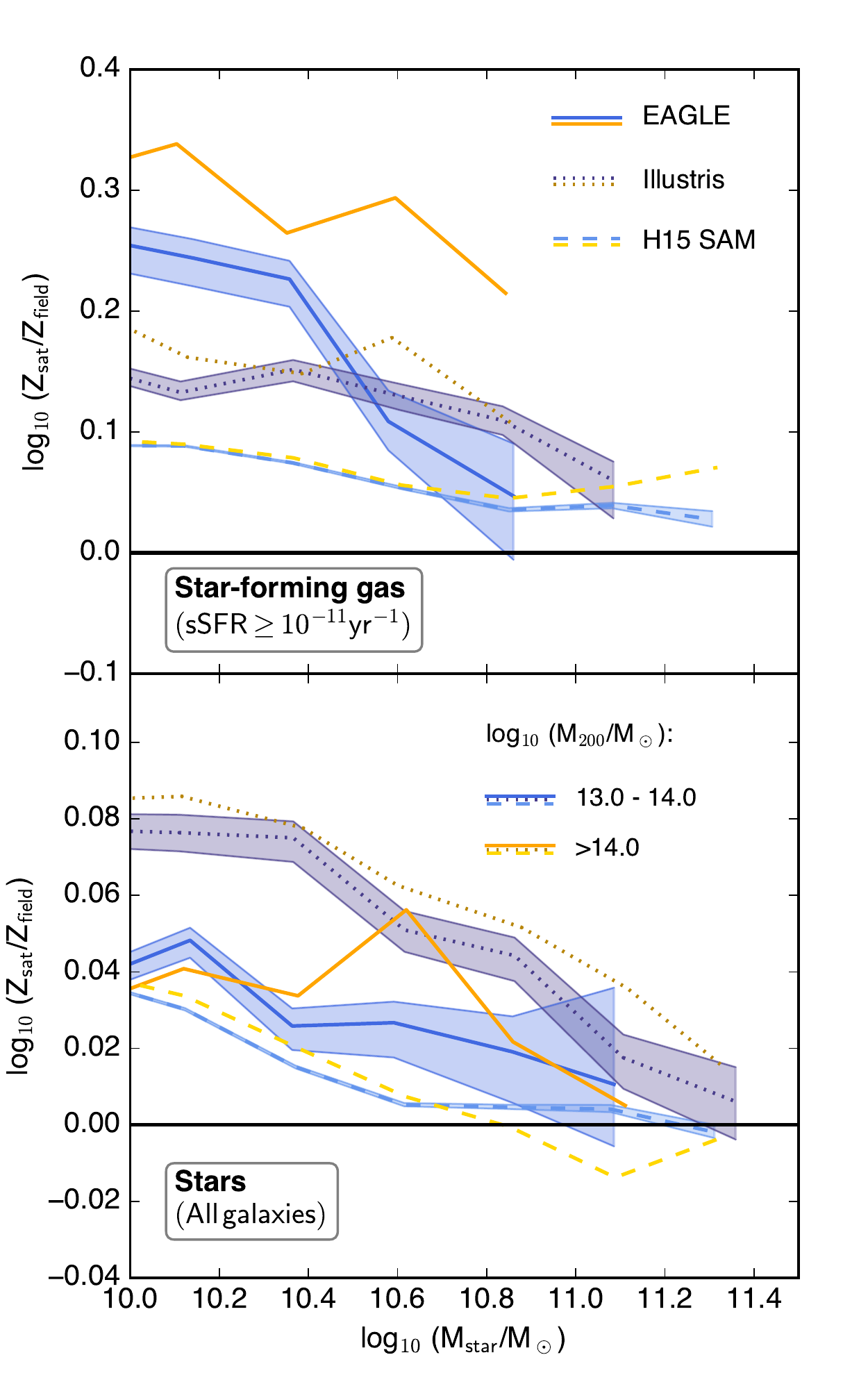}    
       \caption{Comparison of the satellite metallicity excess in three different theoretical models: \eagle{}-Ref (solid lines), the Illustris simulation (dotted), and the semi-analytic model of \citet[H15; dashed]{Henriques_et_al_2015}. Satellites in group haloes ($\mvir = 10^{13}$--$10^{14} \msun$) are shown in shades of blue/purple, shaded bands indicating statistical $1\sigma$ uncertainties, while cluster satellites ($\mvir > 10^{14} \msun$) are represented by orange/yellow lines. The \textbf{top} panel compares predictions for gas metallicity, while stellar metallicity is analysed in the \textbf{bottom} panel. Although all three models broadly agree in predicting a metallicity excess for satellites, there are significant quantitative differences.}
    \label{fig:modelcomp}
  \end{figure}

Predictions of these three models for the excess in metallicity of both star forming gas and stars are compared in the top and bottom panels of Fig.~\ref{fig:modelcomp}, respectively. For \eagle{}, we show mass-weighted metallicities within an aperture of 30 pkpc, whereas for Illustris we take (for simplicity) as gas-phase metallicity the SFR-weighted average over the entire subhalo, and the mass-weighted average within twice the stellar half-mass radius for stellar metallicity, both of which are available from the Illustris \textsc{Subfind} catalogues \citep{Nelson_et_al_2015}. Based on our analysis of \eagle{} above, these differences are not expected to impact significantly on our results. The H15 SAM only makes predictions for the total metal content of cold gas and stars\footnote{In fact, the model distinguishes between the stellar bulge and disc, but for our purpose we simply combine both the total mass and the mass of metals in both components to calculate a mass-weighted overall stellar metallicity.}, respectively, which is what is plotted. For simplicity and consistency, we compare in all cases the predictions for metallicity differences between \emph{all} centrals (most massive subhaloes in a FOF halo), i.e.~not just those far away from groups and clusters, and satellites (subhaloes that are not the most massive one in a group/cluster FOF halo).

Although all three models are in broad qualitative agreement with the observational result of enhanced metallicity in satellites compared to centrals, there are significant quantitative differences. The H15 SAM only predicts a marginal difference between the metallicity of satellites in groups (blue) and clusters (yellow), which is more pronounced in both \eagle{} and Illustris. Likewise, the difference between central and satellite galaxies is generally smallest in the SAM, at a level of 0.09 dex compared to 0.18 dex in Illustris and 0.32 in \eagle{}, for gas metallicity in cluster satellites with $\mstar \approx 10^{10}\, \msun$. Quantitative differences also exist between the two hydrodynamical simulations: \eagle{} predicts a stronger excess in gas metallicity for satellites, especially in clusters (by almost 0.15 dex), whereas the stellar metallicity offset is consistently larger in Illustris. The latter is plausibly connected to the difference in feedback implementation, given that the Illustris prescription is more similar to ``FB$\sigma$'' than to the Reference model of \eagle{}.

As a final remark, we note that none of the three models presented in Fig.~\ref{fig:modelcomp} reproduces the steep increase in the stellar metallicity excess in satellites at $\mstar \lesssim 10^{10.5} \msun$ that is seen in the observational data of \citetalias{Pasquali_et_al_2010}. The increase is strongest at $\mstar < 10^{10} \msun$, where a meaningful comparison to the data is hampered by lack of numerical resolution in case of the \eagle{} Ref-L100 simulation (and plausibly also Illustris), and the small volume in case of the high resolution Recal-L025N0752 run. However, the H15 SAM was also applied to a higher resolution DM-only simulation, Millennium-II \citep{Boylan-Kolchin_et_al_2009}, which is numerically reliable down to $\mstar < 10^9 \msun$, and does still not predict a stellar metallicity excess in satellites of more than 0.05 dex (not shown in Fig.~\ref{fig:modelcomp}). Two potential conclusions from this (tentative) disagreement are discussed in Section \ref{sec:summary}.

We conclude from the comparisons discussed above that predictions about the metallicity offset in satellite galaxies made by current theoretical models are subject to significant systematic uncertainties, in particular due to details in the modelling of star formation feedback, at a level that is comparable to the difference between central and satellite galaxies. Nevertheless, at a qualitative level the prediction of enhanced metallicities in satellites, both in the star forming gas phase and in stars, appears robust. We can therefore still expect to gain relevant qualitative insight into the origin of the metal enhancement in satellites from an in-depth analysis of the \eagle{} Ref-L100 simulation, which is presented in Sections \ref{sec:origin_zgas} and \ref{sec:origin_zstar}, but need to keep these systematic uncertainties in mind.

\subsection{Summary}
The results from this section may be summarised as follows. In qualitative agreement with observations, satellite galaxies in the \eagle{} Ref-L100 simulation exhibit metallicities of both their star forming gas and stars that exceed those in equally massive field galaxies. This difference is somewhat more pronounced for galaxies in more massive haloes and (in the case of stellar metallicity) at smaller halo-centric radii, but already significant for those in poor groups and outside 2 $r_{200}$. Stellar metallicities are sensitive to the adopted efficiency scaling of star formation feedback, and both indicators show significant differences between different theoretical models, although qualitatively the results appear robust.


\section{The drivers of environmental differences in the metallicity of star forming gas}
\label{sec:origin_zgas}
Satellite galaxies may be subject to a multitude of physical processes that could affect, directly or indirectly, the metallicity of their dense star-forming gas. These include the reduction or total cut-off of cosmological accretion \citep{McGee_et_al_2014}, which is expected to dilute the gas reservoir of centrals with metal-poor gas from the inter-galactic medium (e.g.~\citealt{Dave_et_al_2012}), stripping of gas through ram pressure (\citealt{Gunn_Gott_1972, Larson_et_al_1980}), or thermal pressure confinement of galactic gas to prevent metal-rich outflows (\citealt{Mulchaey_Jeltema_2010}, \citetalias{Pasquali_et_al_2012}; but see \citealt{Bahe_et_al_2012}). In this section, we aim to identify which of these effects are key in explaining the elevated metallicity of star forming gas in satellite galaxies. We begin by comparing the radial mass and metallicity profiles of satellite and field galaxies (\S \ref{sec:gas_profiles}), and then analyse the distribution of particle metallicities at fixed radius (\S \ref{sec:gas_histograms}).


\subsection{Metallicity profiles for star-forming gas}
\label{sec:gas_profiles}
\begin{figure*}
\centering
    \includegraphics[width=2.1\columnwidth]{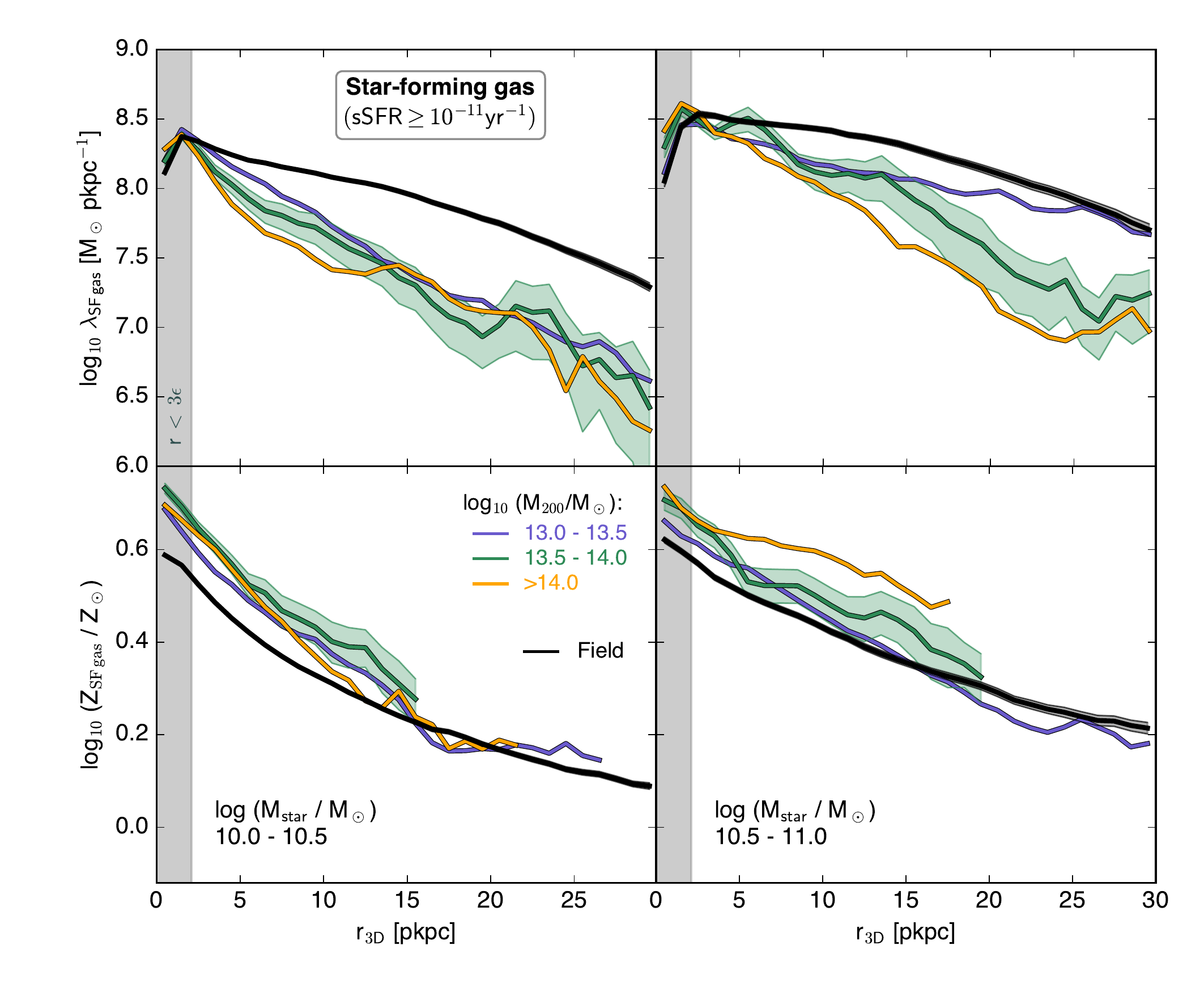}    
       \caption{\textbf{Top row}: mass profiles of star forming gas in the \eagle{} Ref-L100 simulation, for star-forming (sSFR $> 10^{-11}$ yr$^{-1}$) satellite (purple/green/yellow) and field galaxies (black) in two narrow bins of stellar mass. \textbf{Bottom}: the corresponding mass-weighted metallicity profiles of star forming gas, shown for bins in which its total mass exceeds $10^9 \msun$. Shaded bands indicate the $1\sigma$ uncertainty on the mean, and are only shown for the field and one bin in group mass. There is clear evidence for outside-in stripping of star forming gas, but also for enhanced metallicity at a fixed radius within $\sim$15 pkpc.}
    \label{fig:gas_profiles}
  \end{figure*}

It is plausible that satellite galaxies which are still, to some extent, star forming have already lost part of their former gas reservoir, either through direct ram pressure stripping (\citealt{Gunn_Gott_1972}) or unreplenished consumption by star formation and feedback (e.g.~\citealt{Larson_et_al_1980, McGee_et_al_2014}). If gas has been lost predominantly from the galaxy outskirts, where metallicities tend to be lower \citep{Vila-Costas_Edmunds_1992}, this could lead to an increase in galaxy-averaged metallicity. To test this hypothesis, we compare in Fig.~\ref{fig:gas_profiles} the radial mean mass (top row) and mass-weighted mean metallicity profiles\footnote{As we are not directly comparing to observations here, we have chosen to express metallicity here not in terms of the oxygen abundance 12 + log (O/H) as in the top panel of Fig.~\ref{fig:gasz}, but as the mass-weighted fraction of all metals normalised to solar metallicity.} (bottom row) of star-forming gas for field and satellite galaxies within two bins of similar stellar mass (different panels; left: $\log_{10} (\mstar/\msun)$ = [10.0, 10.5], right: [10.5, 11.0]). The profiles combine all galaxies in the appropriate range of $\mstar$ whose sSFR (within 30 pkpc) exceeds $10^{-11}$ yr$^{-1}$. Note that we here distinguish between three bins in halo mass (purple/green/orange lines). Furthermore, in order to highlight differences between the field and satellite populations more clearly, we have chosen to display the mass profiles in the top row not in terms of volume density $\rho$, but mass per unit radius $\lambda$, equivalent to $\rho r^2$. 

Focusing first on the mass profiles (top row), it is evident that even star-forming satellite galaxies in \eagle{} are, on average, depleted in star forming gas compared to the field. This effect shows only mild variation with halo mass, in the sense that the depletion is typically slightly more pronounced for satellites in more massive haloes. It is strongest in the galaxy outskirts, while the densities in the central few kpc are the same as in field galaxies, or even slightly above; ram pressure stripping would explain this `outside-in' loss, because gas in the outskirts is less tightly bound to the galaxy and hence easier to remove. 

Note that all mass profiles exhibit a slight dip within the central $\sim$2 pkpc. This is likely a numerical effect caused by the softening of the gravitational force in the \eagle{} simulations with a (Plummer-equivalent) softening length $\epsilon = 0.7$ pkpc at low redshift. This leads to an unphysical suppression of gas density within $\sim$3$\epsilon$, the range shaded grey in Fig.~\ref{fig:gas_profiles}. The metallicities, however, appear largely unaffected by this, with at best a mild break in the gradient at $r \approx 3\epsilon$.

All galaxies -- field and satellites alike -- show a decline in metallicity with increasing radius, which is marginally steeper in the lower mass bin (-0.5 dex from 0 to 30 pkpc, as opposed to -0.4 dex in the higher mass bin). Observational measurements have similarly found a general anti-correlation of metallicity with galacto-centric radius (e.g.~\citealt{Vila-Costas_Edmunds_1992, Zaritsky_et_al_1994, Ferguson_et_al_1998, Carton_et_al_2015}). The metallicities of satellite galaxies at a given galacto-centric radius are, in general, either similar to what is seen in the field, or moderately higher, by up to $\sim$0.2 dex. An excess is seen particularly in the central galaxy region ($r < 15$ pkpc), while star-forming gas in the outer parts -- in those bins of $\mvir$ and $\mstar$ where enough of it is present to form meaningful metallicity profiles -- is not systematically metal-enriched in satellites, despite the significant removal of gas at these radii. As with the depletion of star forming gas, the metallicity enhancement at fixed radius is typically somewhat stronger in more massive haloes.

Fig.~\ref{fig:gas_profiles} therefore demonstrates that the metallicity of star-forming gas in \eagle{} satellites is raised for at least two different reasons: stripping of (generally metal-poor) gas from the galactic outskirts, but also increased metal abundance at fixed radius near the centre. The former is likely a result of ram pressure stripping \citep{Bahe_McCarthy_2015}; the physical origin of the latter effect is illuminated below.

Finally, we point out that the profiles plotted in Fig.~\ref{fig:gas_profiles} are based on 3D radii, i.e.~they show the mass and metallicity of star-forming gas within concentric shells centred on the potential minimum of each subhalo. For completeness, we have also constructed \emph{projected} profiles based on 2D radii, i.e.~using concentric annuli (not shown). As expected, 2D profiles show slightly lower metallicities near the galaxy centre, due to `dilution' by less metal-rich fore-/background gas, but only by $<$ 0.05 dex. Qualitatively, they agree with the 3D profiles discussed above.

\subsection{Distribution of particle metallicities: which gas is missing?}
\label{sec:gas_histograms}

With the exception of the central few kpc -- which are plausibly affected by the softening of gravitational interactions in the simulation -- Fig.~\ref{fig:gas_profiles} shows that even satellites which are still forming stars are depleted significantly in star forming gas, at least at $\mstar > 10^{10} \msun$. This raises the question whether the increase in metallicity at fixed radius (bottom panel of Fig.~\ref{fig:gas_profiles}) is the result of a preferential absence of low-metallicity gas, or an increased metal-enrichment of the remaining reservoir. 

\begin{figure}
\centering
    \includegraphics[width=\columnwidth]{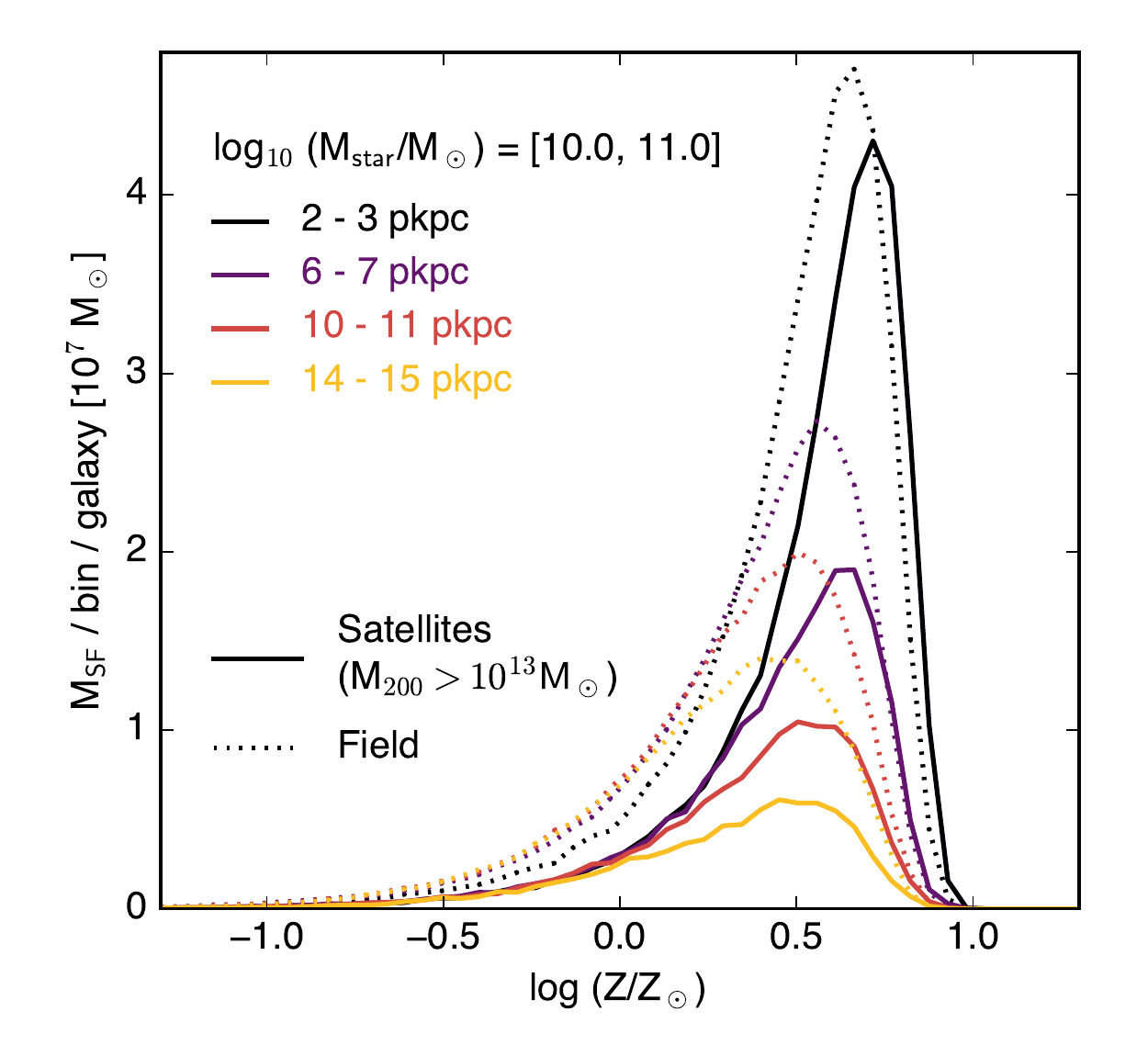}    
       \caption{The distribution of particle metallicities in satellite (solid lines) and field galaxies (dotted lines) of stellar mass in the range $10^{10}\, \msun \leq \mstar < 10^{11}\, \msun$, in four narrow bins of radial distance from the galaxy centre (different colours as indicated near the top left). In each bin, we show the mean mass of star forming gas averaged over all galaxies. At all radii, star forming gas particles have a broad distribution of metallicities in the range $-1.0 \lesssim \log_{10} (Z/Z_\odot) < 1.0$. Satellites are deficient in low-metallicity gas ($\log_{10} (Z/\text{Z}_\odot) \lesssim 0.5$) at all radii, and show an increased abundance of metal-rich gas at small radii (purple/black lines).}
    \label{fig:gas_zmet_histograms}
  \end{figure}

To distinguish between these two scenarios, we plot in Fig.~\ref{fig:gas_zmet_histograms} the mass-weighted metallicity distribution of star forming gas particles, in galaxies with sSFR $> 10^{-11}$ yr$^{-1}$. To eliminate biases arising from different radial distributions in field and satellite galaxies, we concentrate on four narrow radial bins 1 pkpc in width, beginning at a distance of 2, 6, 10, and 14 pkpc from the galaxy centre; these are shown with different colours in Fig.~\ref{fig:gas_zmet_histograms}. We focus on galaxies in the mass range $\log_{10} (\mstar/\msun) = [10.0, 11.0]$, and do not differentiate between satellites in haloes of different mass (as Fig.~\ref{fig:gas_profiles} shows, the differences between different halo mass bins are generally smaller than the overall offset between field and satellites). Field galaxies are plotted as dotted lines, whereas satellites are represented with solid lines in corresponding colours. Note that we here show (non-smoothed) \emph{particle} metallicities, because they are more directly connected to the individual particle histories. 

In both field and satellite galaxies, individual particles cover a wide range of metallicities at $z=0.1$, from $\sim$0.1 to 10 Z$_\odot$ with a peak around log$_{10} (Z/\text{Z}_\odot) \approx 0.5$. This spread is significantly larger than the systematic variation of metallicity with radius, but on closer inspection it is evident that, in field galaxies, both the peak of the distribution shifts slightly (by $\sim$0.2 dex) towards lower metallicities from the innermost to the outermost bin, and that the occurrence of relatively low-metallicity gas ($Z \lesssim 10^{0.2}\, \text{Z}_\odot$) is lowest in the central bin (black). 

The difference between satellites and field galaxies is most pronounced for gas of relatively low metallicity ($Z \lesssim 10^{0.5}\, \text{Z}_\odot$), which is strongly deficient in satellites at all radii, by factors of typically 2--3. This depletion is \emph{much} stronger than the radial variation of metallicity distributions in the field: even in the 2-3 pkpc bin (black solid line), the abundance of $Z < 10^{0.4}\, \text{Z}_\odot$ gas in satellites is below that at 14-15 pkpc in the field (yellow dotted); this clearly shows that it is not a residual bias from the preferential removal of gas at larger radii within individual bins. In contrast, high-metallicity gas is depleted much less severely, and actually \emph{exceeds} the abundance in the field in the central two bins (black and maroon coloured lines). As a consequence, both the peak and the median of the distribution is shifted to higher metallicities. The retention of high-metallicity gas is least strong in the outermost bin (yellow), so that the overall metallicity enhancement is also smallest there.

In principle, it is possible that the depletion of low-metallicity gas, and increased abundance at the high-metallicity end, are two effects of the same process, namely gas being enriched more strongly in satellites than in the field: with a reduced reservoir of star forming gas, metals ejected by stars are swept up by less gas, which is therefore enriched more rapidly (see also \citealt{Segers_et_al_2016a}). However, this explanation is not only inapplicable to the outermost bins -- where no enhancement at the high-metallicity end is seen -- but also in the centre, where the depletion of low-metallicity gas is far stronger than the excess at high metallicity. The rather indiscriminate removal of gas from the outskirts (yellow) is most naturally explained by gas stripping, whose efficiency is unaffected by gas metallicity. Closer to the centre, however, a dominant role of stripping is difficult to reconcile with the substantially unaffected population of metal-rich particles. 

It is conceivable that low-metallicity star-forming gas might be easier to strip than metal-rich gas, for two reasons. First, if its density were lower than that of high-metallicity gas, so would be the gravitational restoring force \citep{Gunn_Gott_1972}. However, we have tested this and found no such correlation between metallicity and density of star-forming gas in our simulation. Secondly, the efficiency of star formation feedback in the \eagle{} Reference model is higher in low-metallicity gas to account for smaller (physical) cooling losses \citep{Crain_et_al_2015}, which could plausibly enhance the efficiency at which this gas is stripped by ram pressure (see \citealt{Bahe_McCarthy_2015}). It is difficult to conclusively assess the significance of this second effect. However, the fact that the difference in metallicity distributions between field and satellites at $Z \lesssim \text{Z}_\odot$ -- where this effect of `feedback-assisted stripping' should be strongest -- are similar in all four radial bins shown in Fig.~\ref{fig:gas_zmet_histograms} suggests that its role is not dominant, because ram pressure stripping should be more effective at larger radii. The depletion of low-metallicity gas evident in Fig.~\ref{fig:gas_zmet_histograms} is therefore most easily interpreted as \emph{caused by suppression of metal-poor inflow of gas into the inner galaxy}, as expected in the `strangulation' scenario. The same process is also a plausible contributor to the enhanced abundance of high-metallicity gas, because the remaining gas inflow is itself expected to be of higher metallicity than in isolated galaxies due to preferential removal of less dense metal-poor gas from the galaxy halo. 

An alternative interpretation for the higher metallicity of the most metal-rich gas in the galaxy centres is that it results from assembly bias (e.g.~\citealt{Gao_et_al_2005, Zentner_et_al_2014}), i.e.~the typically earlier formation of galaxies in and around massive haloes. As a result, the stellar population of satellite galaxies will, at a given time, be more evolved, and hence more metal-enriched, than in equally-massive field galaxies, leading to stronger metal injection  into the star forming ISM through stellar outflows. However, Fig.~\ref{fig:gas_zmet_histograms} suggests that such an effect is sub-dominant to the direct environmental influence of gas stripping and suppression of metal-poor inflows.

\subsection{Summary}
The results of our investigation into the origin of the metallicity enhancement of star forming gas in satellite galaxies may be summarised as follows. The metal enhancement can mostly be attributed to two distinct physical processes: the first is ram pressure stripping of gas from the outer part of the star forming disk, whose metallicity is generally lower than that of gas nearer the galaxy centre. The second effect is a marked reduction of metal-poor inflows into the inner galaxy part, itself plausibly a consequence of the aforementioned ram pressure stripping of gas from the outer disk. A third, though minor, contributor is a stronger enrichment of the most metal-rich gas in the galaxy centre due to continued star formation in a depleting gas reservoir and the increased contribution of stellar ejecta \citep{Segers_et_al_2016a}.


\section{The drivers of the excess metallicity in satellite stars}
\label{sec:origin_zstar}

We now turn to analysing the origin of the enhanced \emph{stellar} metallicity in satellite galaxies. First, we test for correlations between the star formation activity and metallicity enhancement in satellites (\S \ref{sec:zstar_ssfr}), and then compare the metallicity of equally old stellar populations in satellites and the field (\S \ref{sec:zstar_age}). The effect of stellar mass stripping is investigated in \S \ref{sec:stripping}. Finally, we test the extent of differing birth conditions for stars in satellite and field galaxies in \S \ref{sec:birth}.

\subsection{Stellar metallicity of star forming and passive galaxies}
\label{sec:zstar_ssfr}

The properties of stars are naturally connected to the star formation history of a galaxy, which motivates an analysis of how the stellar metallicity excess in satellites depends on the $z=0.1$ sSFR. We have therefore split the galaxy sample into star forming and passive galaxies; in order to obtain a clear separation between these two, we adopt the stricter threshold of sSFR $> 10^{-10.5}$ yr$^{-1}$ for the former, and likewise sSFR $< 10^{-11.5}$ yr$^{-1}$ for the latter and consider `transitional' galaxies with sSFR between these two values separately. To counter the reduction in galaxy numbers resulting from this split by sSFR, satellites in both groups and clusters are combined into a single bin covering halo masses of $\mvir > 10^{13}\, \msun$. In all cases, we compute metallicities as mass-weighted mean of all subhalo star particles within an aperture of $R_\text{3D} \leq 30$ pkpc.

\begin{figure}
   \centering
   \includegraphics[width=\columnwidth]{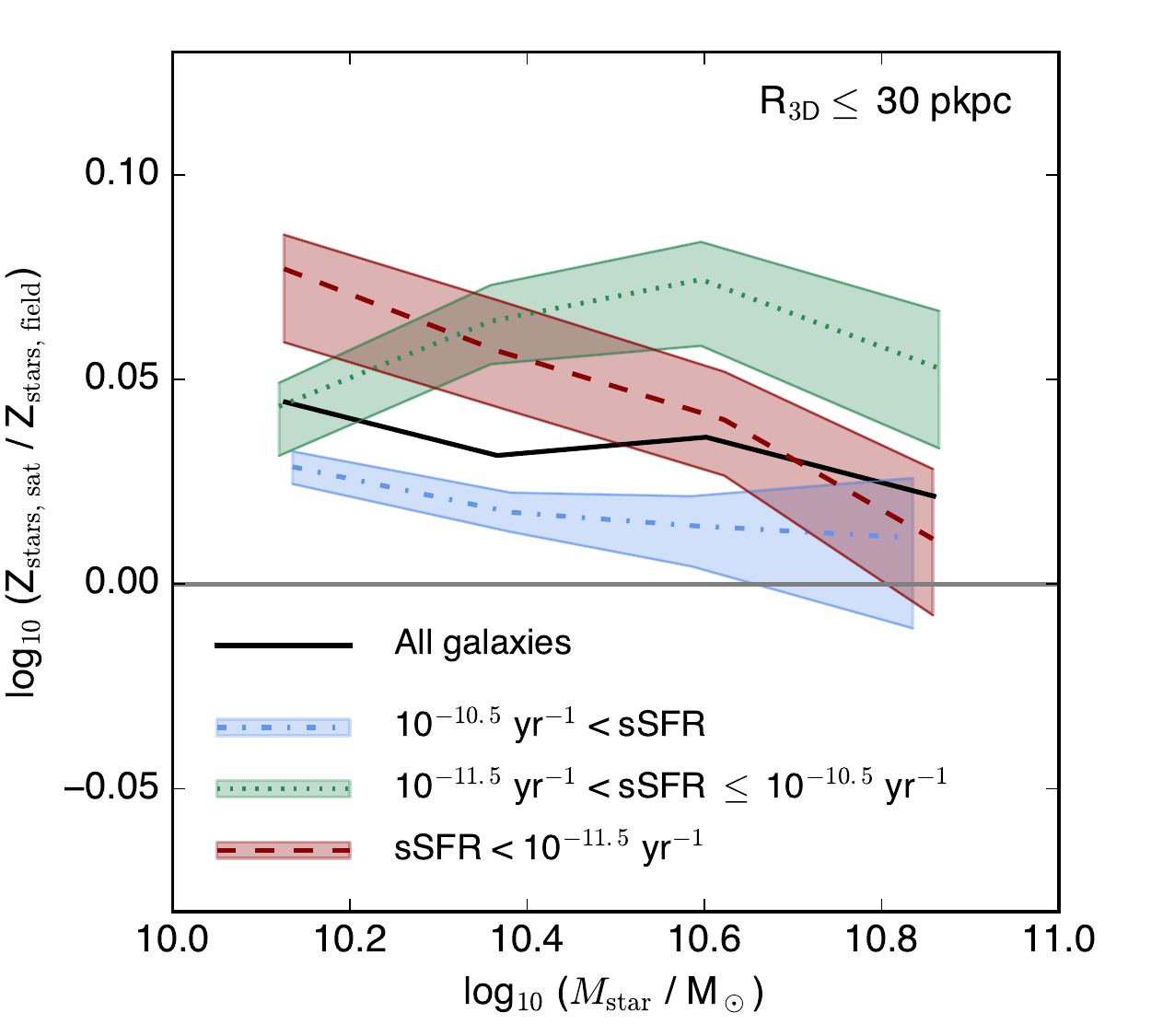}
   \caption{Excess stellar metallicity in satellites compared to the field for star forming (sSFR $> 10^{-10.5}$ yr$^{-1}$, blue dash-dot), passive (sSFR $\leq 10^{-11.5}$ yr$^{-1}$, red dashed), and transitional ($10^{-11.5}$ yr$^{-1} < \text{sSFR} \leq 10^{-10.5}$ yr$^{-1}$, green dotted) galaxies. Shaded regions indicate $1\sigma$ uncertainties on the median; the scatter between individual galaxies (not shown) is several times larger. For reference, the excess in the overall galaxy population (not split by sSFR) is shown as a solid black line. In qualitative agreement with SDSS data, \eagle{} predicts a less severe metallicity excess for star forming satellites. The largest offset is predicted for transitional galaxies.}
    \label{fig:zstar_ssfr}
\end{figure}

The result is plotted in Fig.~\ref{fig:zstar_ssfr}, where star forming, transitional, and passive galaxies are represented by blue dash-dot, green dotted, and red dashed lines, respectively. For ease of comparison, we also include the metallicity excess derived from the full galaxy population, without a split by sSFR, as solid black line. It is evident that the sSFR is indeed correlated with the stellar metallicity excess in satellites: star forming galaxies (blue) show a smaller excess than the full population, in agreement with a similar result obtained from SDSS data by \citetalias{Pasquali_et_al_2012}. While the difference for passive galaxies (red) is broadly consistent with the full sample (albeit with a moderately steeper decline with stellar mass), the perhaps most surprising feature of Fig.~\ref{fig:zstar_ssfr} is the prediction of a consistently stronger environmental effect on the metallicity of transitional galaxies (green), of up to 0.08 dex. We note, however, that the scatter between individual galaxies (not shown) is several times larger than the difference between the median trends.

Although it could, in principle, be possible that all three sSFR bins exhibit smaller metallicity differences than the combined population due to different relative contributions, Fig.~\ref{fig:zstar_ssfr} demonstrates that this is clearly not the case. This implies that the increase in stellar metallicity is not simply the consequence of an enhanced passive fraction amongst satellites, but is the result of an \emph{environment-specific} process. Since this is clearly more effective in transitional than star forming galaxies, it is furthermore likely that the enhancement of stellar metallicity is directly related to the removal of star forming gas, and hence -- following our conclusions from the previous section -- also to the enhancement in \emph{gas} metallicity. This hypothesis is tested in more detail below. A second implication is that star formation quenching itself is driven by different factors in field and satellite galaxies, at least within the \eagle{} galaxy formation model.

\subsection{Accounting for the effect of stellar age}
\label{sec:zstar_age}

In contrast to the metallicity of gas particles, which can change throughout the simulation, the metal content of a star is fixed at the epoch of its birth. The metallicity of the stellar component therefore provides an `archaeological' record of the conditions in the star forming gas across cosmic time, and can therefore give clues to the past evolution of a galaxy. Since the ISM is gradually enriched with heavy elements over time, it is expected that younger stellar populations exhibit higher metallicities, and vice versa. However, \citetalias{Pasquali_et_al_2010} have shown that SDSS satellites are both more metal-rich and \emph{older} than centrals of the same stellar mass; as we have shown in Fig.~\ref{fig:subgridcomp}, this age difference is qualitatively reproduced by \eagle{}.    

\begin{figure}
  \centering
    \includegraphics[width=1.02\columnwidth]{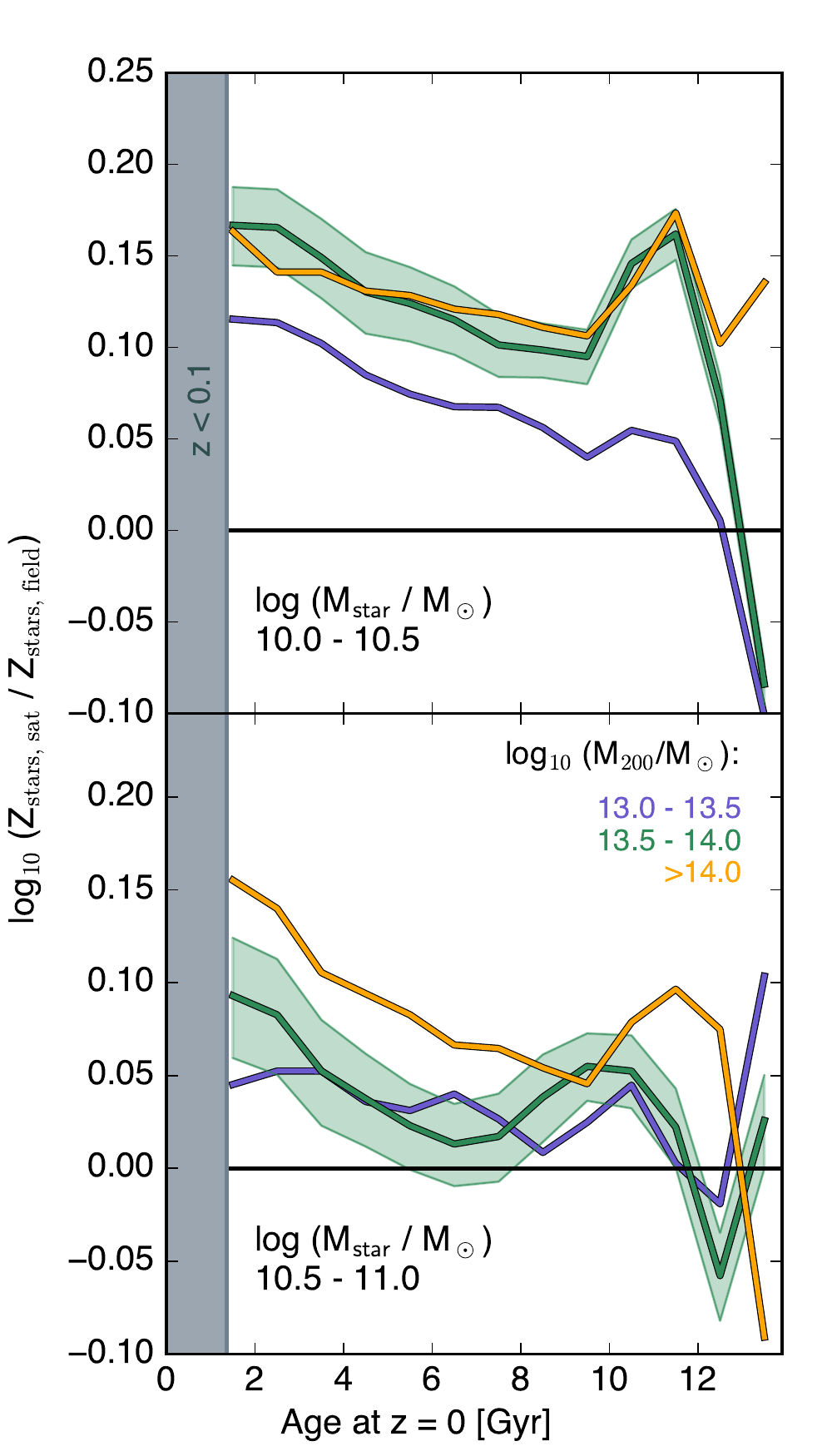}
       \caption{Metallicity excess of stars in simulated satellite galaxies relative to the field, split by galaxy stellar mass (different panels, see bottom-left corners). Along the $x$-axis, stars are separated by their $z=0$ age. In general, stars of a given age are significantly metal-enriched in satellites compared to the field, an effect that persists even to stars formed at $z > 2$ (age $>$ 10 Gyr).}
    \label{fig:stars.age-zmet}
\end{figure}

The age difference cannot therefore be the cause of the excess in metallicity, and instead reduces its intrinsic magnitude. To account for this age bias, we can exploit the fact that both field and satellite galaxies are comprised of multiple stellar populations of different age, and compare stellar metallicities between \emph{equally old} stellar populations in both sets. The result is shown in Fig.~\ref{fig:stars.age-zmet}, where we split galaxies into two panels according to their stellar mass, and plot the (stellar) metallicity excess in satellites relative to the field as a function of the time of star formation, expressed here as the age of the star particle at $z=0$. Note that, to connect to the analysis in the previous sections, we continue to analyse the \eagle{} snapshot at $z=0.1$, which leads to the lack of data points at ages $\lesssim$ 1 Gyr.

Comparing stellar populations at the same age in this way reveals an almost ubiquitous enhancement of stellar metallicity in satellites, at a level that is typically higher than the $\sim$0.04 dex offset obtained without accounting for age differences (see Fig.~\ref{fig:stars.total-zmet}): for the youngest stellar populations, satellites are metal-enriched by up to $0.16$ dex, in line with expectations from the metallicity difference in the star forming gas (see Fig.~\ref{fig:gasz}). Perhaps more surprising is the fact that the difference between field and satellites persists even to stars formed at $z > 2$: at this high redshift, only a small fraction of galaxies that are satellites today were already part of the still assembling group or cluster halo (e.g.~\citealt{Wetzel_et_al_2013, Bahe_McCarthy_2015}). At face value, Fig.~\ref{fig:stars.age-zmet} would therefore suggest that galaxies destined to become a satellite at late times were already special in the early Universe, even before they could be shaped by their environment. However, we show below that this interpretation does not hold up to closer scrutiny.

\subsection{Effect of mass loss in satellites}
\label{sec:stripping}

\citetalias{Pasquali_et_al_2010} suggest that the excess stellar metallicity observed in satellite galaxies can be explained as an indirect effect driven by stellar mass stripping from satellites. Because stellar metallicity reflects the conditions at the time a star is born, rather than when it is observed, late-time stellar mass loss driven by tidal stripping is not expected to reduce a galaxy's metallicity\footnote{The average galaxy metallicity may increase somewhat if material that is stripped is preferentially of low metallicity, e.g.~from the galactic outskirts.}, but does reduce its stellar mass. Combined with the underlying trend towards higher metallicity in more massive galaxies, this would naturally explain the enhanced metallicity of satellites. We now test to what extent this scenario is corroborated by the \eagle{} simulation.

From Fig.~\ref{fig:stars.total-zmet}, a field galaxy with $\mstar \approx 10^{10.5} \msun$ would need to reduce its mass by $\sim$0.4 dex to move from the field mass--metallicity relation to that of satellites; \citetalias{Pasquali_et_al_2010} inferred a requirement for a similarly strong mass loss from their observational analysis. To test whether \eagle{} satellites actually experience such strong mass loss, we use the galaxy progenitor information derived as described in Section \ref{sec:tracing}. For each galaxy identified at $z=0.1$, we look up the stellar mass of its main progenitor in previous snapshots, and then find the maximum of these values ($M_\text{star}^\text{max}$) and the redshift at which it is reached ($z_\text{max}$).

\begin{figure}
  \centering
  
      \includegraphics[width=\columnwidth]{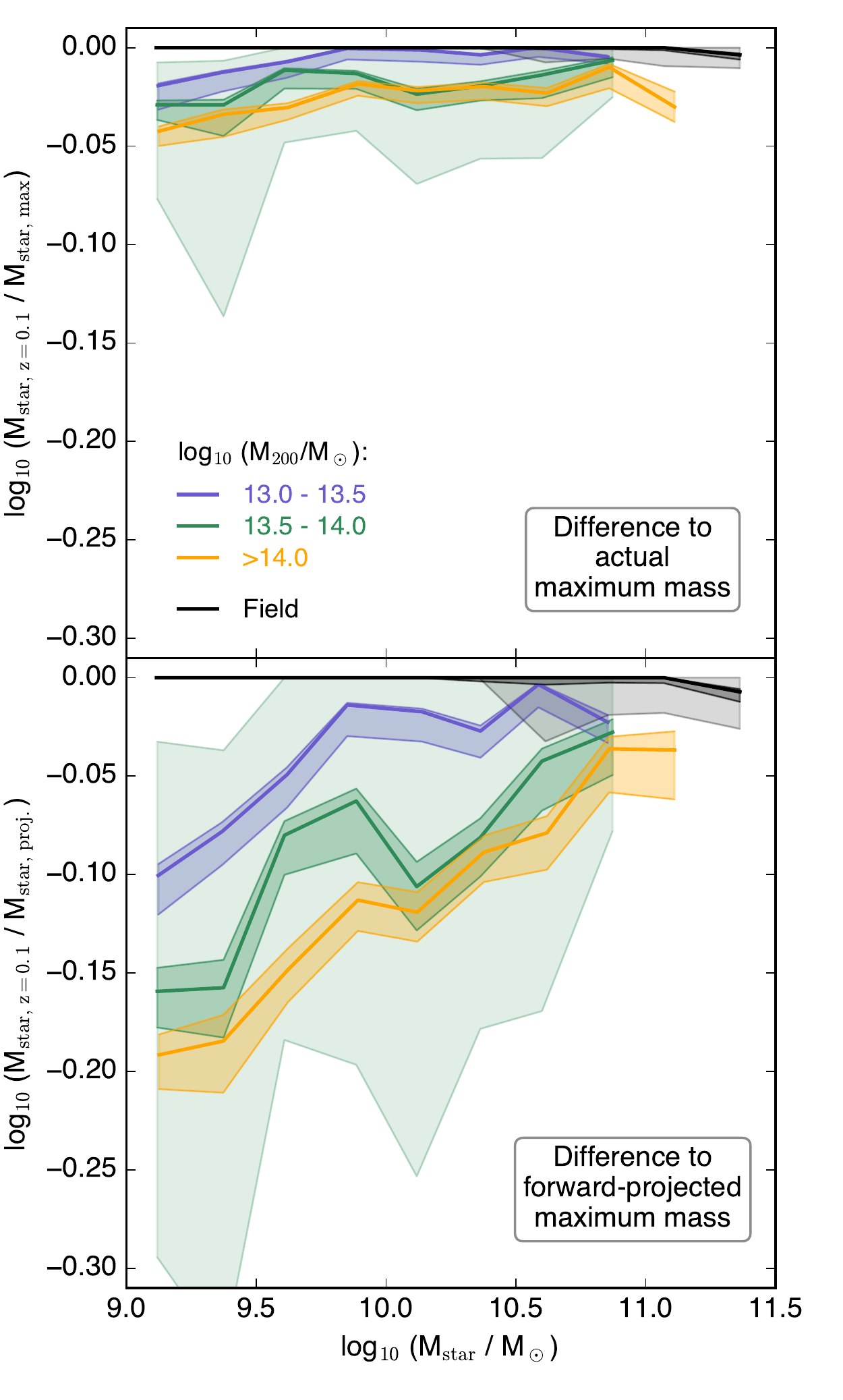}
       \caption{Effect of stellar mass loss in simulated galaxies. \textbf{Top:} difference between the stellar mass at $z=0.1$ and the maximum stellar mass of the galaxy's main progenitor (at $z_\text{max}$), i.e.~the amount of (net) mass loss since $z_\text{max}$. \textbf{Bottom:} difference between $\mstar (z=0.1)$ and the ``forward-projected'' maximum stellar mass, which takes into account the additional mass growth in field galaxies at $z < z_\text{max}$. The actual stellar mass loss of surviving satellites in \eagle{} is typically small ($\lesssim 0.05$ dex), and even when the missed growth is taken into account (bottom panel), the effect is not sufficient to account for the discrepancy in stellar metallicities between the field and satellites (see text for details).}
    \label{fig:mstar_diff}
\end{figure}

The logarithmic ratio between $M_\text{star}^\text{max}$ and the mass at $z=0.1$ is shown in the top panel of Fig.~\ref{fig:mstar_diff}. Because this analysis does not involve any metallicity measurement, and may be of relevance to other galaxy properties as well, we include galaxies with stellar masses down to $\mstar = 10^9\, \msun$. As expected, field galaxies show essentially no net mass loss, indicating that they have grown continuously throughout cosmic history despite the continuous mass loss from individual star particles to model the effect of stellar winds (\citealt{Wiersma_et_al_2009b}; see also \citealt{Segers_et_al_2016a}). In contrast, the current mass of satellites is typically somewhat lower than their maximum, particularly for the lowest mass galaxies ($\mstar \approx 10^9\, \msun$) in clusters ($\mvir \approx 10^{14}\, \msun$). The median mass loss reaches $\sim$0.05 dex, and a non-negligible fraction of galaxies (25 per cent) may lose in excess of 0.1 dex of their stellar mass, as indicated by the light shaded green band enclosing 50 per cent of galaxies in the middle halo mass bin ($\mvir = 10^{13.5}$--$10^{14}\, \msun$). This loss is due to a combination of tidal stripping and mass loss from stellar winds, with a larger contribution from the second effect. It is evident that the combination of these two mechanisms does not lead to mass loss as large as required to explain the stellar metallicity excess in satellites (see also \citealt{Barber_et_al_2016}).

There is, however, a second effect that also needs to be considered: while satellite galaxies experience a (mild) reduction in their stellar mass, field galaxies continue to grow. If the relation between stellar mass and metallicity is driven primarily by the varying efficiency of outflows at removing metals from galaxies with different mass (\citealt{Tremonti_et_al_2004, Gallazzi_et_al_2005}) then the more fundamental galaxy parameter determining its metallicity is its total (halo), rather than stellar, mass (see also \citealt{Gallazzi_et_al_2006}). It is therefore necessary to account for the additional late-time growth of field galaxies in order to remove mass-induced biases in the comparison between field and satellite galaxies. To accomplish this, we identify for each galaxy with $z_\text{max} > 0.1$ a set of similar field galaxies in the snapshot at $z_\text{max}$ (specifically, those differing in stellar mass by $|\Delta \mstar | < 0.1$ dex) and compute the median $z=0.1$ stellar mass \emph{of these matched field galaxies} as a hypothetical mass that the satellite would have reached had it remained a field galaxy. Below, we will refer to masses obtained in this way as ``forward-projected maximum'' stellar masses.

As the bottom panel of Fig.~\ref{fig:mstar_diff} shows, the effect of the missed growth of satellite galaxies is considerably larger than that of mass loss alone. Compared to field galaxies of similar initial evolution, satellites today typically fall short by $\sim$0.1--0.2 dex in stellar mass at the low-mass end, and by more than 0.3 dex for the most affected quartile. At higher masses, however, the discrepancy eases, and is no larger than 0.1 dex for a typical Milky Way analogue ($\mstar \approx 10^{10.5} \msun$). Although the effect of mass deficiency is clearly non-negligible, it is therefore not strong enough to fully account for the increase in stellar metallicity of satellite galaxies, as we have verified by repeating the analysis in Fig.~\ref{fig:stars.age-zmet} with field and satellites matched by forward-projected maximum stellar mass (not shown).

\subsection{Birth conditions of stars}
\label{sec:birth}

We have shown above that differences in the mass evolution of field and satellite galaxies cannot explain the raised stellar metallicities in the latter. As a final indirect effect, we now test the influence of different merger histories between the two populations. It is well-established that only part of the stars inhabiting a galaxy at low redshift were formed in the galaxy's main progenitor itself, with the rest having been accreted from smaller galaxies through mergers (e.g.~\citealt{van_Dokkum_et_al_2010, Oser_et_al_2010, Font_et_al_2011, Lackner_et_al_2012, D_Souza_et_al_2014, Rodriguez-Gomez_et_al_2015}).

\begin{figure}
  \centering
  
      \includegraphics[width=\columnwidth]{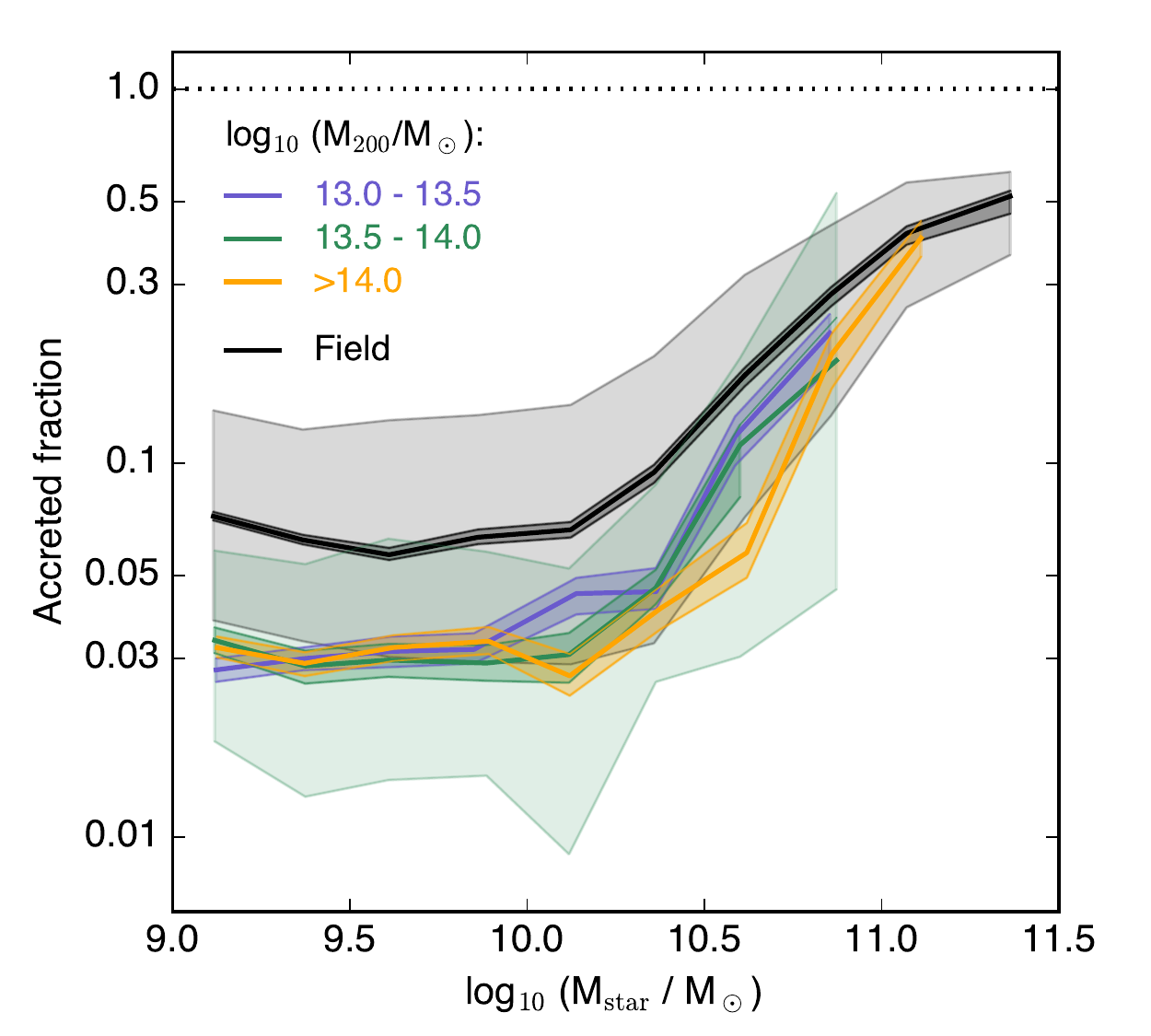}
       \caption{The fraction of stellar mass of \eagle{} galaxies that was accreted from other (smaller) galaxies. The accreted fraction increases strongly with the mass of the galaxy at $\mstar \gtrsim 10^{10} \msun$, and is also systematically lower in satellites than equally massive field galaxies.}
    \label{fig:accreted_fraction}
\end{figure}

From our galaxy progenitor histories, we determine the subhalo to which each star particle belonged in the first snapshot after its formation. If this subhalo was the main progenitor of the subhalo hosting the particle at $z=0.1$, we identify the star as `born in-situ', and otherwise as `accreted'\footnote{In principle, `accreted' star particles can be born in a subhalo that is neither the main nor a minor progenitor of the subhalo hosting the star at late times, for example if a star particle is stripped from an infalling, but surviving, satellite and accreted by the central galaxy in the halo. In \eagle{}, such `stolen' stars contribute at most a few per cent even in massive galaxies, with negligible environmental variation. Their contribution is therefore not considered separately here.}. In Fig.~\ref{fig:accreted_fraction}, the accreted mass fraction of stars in \eagle{} galaxies is shown as a function of stellar mass, split between field galaxies and satellites in haloes of different mass. For both sets of galaxies, the accreted fraction is relatively small ($< 10$ per cent) at stellar masses $\mstar < 10^{10} \msun$, and increases with $\mstar$ for more massive systems, up to $\sim$50 per cent in the most massive galaxies. This is qualitatively similar to results reported by \citet{Oser_et_al_2010} and \citet{Lackner_et_al_2012} based on older simulations.\footnote{\citet{Oser_et_al_2010} found an accreted fraction as high as 80 per cent for $\mstar \gtrsim 2 \times 10^{11} \msun$, whereas this fraction only reaches $\sim$40 per cent in the simulations analysed by \citet{Lackner_et_al_2012}. The difference is plausibly due to differing strength of feedback from star formation in the different simulations analysed by these authors.}. The distribution of accreted and in-situ stars in central \eagle{} galaxies will be analysed in greater detail by Qu et al. (in prep.).

Besides the trend with stellar mass, Fig.~\ref{fig:accreted_fraction} also shows a clear influence of environment on the accreted fraction, in the sense that accretion contributes less stellar mass to satellites; we note that this is in contrast to what \citet{Lackner_et_al_2012} concluded from a similar analysis of their simulations. At $\mstar \lesssim 10^{10}\, \msun$, this effect is nearly independent of host halo mass, with an accreted fraction of $\sim$3 per cent in satellites compared to $\sim$7 per cent in the field. In more massive galaxies, Fig.~\ref{fig:accreted_fraction} hints at accretion of stars being more strongly suppressed in more massive haloes, but the relatively small number of galaxies precludes a more robust conclusion. One possible origin of this environmental difference is that the strong tidal forces within massive haloes increase the efficiency of mergers between satellites and the central galaxy, rather than mergers between two satellites (see also \citealt{Moreno_et_al_2013}); another factor is the aforementioned bias of satellite galaxies to older ages. The potentially important connection of this difference to the stellar metallicity of satellite galaxies is that accreted stars were born in less massive galaxies than the main progenitor, and are therefore generally more metal-poor. Differences in the accretion efficiency can therefore lead to different stellar metallicities at late times, even for stars that were born in the early Universe.

We account for this potential bias by explicitly considering the (stellar) mass of the galaxy in which a star (particle) was born. First, we bin all stars that belong to field galaxies at $z = 0.1$ into a 2D-grid by birth mass (bin size 0.06 dex) and stellar formation time (bin size 500 Myr), and calculate the mass-weighted average metallicity of star particles in each of these cells. We then assign to each star particle residing in a satellite galaxy as its `field-equivalent' metallicity the average in its respective grid cell. By comparing these to the actual metallicity of satellite stars, we can test whether the differences seen above are indeed explicable by the indirect effects of different mass growth.

\begin{figure}
  \centering
  
      \includegraphics[width=1.02\columnwidth]{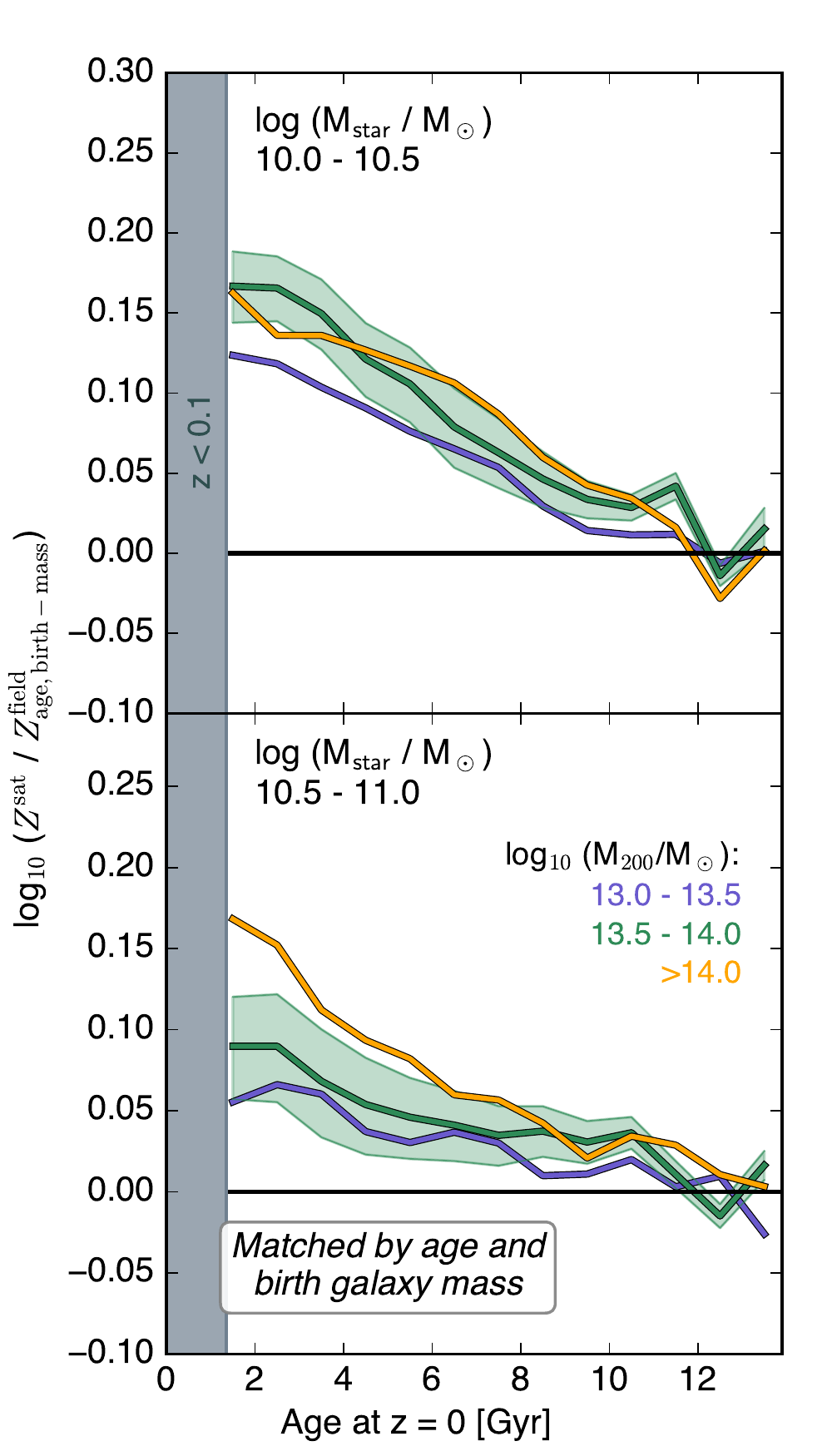}
       \caption{Metallicity excess in satellite galaxies compared to field star particles formed at similar times in similarly massive galaxies. The two panels show galaxies of different z = 0.1 stellar mass as indicated in the top left, the different lines represent satellites in differently massive haloes. The black line (zero offset) represents field galaxies. The metallicity excess in satellites that cannot be attributed to differing age and birth mass vanishes for the oldest stars, and increases steadily towards later times (towards the left).}
    \label{fig:zmet_mbirthmatched}
\end{figure}

This comparison is presented in Fig.~\ref{fig:zmet_mbirthmatched}, the setup of which is identical to Fig.~\ref{fig:stars.age-zmet} above, except that satellite (stellar) metallicities are now compared to the field-equivalent values that match birth mass and age simultaneously. Solid lines of purple, green, and yellow represent satellites in increasingly massive haloes, while the black line (at an excess of zero) represents field galaxies. The main result from this exercise is that the metallicity excess in the oldest satellite stars that was visible in Fig.~\ref{fig:stars.age-zmet} can be fully attributed to the indirect effect of differing stellar birth masses: for stars born $\gtrsim 10$ Gyr ago ($z \gtrsim 2$), the metallicity in satellites is not significantly raised above the matched field value. Younger stars, however, do show a remaining metallicity excess that increases with decreasing age and is -- generally -- larger in more massive haloes. We can therefore conclude that environment does indeed have a \emph{direct} influence on the stellar metallicity of galaxies, even at fairly early times. This is not unexpected in light of our findings in \S \ref{sec:zgas}: a higher average metallicity of star forming gas naturally leads to the formation of stars with enhanced metallicity.

\begin{figure}
  \centering
  
      \includegraphics[width=\columnwidth]{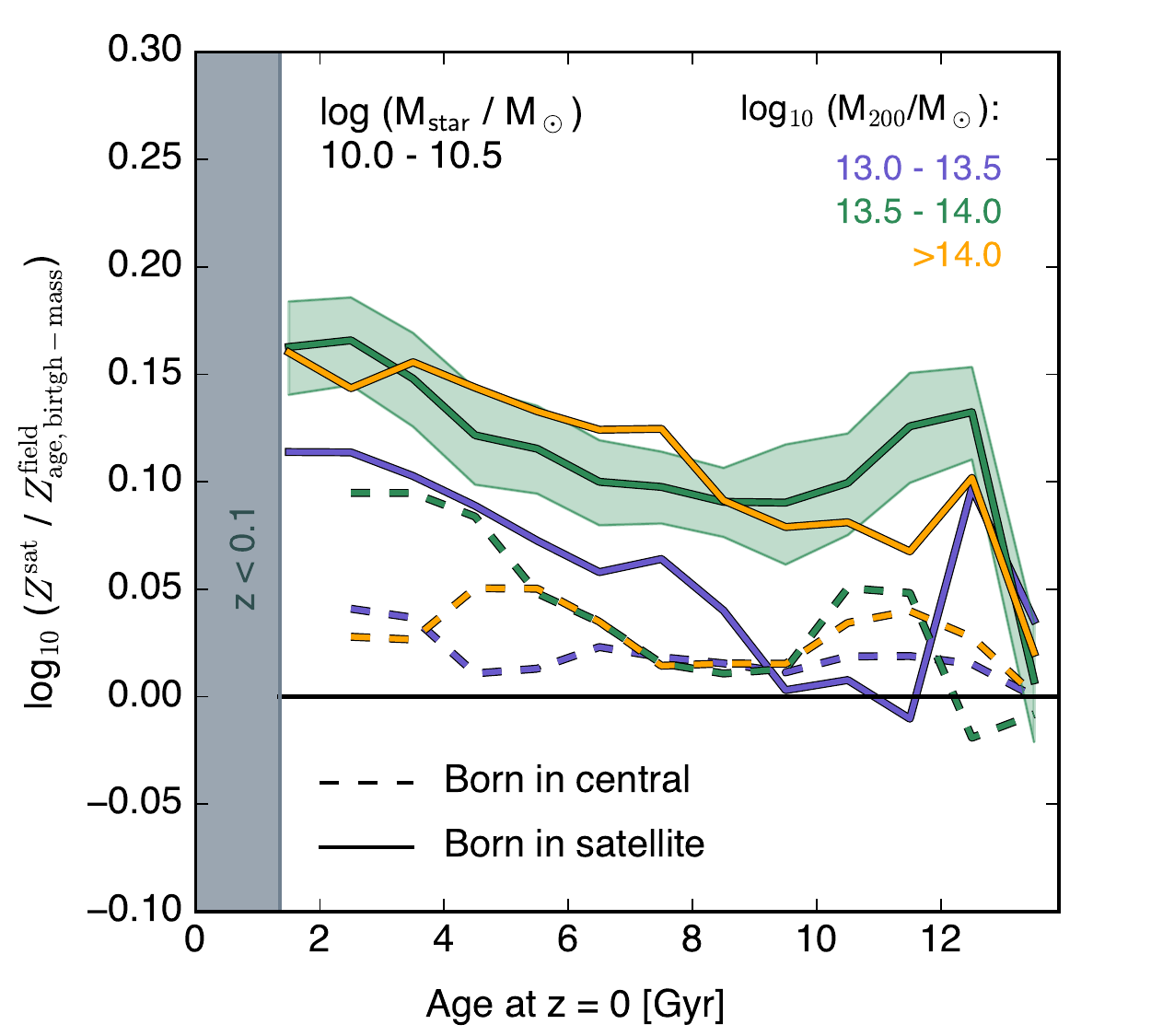}
       \caption{Breakdown of relative satellite metallicity for one narrow bin of stellar mass ($10.0 \leq \log_{10}(\mstar/\msun) < 10.5$, top panel of Fig.~\ref{fig:zmet_mbirthmatched}) into stars formed in central galaxies (dashed) and in satellites (solid). The dashed lines do not extend to the youngest age snapshot because by definition all these galaxies were satellites at this point ($z=0.1$). Stars born in a satellite galaxy are significantly more metal-enhanced than those formed when the galaxy was still a central.}
    \label{fig:censat}
\end{figure}

The increase in the metallicity excess towards later times can be due to two effects: on the one hand, the fraction of galaxies that were already a satellite -- and hence affected by their environment -- at the time the stars were formed increases with decreasing age of the stars. On the other hand, it is also conceivable that the strength of the environmental impact on the metallicity of the star forming gas has increased with time as the host halo grew. To disentangle these, we split, for one bin of $z=0.1$ stellar mass ($10^{10} \leq \mstar < 10^{10.5} \msun$), the stars of satellite galaxies into those born in a central and those born in a satellite subhalo. Fig.~\ref{fig:censat} shows the former as dashed and the latter as solid lines. Not unexpectedly, stars born in satellites show a larger metallicity excess, which already reaches $\sim$0.15 dex at $z \approx 2$ and is relatively constant after this point. We note, however, that this behaviour excludes the lowest mass groups with $\mvir$ in the range $10^{13}$ to $10^{13.5} \msun$ (black line), whose satellite-born stars show a more gradual increase in metallicity over time, possibly as a consequence of these haloes not having been massive enough at earlier times to lead to significant excess metal-enrichment of their star forming gas. Although stars born in centrals (but residing in group satellites at $z=0.1$) are less metal-enriched, a small excess is visible even for this set (typically 0.03 dex), in agreement with the enhanced metallicity at $r > 2 r_{200}$ that we had noted in Fig.~\ref{fig:rrel}. This is indicative of a small, but non-negligible contribution of assembly bias and/or large-scale `direct' environmental influence to the overall (stellar) metallicity excess in satellites at $z=0.1$.

\begin{figure*}
\centering
    \includegraphics[width=2.1\columnwidth]{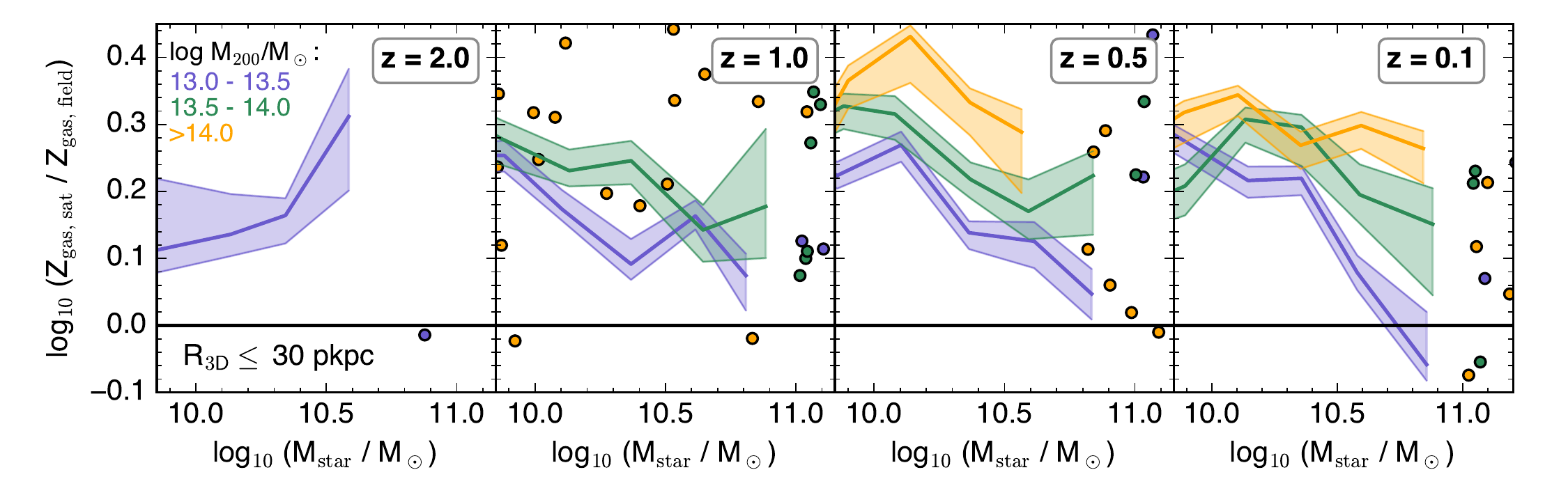}    
       \caption{The excess metallicity of star forming gas in satellites compared to the field at four different redshifts, decreasing from left to right. The layout of each panel is identical to the bottom half of Fig.~\ref{fig:gasz}, except that we show total metallicities averaged over the full galaxy (30 pkpc), instead of oxygen abundances within an aperture matched to the SDSS fibre size. In bins with fewer than ten galaxies, individual galaxies are shown as circles in the respective colour. In \eagle{}, metallicity is predicted to be enhanced in groups as early as $z=2$, at a magnitude comparable to the offset at the present epoch. The small number of cluster satellites at $z=1.0$, and total absence thereof at $z=2$, is a consequence of the limited volume of the \eagle{} Ref-L100 simulation.}
    \label{fig:gas_redshift}
  \end{figure*}

To explicitly test the connection between the metallicity of stars at $z=0.1$ and star forming gas at higher redshift, we show in Fig.~\ref{fig:gas_redshift} the evolution of the gas-phase metallicity excess in \eagle{} satellites. Individual panels represent redshifts of 2.0, 1.0, 0.5, and 0.1; in each case we plot the difference between satellite and field metallicities. The setup of each panel is identical to the bottom half of Fig.~\ref{fig:gasz}, except that we only show metallicities averaged over a larger aperture of 30 pkpc, since we are not comparing directly to SDSS data. In the highest redshift panel, only the lowest bin in halo mass is occupied (blue). This is because galaxy clusters at $z > 1$ are too rare to be sampled by the relatively small volume of the \eagle{} Ref-L100 simulation.

The (gas-phase) metallicity of satellites is consistently offset from the field, even at $z=2$, by an overall similar amount as at $z = 0.1$ ($\sim$0.2 dex). This is consistent with our inference from the analysis of differently old stellar populations: the environmental impact on galaxy metallicities in \eagle{} reaches back to $z \approx 2$.

In contrast to the local Universe, where SDSS observations of tens of thousands of galaxies enable relatively robust determinations of second order effects such as the influence of environment on the mass-metallicity relation, observations at higher redshift are still limited to much smaller samples of typically at most a few hundred galaxies. In two recent studies, \citet{Maier_et_al_2016} and \citet{Wuyts_et_al_2016} compared central and satellite galaxies at $z=0.4$ and in the range $0.6 \leq z < 2.7$, respectively; neither study reports a significant difference between the two samples, with the latter even suggesting a (mildly significant) \emph{deficiency} of metallicity in satellite galaxies at $z \approx 0.9$. Earlier observational work based on smaller samples has likewise found no difference between cluster and field galaxies at $z \approx 2$ in terms of their gas metallicity \citep{Kacprzak_et_al_2015} or hints of metal-deficiency in dense environments \citep{Valentino_et_al_2015}. While pointing out that we have made no attempt to match the specific characteristics of either of these observational studies, this comparison indicates a possible shortcoming of the \eagle{} simulation, which merits further investigation.


\section{Summary and Discussion}
\label{sec:summary}

Motivated by observational reports that satellite galaxies in groups and clusters have metallicities that are higher than those of central galaxies of the same stellar mass, we have compared the gas-phase and stellar metallicities of $> 3600$ field and group/cluster satellite galaxies (host halo mass $\log_{10} (\mvir/\msun) = 10^{13}$--$10^{14.5}$, galaxy stellar mass $\log_{10} (\mstar/\msun) > 10^{10}$) in the 100 cMpc \eagle{} `Reference' simulation (Ref-L100), and have also compared to alternative theoretical models. After confirming that \eagle{} broadly reproduces the observed environmental difference in both gas and stellar metallicities, we have tested several mechanisms that could cause this effect, including gas stripping, suppression of gas inflows, differing stellar age, stripping of stars, and differences in accretion of stars from other galaxies. The main results of our study may be summarised as follows:

\begin{enumerate}
\item The \eagle{} simulation generally reproduces the observed enhancement of metallicity in both the star-forming gas and the stellar components. For gas metallicity, an approximate match to the observational galaxy selection (specific star formation rate sSFR $\equiv$ SFR/$\mstar > 10^{-10.5}$ yr$^{-1}$), fibre size ($R_\text{2D} \leq 3$ pkpc), and weighting (by star formation rate) leads to quantitative agreement within the statistical uncertainties (Fig.~\ref{fig:gasz}). The stellar metallicity enhancement of satellites with $\mstar \gtrsim10^{10.5} \msun$ is also matched quantitatively if simulated metallicities are weighted by stellar mass, while weighting by luminosity underpredicts the observed excess. At lower masses, the simulations predict a smaller stellar metallicity excess than observed regardless of the weighting scheme, which is only partly ameliorated at higher resolution (Fig.~\ref{fig:stars.total-zmet}).

\item The stellar metallicity enhancement is sensitive to the subgrid efficiency of star formation feedback, with alternative \eagle{} models (which produce galaxies that are too compact) generally predicting a larger excess than the Reference implementation, in particular for satellites in galaxy clusters (Fig.~\ref{fig:subgridcomp}). A comparison to other simulations taken from the literature has shown qualitative agreement on enhanced gas and stellar metallicity in satellites, but with significant differences at a quantitative level (Fig.~\ref{fig:modelcomp}).

\item Satellites in \eagle{} show evidence of a significant removal of star-forming gas from their outskirts. This explains the elevated level of metallicity of the star-forming gas only partly, however: even at fixed radius ($r \lesssim 15$ pkpc), satellites are metal-enriched compared to the field. This is predominantly the result of suppressed metal-poor inflows, but to a lesser extent also of enhanced enrichment due to a larger relative contribution of recycled stellar outflows, from more metal-rich stars (Figs.~\ref{fig:gas_profiles} and \ref{fig:gas_zmet_histograms}). 

\item As observed, the stellar metallicity enhancement in \eagle{} satellites is less strong amongst star-forming galaxies than the general population. Furthermore, our analysis predicts a significantly stronger enhancement for transitional galaxies (sSFR $\approx  10^{-11}$ yr$^{-1}$) compared to those with higher star formation rates. This suggests a tight causal link between star formation quenching and metallicity enhancement in satellite galaxies (Fig.~\ref{fig:zstar_ssfr}).  

\item Stellar mass loss through e.g.~tidal forces cannot account for the stellar metallicity offset, because galaxies surviving until $z  = 0.1$ have typically only lost $< 0.05$ dex in stellar mass since reaching their maximum $\mstar$. Taking into account the missed stellar growth in satellites as a consequence of star formation quenching, this difference increases to only $\lesssim 0.2$ dex even for galaxies of $\mstar \approx 10^9 \msun$ in clusters. Mass loss of $\sim$0.4 dex would be required to explain the metallicity offset, both in \eagle{} and in the observations of \citet{Pasquali_et_al_2010} (Fig.~\ref{fig:mstar_diff}).

\item \eagle{} satellites accrete a smaller fraction of their stars from other galaxies than field galaxies (3 per cent vs.~7 per cent at $\mstar \approx 10^9 \msun$; Fig.~\ref{fig:accreted_fraction}). Taking this difference into account by comparing stellar populations in centrals and satellites that were formed at the same time in galaxies of the same stellar mass, satellites show no increase in metallicity for stars formed at $z \gtrsim 2$ (Fig.~\ref{fig:zmet_mbirthmatched}). 

\item A metallicity offset due to `direct' environmental contributions remains for stars born at $z \lesssim 2$; this increases towards later star formation times when a larger fraction of satellites had already fallen into their host halo (Fig.~\ref{fig:censat}). We confirm that, in \eagle{}, this is due to excess metallicity in satellites compared to the field even at $z=2$ (Fig.~\ref{fig:gas_redshift}).

\end{enumerate}

The salient conclusion of this analysis is that the excess stellar and gas-phase metallicities in satellite galaxies are both directly linked to environmental star formation quenching, and are not symptoms of two different physical processes, as was suggested by \citet{Pasquali_et_al_2012}. Stellar metallicities in satellites are raised predominantly because they formed from relatively highly metal-enriched gas. In turn, this excess gas enrichment results from the removal of relatively metal-poor gas from galaxy outskirts -- likely due to ram pressure stripping -- and suppression of metal-poor gas inflows, which is plausibly a consequence of the removal of less dense gas from the galaxy halo. A testable prediction of this scenario is that the stellar metallicity of transitional galaxies (sSFR $\approx 10^{-11}$ yr$^{-1}$, which are likely in the process of being quenched) should be significantly raised in satellites compared to the field.

The key problem of this general picture is its prediction of, and indeed reliance upon, an enhancement of satellite gas-phase metallicity not only at $z \approx 0$, but also at earlier epochs, at least as far back as $z \approx 0.5$ when a significant fraction of the stars making up present-day galaxies were yet to form. What limited observational evidence is available on this topic instead suggests that any difference between the metallicity of satellites and centrals is insignificant \citep{Kacprzak_et_al_2015,Maier_et_al_2016}, with some studies even presenting evidence for a \emph{lower} metallicity in satellites \citep{Valentino_et_al_2015, Wuyts_et_al_2016}. 

Our analysis suggests that environmental differences in gas metallicity are highly sensitive to both galaxy selection and analysis details such as aperture and weighting scheme, and that observations may significantly underestimate the `true' metallicity enhancement of satellites. While it is unclear at present to what extent this conclusion is also applicable to $z \gg 0$, it nevertheless highlights the need for careful like-with-like comparisons tailored to the characteristics of a given observation to draw meaningful conclusions about the success or failure of theoretical galaxy formation models in this respect. 

A second potential discrepancy between not just \eagle{}, but also the Illustris simulation and the \citet{Henriques_et_al_2015} semi-analytic galaxy formation model and observations, is their collective failure to reproduce the strong rise in satellite stellar metallicity enhancement with decreasing stellar mass at $\mstar \lesssim 10^{10.5} \msun$ \citep{Pasquali_et_al_2010}. Although the severity of this discrepancy cannot be authoritatively assessed without recourse to larger high-resolution hydrodynamical simulations that adequately sample the satellite galaxy population, it is nevertheless interesting to speculate on two potential causes. 

First, it might hint at some physical process whose importance is \emph{fundamentally} underestimated in current theoretical models, for example thermal conduction, (physical) viscosity, or magnetic fields. Alternatively, it is at least possible that the effect is actually overestimated in the observational data: its primary driver is not an actual rise of satellite metallicity, but rather a strong decline in the stellar metallicity of central galaxies. As discussed by \citet{Gallazzi_et_al_2005}, estimating stellar metallicities from absorption lines in SDSS spectra requires prior subtraction of (often much stronger) emission lines, which has a larger impact on star forming than passive galaxies. Towards lower mass, most field galaxies are star forming, but a significant fraction of satellites are not (e.g.~\citealt{Wetzel_et_al_2012}), which might lead to subtle biases in the derived metallicities of these two populations.  The fact that \citet{Pasquali_et_al_2012} demonstrate a lack of strong stellar metallicity enhancement in \emph{star-forming} low-mass satellites is consistent with this hypothesis, but would also arise naturally from a causal connection between star formation quenching and metallicity enhancement in satellites, as advocated by \eagle{}. We note that a recent study of \citet{Peng_et_al_2015} reports only a small environmental difference between the stellar metallicities of \emph{passive} galaxies in SDSS, which indicates that a varying star-forming fraction is indeed the main driver behind the metallicity excess observed in the overall satellite population \citepalias{Pasquali_et_al_2012}.

Another important area of progress from the observational side is the ability to measure metallicity across entire galaxies, as opposed to only the innermost few kpc, with integral-field-units (IFUs) such as CALIFA \citep{Sanchez_et_al_2013}, MaNGA \citep{Bundy_et_al_2015}, and MUSE \citep{Bacon_et_al_2010}. IFU observations of a representative number of group/cluster galaxies in the local Universe will be able to directly test our prediction that the metallicity of star forming gas is enhanced in satellites even after accounting for the removal of low-metallicity gas from the galaxy outskirts. Furthermore, combining such data with planned large \hi{} surveys such as \textsc{Apertif} or eventually the SKA could directly link the stripping of low-density gas with the enhancement of metallicity in the remaining dense, star forming gas, and thus shed new light onto the effects of environment on galaxy evolution.

\section*{Acknowledgments}
RAC is a Royal Society University Research Fellow. IGM is an STFC Advanced Fellow. We thank the anonymous referee for a constructive report, Simon White for helpful discussions, Anna Pasquali for providing her data, and Lydia Heck for expert computational support with the Cosma machine in Durham, where part of the analysis presented here was performed. This work used the DiRAC Data Centric system at Durham University, operated by the Institute for Computational Cosmology on behalf of the STFC DiRAC HPC Facility (www.dirac.ac.uk). This equipment was funded by BIS National E-infrastructure capital grant ST/K00042X/1, STFC capital grant ST/H008519/1, and STFC DiRAC Operations grant ST/K003267/1 and Durham University. DiRAC is part of the National E-Infrastructure. We also gratefully acknowledge PRACE for awarding the EAGLE project with access to the Curie facility based in France at Tr\`{e}s Grand Centre de Calcul. Support was also received via the Interuniversity Attraction Poles Programme initiated by the Belgian Science Policy Office ([AP P7/08 CHARM]), the National Science Foundation under Grant No. NSF PHY11-25915, and the UK Science and Technology Facilities Council (grant numbers ST/F001166/1 and ST/I000976/1) via rolling and consolidated grants awarded to the ICC. The research was supported by the Netherlands Organisation for Scientific Research (NWO), through VICI grant 639.043.409, and the European Research Council under the European Union's Seventh Framework Programme (FP7/2007-2013) / ERC Grant agreement 278594-GasAroundGalaxies. All figures in this paper were produced using the \textsc{Astropy} \citep{Astropy_2013} and \textsc{Matplotlib} \citep{Hunter_2007} Python packages.
\bibliographystyle{mn2e}
\bibliography{Metallicity}

\begin{thebibliography}{128}
\expandafter\ifx\csname natexlab\endcsname\relax\def\natexlab#1{#1}\fi

\bibitem[{{Agertz} {et~al}\mbox{.}(2007){Agertz}, {Moore}, {Stadel}, {Potter},
  {Miniati}, {Read}, {Mayer}, {Gawryszczak}, {Kravtsov}, {Nordlund}, {Pearce},
  {Quilis}, {Rudd}, {Springel}, {Stone}, {Tasker}, {Teyssier}, {Wadsley}, \&
  {Walder}}]{Agertz_et_al_2007}
{Agertz} O. {et~al.}, 2007, \mnras, 380, 963

\bibitem[{{Allende Prieto}, {Lambert} \& {Asplund}(2001){Allende Prieto},
  {Lambert}, \& {Asplund}}]{Allende-Prieto_et_al_2001}
{Allende Prieto} C., {Lambert} D.~L., {Asplund} M., 2001, \apjl, 556, L63

\bibitem[{{Andrews} \& {Martini}(2013)}]{Andrews_Martini_2013}
{Andrews} B.~H., {Martini} P., 2013, \apj, 765, 140

\bibitem[{{Asplund} {et~al}\mbox{.}(2009){Asplund}, {Grevesse}, {Sauval}, \&
  {Scott}}]{Asplund_et_al_2009}
{Asplund} M., {Grevesse} N., {Sauval} A.~J., {Scott} P., 2009, \araa, 47, 481

\bibitem[{{Astropy Collaboration}(2013){Astropy Collaboration},
  {Robitaille}, {Tollerud}, {Greenfield}, {Droettboom}, {Bray}, {Aldcroft},
  {Davis}, {Ginsburg}, {Price-Whelan}, {Kerzendorf}, {Conley}, {Crighton},
  {Barbary}, {Muna}, {Ferguson}, {Grollier}, {Parikh}, {Nair}, {Unther},
  {Deil}, {Woillez}, {Conseil}, {Kramer}, {Turner}, {Singer}, {Fox}, {Weaver},
  {Zabalza}, {Edwards}, {Azalee Bostroem}, {Burke}, {Casey}, {Crawford},
  {Dencheva}, {Ely}, {Jenness}, {Labrie}, {Lim}, {Pierfederici}, {Pontzen},
  {Ptak}, {Refsdal}, {Servillat}, \& {Streicher}}]{Astropy_2013}
{Astropy Collaboration}, 2013, \aap, 558, A33

\bibitem[{{Bacon} {et~al}\mbox{.}(2010){Bacon}, {Accardo}, {Adjali}, {Anwand},
  {Bauer}, {Biswas}, {Blaizot}, {Boudon}, {Brau-Nogue}, {Brinchmann},
  {Caillier}, {Capoani}, {Carollo}, {Contini}, {Couderc}, {Daguis{\'e}},
  {Deiries}, {Delabre}, {Dreizler}, {Dubois}, {Dupieux}, {Dupuy}, {Emsellem},
  {Fechner}, {Fleischmann}, {Fran{\c c}ois}, {Gallou}, {Gharsa}, {Glindemann},
  {Gojak}, {Guiderdoni}, {Hansali}, {Hahn}, {Jarno}, {Kelz}, {Koehler},
  {Kosmalski}, {Laurent}, {Le Floch}, {Lilly}, {Lizon}, {Loupias}, {Manescau},
  {Monstein}, {Nicklas}, {Olaya}, {Pares}, {Pasquini}, {P{\'e}contal-Rousset},
  {Pell{\'o}}, {Petit}, {Popow}, {Reiss}, {Remillieux}, {Renault}, {Roth},
  {Rupprecht}, {Serre}, {Schaye}, {Soucail}, {Steinmetz}, {Streicher}, {Stuik},
  {Valentin}, {Vernet}, {Weilbacher}, {Wisotzki}, \&
  {Yerle}}]{Bacon_et_al_2010}
{Bacon} R. {et~al.}, 2010, in \procspie, Vol. 7735, Ground-based and Airborne
  Instrumentation for Astronomy III, p. 773508

\bibitem[{{Bah{\'e}} \& {McCarthy}(2015)}]{Bahe_McCarthy_2015}
{Bah{\'e}} Y.~M., {McCarthy} I.~G., 2015, \mnras, 447, 969

\bibitem[{{Bah{\'e}} {et~al}\mbox{.}(2013){Bah{\'e}}, {McCarthy}, {Balogh}, \&
  {Font}}]{Bahe_et_al_2013}
{Bah{\'e}} Y.~M., {McCarthy} I.~G., {Balogh} M.~L., {Font} A.~S., 2013, \mnras,
  430, 3017

\bibitem[{{Bah{\'e}} {et~al}\mbox{.}(2012){Bah{\'e}}, {McCarthy}, {Crain}, \&
  {Theuns}}]{Bahe_et_al_2012}
{Bah{\'e}} Y.~M., {McCarthy} I.~G., {Crain} R.~A., {Theuns} T., 2012, \mnras,
  424, 1179

\bibitem[{{Baldwin}, {Phillips} \& {Terlevich}(1981){Baldwin}, {Phillips}, \&
  {Terlevich}}]{Baldwin_et_al_1981}
{Baldwin} J.~A., {Phillips} M.~M., {Terlevich} R., 1981, \pasp, 93, 5

\bibitem[{{Balogh} {et~al}\mbox{.}(1999){Balogh}, {Morris}, {Yee}, {Carlberg},
  \& {Ellingson}}]{Balogh_et_al_1999}
{Balogh} M.~L., {Morris} S.~L., {Yee} H.~K.~C., {Carlberg} R.~G., {Ellingson}
  E., 1999, \apj, 527, 54

\bibitem[{{Barber} {et~al}\mbox{.}(2016){Barber}, {Schaye}, {Bower}, {Crain},
  {Schaller}, \& {Theuns}}]{Barber_et_al_2016}
{Barber} C., {Schaye} J., {Bower} R.~G., {Crain} R.~A., {Schaller} M., {Theuns}
  T., 2016, \mnras, 460, 1147

\bibitem[{{Bothwell} {et~al}\mbox{.}(2013){Bothwell}, {Smail}, {Chapman},
  {Genzel}, {Ivison}, {Tacconi}, {Alaghband-Zadeh}, {Bertoldi}, {Blain},
  {Casey}, {Cox}, {Greve}, {Lutz}, {Neri}, {Omont}, \&
  {Swinbank}}]{Bothwell_et_al_2013}
{Bothwell} M.~S. {et~al.}, 2013, \mnras, 429, 3047

\bibitem[{{Boylan-Kolchin} {et~al}\mbox{.}(2009){Boylan-Kolchin}, {Springel},
  {White}, {Jenkins}, \& {Lemson}}]{Boylan-Kolchin_et_al_2009}
{Boylan-Kolchin} M., {Springel} V., {White} S.~D.~M., {Jenkins} A., {Lemson}
  G., 2009, \mnras, 398, 1150

\bibitem[{{Brinchmann} {et~al}\mbox{.}(2004){Brinchmann}, {Charlot}, {White},
  {Tremonti}, {Kauffmann}, {Heckman}, \& {Brinkmann}}]{Brinchmann_et_al_2004}
{Brinchmann} J., {Charlot} S., {White} S.~D.~M., {Tremonti} C., {Kauffmann} G.,
  {Heckman} T., {Brinkmann} J., 2004, \mnras, 351, 1151

\bibitem[{{Bruzual} \& {Charlot}(2003)}]{Bruzual_Charlot_2003}
{Bruzual} G., {Charlot} S., 2003, \mnras, 344, 1000

\bibitem[{{Bundy} {et~al}\mbox{.}(2015){Bundy}, {Bershady}, {Law}, {Yan},
  {Drory}, {MacDonald}, {Wake}, {Cherinka}, {S{\'a}nchez-Gallego}, {Weijmans},
  {Thomas}, {Tremonti}, {Masters}, {Coccato}, {Diamond-Stanic},
  {Arag{\'o}n-Salamanca}, {Avila-Reese}, {Badenes}, {Falc{\'o}n-Barroso},
  {Belfiore}, {Bizyaev}, {Blanc}, {Bland-Hawthorn}, {Blanton}, {Brownstein},
  {Byler}, {Cappellari}, {Conroy}, {Dutton}, {Emsellem}, {Etherington},
  {Frinchaboy}, {Fu}, {Gunn}, {Harding}, {Johnston}, {Kauffmann}, {Kinemuchi},
  {Klaene}, {Knapen}, {Leauthaud}, {Li}, {Lin}, {Maiolino}, {Malanushenko},
  {Malanushenko}, {Mao}, {Maraston}, {McDermid}, {Merrifield}, {Nichol},
  {Oravetz}, {Pan}, {Parejko}, {Sanchez}, {Schlegel}, {Simmons}, {Steele},
  {Steinmetz}, {Thanjavur}, {Thompson}, {Tinker}, {van den Bosch}, {Westfall},
  {Wilkinson}, {Wright}, {Xiao}, \& {Zhang}}]{Bundy_et_al_2015}
{Bundy} K. {et~al.}, 2015, \apj, 798, 7

\bibitem[{{Carton} {et~al}\mbox{.}(2015){Carton}, {Brinchmann}, {Wang},
  {Bigiel}, {Cormier}, {van der Hulst}, {J{\'o}zsa}, {Serra}, \&
  {Verheijen}}]{Carton_et_al_2015}
{Carton} D. {et~al.}, 2015, \mnras, 451, 210

\bibitem[{{Chabrier}(2003)}]{Chabrier_2003}
{Chabrier} G., 2003, \pasp, 115, 763

\bibitem[{{Chang} {et~al}\mbox{.}(2015){Chang}, {van der Wel}, {da Cunha}, \&
  {Rix}}]{Chang_et_al_2015}
{Chang} Y.-Y., {van der Wel} A., {da Cunha} E., {Rix} H.-W., 2015, \apjs, 219,
  8

\bibitem[{{Conroy}, {Gunn} \& {White}(2009){Conroy}, {Gunn}, \&
  {White}}]{Conroy_et_al_2009}
{Conroy} C., {Gunn} J.~E., {White} M., 2009, \apj, 699, 486

\bibitem[{{Cooper} {et~al}\mbox{.}(2008){Cooper}, {Tremonti}, {Newman}, \&
  {Zabludoff}}]{Cooper_et_al_2008}
{Cooper} M.~C., {Tremonti} C.~A., {Newman} J.~A., {Zabludoff} A.~I., 2008,
  \mnras, 390, 245

\bibitem[{{Crain} {et~al}\mbox{.}(2015){Crain}, {Schaye}, {Bower}, {Furlong},
  {Schaller}, {Theuns}, {Dalla Vecchia}, {Frenk}, {McCarthy}, {Helly},
  {Jenkins}, {Rosas-Guevara}, {White}, \& {Trayford}}]{Crain_et_al_2015}
{Crain} R.~A. {et~al.}, 2015, \mnras, 450, 1937

\bibitem[{{Dalla Vecchia} \& {Schaye}(2012)}]{DallaVecchia_Schaye_2012}
{Dalla Vecchia} C., {Schaye} J., 2012, \mnras, 426, 140

\bibitem[{{Dav{\'e}}, {Finlator} \& {Oppenheimer}(2011){Dav{\'e}}, {Finlator},
  \& {Oppenheimer}}]{Dave_et_al_2011}
{Dav{\'e}} R., {Finlator} K., {Oppenheimer} B.~D., 2011, \mnras, 416, 1354

\bibitem[{{Dav{\'e}}, {Finlator} \& {Oppenheimer}(2012){Dav{\'e}}, {Finlator},
  \& {Oppenheimer}}]{Dave_et_al_2012}
---, 2012, \mnras, 421, 98

\bibitem[{{De Lucia} {et~al}\mbox{.}(2012){De Lucia}, {Weinmann}, {Poggianti},
  {Arag{\'o}n-Salamanca}, \& {Zaritsky}}]{de_Lucia_et_al_2012}
{De Lucia} G., {Weinmann} S., {Poggianti} B.~M., {Arag{\'o}n-Salamanca} A.,
  {Zaritsky} D., 2012, \mnras, 423, 1277

\bibitem[{{De Rossi} {et~al}\mbox{.}(2015){De Rossi}, {Theuns}, {Font}, \&
  {McCarthy}}]{DeRossi_et_al_2015}
{De Rossi} M.~E., {Theuns} T., {Font} A.~S., {McCarthy} I.~G., 2015, ArXiv
  e-prints

\bibitem[{{Dekel} \& {Silk}(1986)}]{Dekel_Silk_1986}
{Dekel} A., {Silk} J., 1986, \apj, 303, 39

\bibitem[{{Dolag} {et~al}\mbox{.}(2009){Dolag}, {Borgani}, {Murante}, \&
  {Springel}}]{Dolag_et_al_2009}
{Dolag} K., {Borgani} S., {Murante} G., {Springel} V., 2009, \mnras, 399, 497

\bibitem[{{Dressler}(1980)}]{Dressler_1980}
{Dressler} A., 1980, \apj, 236, 351

\bibitem[{{D'Souza} {et~al}\mbox{.}(2014){D'Souza}, {Kauffman}, {Wang}, \&
  {Vegetti}}]{D_Souza_et_al_2014}
{D'Souza} R., {Kauffman} G., {Wang} J., {Vegetti} S., 2014, \mnras, 443, 1433

\bibitem[{{D'Souza}, {Vegetti} \& {Kauffmann}(2015){D'Souza}, {Vegetti}, \&
  {Kauffmann}}]{DSouza_et_al_2015}
{D'Souza} R., {Vegetti} S., {Kauffmann} G., 2015, \mnras, 454, 4027

\bibitem[{{Ellison} {et~al}\mbox{.}(2008{\natexlab{a}}){Ellison}, {Patton},
  {Simard}, \& {McConnachie}}]{Ellison_et_al_2008a}
{Ellison} S.~L., {Patton} D.~R., {Simard} L., {McConnachie} A.~W.,
  2008{\natexlab{a}}, \apjl, 672, L107

\bibitem[{{Ellison} {et~al}\mbox{.}(2008{\natexlab{b}}){Ellison}, {Patton},
  {Simard}, \& {McConnachie}}]{Ellison_et_al_2008b}
---, 2008{\natexlab{b}}, \aj, 135, 1877

\bibitem[{{Fabello} {et~al}\mbox{.}(2012){Fabello}, {Kauffmann}, {Catinella},
  {Li}, {Giovanelli}, \& {Haynes}}]{Fabello_et_al_2012}
{Fabello} S., {Kauffmann} G., {Catinella} B., {Li} C., {Giovanelli} R.,
  {Haynes} M.~P., 2012, \mnras, 427, 2841

\bibitem[{{Ferguson}, {Gallagher} \& {Wyse}(1998){Ferguson}, {Gallagher}, \&
  {Wyse}}]{Ferguson_et_al_1998}
{Ferguson} A.~M.~N., {Gallagher} J.~S., {Wyse} R.~F.~G., 1998, \aj, 116, 673

\bibitem[{{Finlator} \& {Dav{\'e}}(2008)}]{Finlator_Dave_2008}
{Finlator} K., {Dav{\'e}} R., 2008, \mnras, 385, 2181

\bibitem[{{Font} {et~al}\mbox{.}(2011){Font}, {McCarthy}, {Crain}, {Theuns},
  {Schaye}, {Wiersma}, \& {Dalla Vecchia}}]{Font_et_al_2011}
{Font} A.~S., {McCarthy} I.~G., {Crain} R.~A., {Theuns} T., {Schaye} J.,
  {Wiersma} R.~P.~C., {Dalla Vecchia} C., 2011, \mnras, 416, 2802

\bibitem[{{Gallazzi} {et~al}\mbox{.}(2006){Gallazzi}, {Charlot}, {Brinchmann},
  \& {White}}]{Gallazzi_et_al_2006}
{Gallazzi} A., {Charlot} S., {Brinchmann} J., {White} S.~D.~M., 2006, \mnras,
  370, 1106

\bibitem[{{Gallazzi} {et~al}\mbox{.}(2005){Gallazzi}, {Charlot}, {Brinchmann},
  {White}, \& {Tremonti}}]{Gallazzi_et_al_2005}
{Gallazzi} A., {Charlot} S., {Brinchmann} J., {White} S.~D.~M., {Tremonti}
  C.~A., 2005, \mnras, 362, 41

\bibitem[{{Gao}, {Springel} \& {White}(2005){Gao}, {Springel}, \&
  {White}}]{Gao_et_al_2005}
{Gao} L., {Springel} V., {White} S.~D.~M., 2005, \mnras, 363, L66

\bibitem[{{Garnett}(2002)}]{Garnett_2002}
{Garnett} D.~R., 2002, \apj, 581, 1019

\bibitem[{{Genel}(2016)}]{Genel_2016}
{Genel} S., 2016, \apj, 822, 107

\bibitem[{{Guidi}, {Scannapieco} \& {Walcher}(2015){Guidi}, {Scannapieco}, \&
  {Walcher}}]{Guidi_et_al_2015}
{Guidi} G., {Scannapieco} C., {Walcher} C.~J., 2015, \mnras, 454, 2381

\bibitem[{{Gunn} \& {Gott}(1972)}]{Gunn_Gott_1972}
{Gunn} J.~E., {Gott}, III J.~R., 1972, \apj, 176, 1

\bibitem[{{Haardt} \& {Madau}(2001)}]{Haardt_Madau_2001}
{Haardt} F., {Madau} P., 2001, in Clusters of Galaxies and the High Redshift
  Universe Observed in X-rays, {Neumann} D.~M., {Tran} J.~T.~V., eds.

\bibitem[{{Henriques} {et~al}\mbox{.}(2015){Henriques}, {White}, {Thomas},
  {Angulo}, {Guo}, {Lemson}, {Springel}, \& {Overzier}}]{Henriques_et_al_2015}
{Henriques} B.~M.~B., {White} S.~D.~M., {Thomas} P.~A., {Angulo} R., {Guo} Q.,
  {Lemson} G., {Springel} V., {Overzier} R., 2015, \mnras, 451, 2663

\bibitem[{{Henriques} {et~al}\mbox{.}(2013){Henriques}, {White}, {Thomas},
  {Angulo}, {Guo}, {Lemson}, \& {Springel}}]{Henriques_et_al_2013}
{Henriques} B.~M.~B., {White} S.~D.~M., {Thomas} P.~A., {Angulo} R.~E., {Guo}
  Q., {Lemson} G., {Springel} V., 2013, \mnras, 431, 3373

\bibitem[{{Hess} \& {Wilcots}(2013)}]{Hess_Wilcots_2013}
{Hess} K.~M., {Wilcots} E.~M., 2013, \aj, 146, 124

\bibitem[{{Hopkins}(2013)}]{Hopkins_2013}
{Hopkins} P.~F., 2013, \mnras, 428, 2840

\bibitem[{{Hu} {et~al}\mbox{.}(2016){Hu}, {Naab}, {Walch}, {Glover}, \&
  {Clark}}]{Hu_et_al_2016}
{Hu} C.-Y., {Naab} T., {Walch} S., {Glover} S.~C.~O., {Clark} P.~C., 2016,
  \mnras, 458, 3528

\bibitem[{Hunter(2007)}]{Hunter_2007}
Hunter J.~D., 2007, Computing In Science \& Engineering, 9, 90

\bibitem[{{Jenkins}(2013)}]{Jenkins_2013}
{Jenkins} A., 2013, \mnras, 434, 2094

\bibitem[{{Kacprzak} {et~al}\mbox{.}(2015){Kacprzak}, {Yuan}, {Nanayakkara},
  {Kobayashi}, {Tran}, {Kewley}, {Glazebrook}, {Spitler}, {Taylor}, {Cowley},
  {Labbe}, {Straatman}, \& {Tomczak}}]{Kacprzak_et_al_2015}
{Kacprzak} G.~G. {et~al.}, 2015, \apjl, 802, L26

\bibitem[{{Kauffmann} {et~al}\mbox{.}(2004){Kauffmann}, {White}, {Heckman},
  {M{\'e}nard}, {Brinchmann}, {Charlot}, {Tremonti}, \&
  {Brinkmann}}]{Kauffmann_et_al_2004}
{Kauffmann} G., {White} S.~D.~M., {Heckman} T.~M., {M{\'e}nard} B.,
  {Brinchmann} J., {Charlot} S., {Tremonti} C., {Brinkmann} J., 2004, \mnras,
  353, 713

\bibitem[{{Kennicutt}, {Bresolin} \& {Garnett}(2003){Kennicutt}, {Bresolin}, \&
  {Garnett}}]{Kennicutt_et_al_2003}
{Kennicutt}, Jr. R.~C., {Bresolin} F., {Garnett} D.~R., 2003, \apj, 591, 801

\bibitem[{{Kewley} \& {Ellison}(2008)}]{Kewley_Ellison_2008}
{Kewley} L.~J., {Ellison} S.~L., 2008, \apj, 681, 1183

\bibitem[{{Lackner} {et~al}\mbox{.}(2012){Lackner}, {Cen}, {Ostriker}, \&
  {Joung}}]{Lackner_et_al_2012}
{Lackner} C.~N., {Cen} R., {Ostriker} J.~P., {Joung} M.~R., 2012, \mnras, 425,
  641

\bibitem[{{Lara-L{\'o}pez} {et~al}\mbox{.}(2010){Lara-L{\'o}pez}, {Cepa},
  {Bongiovanni}, {P{\'e}rez Garc{\'{\i}}a}, {Ederoclite}, {Casta{\~n}eda},
  {Fern{\'a}ndez Lorenzo}, {Povi{\'c}}, \&
  {S{\'a}nchez-Portal}}]{Lara-Lopez_et_al_2010}
{Lara-L{\'o}pez} M.~A. {et~al.}, 2010, \aap, 521, L53

\bibitem[{{Lara-L{\'o}pez} {et~al}\mbox{.}(2013){Lara-L{\'o}pez}, {Hopkins},
  {L{\'o}pez-S{\'a}nchez}, {Brough}, {Gunawardhana}, {Colless}, {Robotham},
  {Bauer}, {Bland-Hawthorn}, {Cluver}, {Driver}, {Foster}, {Kelvin}, {Liske},
  {Loveday}, {Owers}, {Ponman}, {Sharp}, {Steele}, {Taylor}, \&
  {Thomas}}]{Lara-Lopez_et_al_2013}
---, 2013, \mnras, 434, 451

\bibitem[{{Larson}(1974)}]{Larson_1974}
{Larson} R.~B., 1974, \mnras, 169, 229

\bibitem[{{Larson}, {Tinsley} \& {Caldwell}(1980){Larson}, {Tinsley}, \&
  {Caldwell}}]{Larson_et_al_1980}
{Larson} R.~B., {Tinsley} B.~M., {Caldwell} C.~N., 1980, \apj, 237, 692

\bibitem[{{Lequeux} {et~al}\mbox{.}(1979){Lequeux}, {Peimbert}, {Rayo},
  {Serrano}, \& {Torres-Peimbert}}]{Lequeux_et_al_1979}
{Lequeux} J., {Peimbert} M., {Rayo} J.~F., {Serrano} A., {Torres-Peimbert} S.,
  1979, \aap, 80, 155

\bibitem[{{Lu} {et~al}\mbox{.}(2012){Lu}, {Gilbank}, {McGee}, {Balogh}, \&
  {Gallagher}}]{Lu_et_al_2012}
{Lu} T., {Gilbank} D.~G., {McGee} S.~L., {Balogh} M.~L., {Gallagher} S., 2012,
  \mnras, 420, 126

\bibitem[{{Lu}, {Blanc} \& {Benson}(2015){Lu}, {Blanc}, \&
  {Benson}}]{Lu_et_al_2015}
{Lu} Y., {Blanc} G.~A., {Benson} A., 2015, \apj, 808, 129

\bibitem[{{Maier} {et~al}\mbox{.}(2016){Maier}, {Kuchner}, {Ziegler},
  {Verdugo}, {Balestra}, {Girardi}, {Mercurio}, {Rosati}, {Fritz}, {Grillo},
  {Nonino}, \& {Sartoris}}]{Maier_et_al_2016}
{Maier} C. {et~al.}, 2016, \aap, 590, A108

\bibitem[{{Mannucci} {et~al}\mbox{.}(2010){Mannucci}, {Cresci}, {Maiolino},
  {Marconi}, \& {Gnerucci}}]{Mannucci_et_al_2010}
{Mannucci} F., {Cresci} G., {Maiolino} R., {Marconi} A., {Gnerucci} A., 2010,
  \mnras, 408, 2115

\bibitem[{{Mattsson} \& {Andersen}(2012)}]{Mattsson_Andersen_2012}
{Mattsson} L., {Andersen} A.~C., 2012, \mnras, 423, 38

\bibitem[{{McGee}, {Bower} \& {Balogh}(2014){McGee}, {Bower}, \&
  {Balogh}}]{McGee_et_al_2014}
{McGee} S.~L., {Bower} R.~G., {Balogh} M.~L., 2014, ArXiv e-prints

\bibitem[{{Mitchell} {et~al}\mbox{.}(2009){Mitchell}, {McCarthy}, {Bower},
  {Theuns}, \& {Crain}}]{Mitchell_et_al_2009}
{Mitchell} N.~L., {McCarthy} I.~G., {Bower} R.~G., {Theuns} T., {Crain} R.~A.,
  2009, \mnras, 395, 180

\bibitem[{{Moore} {et~al}\mbox{.}(1996){Moore}, {Katz}, {Lake}, {Dressler}, \&
  {Oemler}}]{Moore_et_al_1996}
{Moore} B., {Katz} N., {Lake} G., {Dressler} A., {Oemler} A., 1996, \nat, 379,
  613

\bibitem[{{Moore}, {Lake} \& {Katz}(1998){Moore}, {Lake}, \&
  {Katz}}]{Moore_et_al_1998}
{Moore} B., {Lake} G., {Katz} N., 1998, \apj, 495, 139

\bibitem[{{Moreno} {et~al}\mbox{.}(2013){Moreno}, {Bluck}, {Ellison}, {Patton},
  {Torrey}, \& {Moster}}]{Moreno_et_al_2013}
{Moreno} J., {Bluck} A.~F.~L., {Ellison} S.~L., {Patton} D.~R., {Torrey} P.,
  {Moster} B.~P., 2013, \mnras, 436, 1765

\bibitem[{{Mulchaey} \& {Jeltema}(2010)}]{Mulchaey_Jeltema_2010}
{Mulchaey} J.~S., {Jeltema} T.~E., 2010, \apjl, 715, L1

\bibitem[{{Nelson} {et~al}\mbox{.}(2015){Nelson}, {Pillepich}, {Genel},
  {Vogelsberger}, {Springel}, {Torrey}, {Rodriguez-Gomez}, {Sijacki}, {Snyder},
  {Griffen}, {Marinacci}, {Blecha}, {Sales}, {Xu}, \&
  {Hernquist}}]{Nelson_et_al_2015}
{Nelson} D. {et~al.}, 2015, Astronomy and Computing, 13, 12

\bibitem[{{Okamoto} {et~al}\mbox{.}(2005){Okamoto}, {Eke}, {Frenk}, \&
  {Jenkins}}]{Okamoto_et_al_2005}
{Okamoto} T., {Eke} V.~R., {Frenk} C.~S., {Jenkins} A., 2005, \mnras, 363, 1299

\bibitem[{{Oppenheimer} \& {Dav{\'e}}(2006)}]{Oppenheimer_Dave_2006}
{Oppenheimer} B.~D., {Dav{\'e}} R., 2006, \mnras, 373, 1265

\bibitem[{{Oppenheimer} \& {Dav{\'e}}(2008)}]{Oppenheimer_Dave_2008}
---, 2008, \mnras, 387, 577

\bibitem[{{Oser} {et~al}\mbox{.}(2010){Oser}, {Ostriker}, {Naab}, {Johansson},
  \& {Burkert}}]{Oser_et_al_2010}
{Oser} L., {Ostriker} J.~P., {Naab} T., {Johansson} P.~H., {Burkert} A., 2010,
  \apj, 725, 2312

\bibitem[{{Pasquali} {et~al}\mbox{.}(2010){Pasquali}, {Gallazzi}, {Fontanot},
  {van den Bosch}, {De Lucia}, {Mo}, \& {Yang}}]{Pasquali_et_al_2010}
{Pasquali} A., {Gallazzi} A., {Fontanot} F., {van den Bosch} F.~C., {De Lucia}
  G., {Mo} H.~J., {Yang} X., 2010, \mnras, 407, 937

\bibitem[{{Pasquali}, {Gallazzi} \& {van den Bosch}(2012){Pasquali},
  {Gallazzi}, \& {van den Bosch}}]{Pasquali_et_al_2012}
{Pasquali} A., {Gallazzi} A., {van den Bosch} F.~C., 2012, \mnras, 425, 273

\bibitem[{{Peng}, {Maiolino} \& {Cochrane}(2015){Peng}, {Maiolino}, \&
  {Cochrane}}]{Peng_et_al_2015}
{Peng} Y., {Maiolino} R., {Cochrane} R., 2015, \nat, 521, 192

\bibitem[{{Peng} {et~al}\mbox{.}(2010){Peng}, {Lilly}, {Kova{\v c}},
  {Bolzonella}, {Pozzetti}, {Renzini}, {Zamorani}, {Ilbert}, {Knobel},
  {Iovino}, {Maier}, {Cucciati}, {Tasca}, {Carollo}, {Silverman}, {Kampczyk},
  {de Ravel}, {Sanders}, {Scoville}, {Contini}, {Mainieri}, {Scodeggio},
  {Kneib}, {Le F{\`e}vre}, {Bardelli}, {Bongiorno}, {Caputi}, {Coppa}, {de la
  Torre}, {Franzetti}, {Garilli}, {Lamareille}, {Le Borgne}, {Le Brun},
  {Mignoli}, {Perez Montero}, {Pello}, {Ricciardelli}, {Tanaka}, {Tresse},
  {Vergani}, {Welikala}, {Zucca}, {Oesch}, {Abbas}, {Barnes}, {Bordoloi},
  {Bottini}, {Cappi}, {Cassata}, {Cimatti}, {Fumana}, {Hasinger}, {Koekemoer},
  {Leauthaud}, {Maccagni}, {Marinoni}, {McCracken}, {Memeo}, {Meneux}, {Nair},
  {Porciani}, {Presotto}, \& {Scaramella}}]{Peng_et_al_2010}
{Peng} Y.-j. {et~al.}, 2010, \apj, 721, 193

\bibitem[{{Peng} {et~al}\mbox{.}(2012){Peng}, {Lilly}, {Renzini}, \&
  {Carollo}}]{Peng_et_al_2012}
{Peng} Y.-j., {Lilly} S.~J., {Renzini} A., {Carollo} M., 2012, \apj, 757, 4

\bibitem[{{Peng} \& {Maiolino}(2014)}]{Peng_Maiolino_2014}
{Peng} Y.-j., {Maiolino} R., 2014, \mnras, 438, 262

\bibitem[{{Petropoulou}, {V{\'{\i}}lchez} \&
  {Iglesias-P{\'a}ramo}(2012){Petropoulou}, {V{\'{\i}}lchez}, \&
  {Iglesias-P{\'a}ramo}}]{Petropoulou_et_al_2012}
{Petropoulou} V., {V{\'{\i}}lchez} J., {Iglesias-P{\'a}ramo} J., 2012, \apj,
  749, 133

\bibitem[{{Planck Collaboration XVI}(2014){Planck Collaboration},
  {Ade}, {Aghanim}, {Armitage-Caplan}, {Arnaud}, {Ashdown}, {Atrio-Barandela},
  {Aumont}, {Baccigalupi}, {Banday}, \& et~al.}]{Planck_2014}
{Planck Collaboration XVI}, 2014, \aap, 571, A16

\bibitem[{{Puchwein} \& {Springel}(2013)}]{Puchwein_Springel_2013}
{Puchwein} E., {Springel} V., 2013, \mnras, 428, 2966

\bibitem[{{Rodriguez-Gomez} {et~al}\mbox{.}(2015){Rodriguez-Gomez},
  {Pillepich}, {Sales}, {Genel}, {Vogelsberger}, {Zhu}, {Wellons}, {Nelson},
  {Torrey}, {Springel}, {Ma}, \& {Hernquist}}]{Rodriguez-Gomez_et_al_2015}
{Rodriguez-Gomez} V. {et~al.}, 2015, ArXiv e-prints

\bibitem[{{Rosas-Guevara} {et~al}\mbox{.}(2015){Rosas-Guevara}, {Bower},
  {Schaye}, {Furlong}, {Frenk}, {Booth}, {Crain}, {Dalla Vecchia}, {Schaller},
  \& {Theuns}}]{Rosas-Guevara_et_al_2015}
{Rosas-Guevara} Y.~M. {et~al.}, 2015, \mnras, 454, 1038

\bibitem[{{S{\'a}nchez} {et~al}\mbox{.}(2013){S{\'a}nchez}, {Rosales-Ortega},
  {Jungwiert}, {Iglesias-P{\'a}ramo}, {V{\'{\i}}lchez}, {Marino}, {Walcher},
  {Husemann}, {Mast}, {Monreal-Ibero}, {Cid Fernandes}, {P{\'e}rez},
  {Gonz{\'a}lez Delgado}, {Garc{\'{\i}}a-Benito}, {Galbany}, {van de Ven},
  {Jahnke}, {Flores}, {Bland-Hawthorn}, {L{\'o}pez-S{\'a}nchez}, {Stanishev},
  {Miralles-Caballero}, {D{\'{\i}}az}, {S{\'a}nchez-Blazquez}, {Moll{\'a}},
  {Gallazzi}, {Papaderos}, {Gomes}, {Gruel}, {P{\'e}rez}, {Ruiz-Lara},
  {Florido}, {de Lorenzo-C{\'a}ceres}, {Mendez-Abreu}, {Kehrig}, {Roth},
  {Ziegler}, {Alves}, {Wisotzki}, {Kupko}, {Quirrenbach}, {Bomans}, \& {Califa
  Collaboration}}]{Sanchez_et_al_2013}
{S{\'a}nchez} S.~F. {et~al.}, 2013, \aap, 554, A58

\bibitem[{{Schaller} {et~al}\mbox{.}(2015){Schaller}, {Dalla Vecchia},
  {Schaye}, {Bower}, {Theuns}, {Crain}, {Furlong}, \&
  {McCarthy}}]{Schaller_et_al_2015}
{Schaller} M., {Dalla Vecchia} C., {Schaye} J., {Bower} R.~G., {Theuns} T.,
  {Crain} R.~A., {Furlong} M., {McCarthy} I.~G., 2015, \mnras, 454, 2277

\bibitem[{{Schaye}(2004)}]{Schaye_2004}
{Schaye} J., 2004, \apj, 609, 667

\bibitem[{{Schaye} {et~al}\mbox{.}(2015){Schaye}, {Crain}, {Bower}, {Furlong},
  {Schaller}, {Theuns}, {Dalla Vecchia}, {Frenk}, {McCarthy}, {Helly},
  {Jenkins}, {Rosas-Guevara}, {White}, {Baes}, {Booth}, {Camps}, {Navarro},
  {Qu}, {Rahmati}, {Sawala}, {Thomas}, \& {Trayford}}]{Schaye_et_al_2015}
{Schaye} J. {et~al.}, 2015, \mnras, 446, 521

\bibitem[{{Schaye} \& {Dalla Vecchia}(2008)}]{Schaye_DallaVecchia_2008}
{Schaye} J., {Dalla Vecchia} C., 2008, \mnras, 383, 1210

\bibitem[{{Segers} {et~al}\mbox{.}(2016){Segers}, {Crain}, {Schaye}, {Bower},
  {Furlong}, {Schaller}, \& {Theuns}}]{Segers_et_al_2016a}
{Segers} M.~C., {Crain} R.~A., {Schaye} J., {Bower} R.~G., {Furlong} M.,
  {Schaller} M., {Theuns} T., 2016, \mnras, 456, 1235

\bibitem[{{Springel}(2005)}]{Springel_2005}
{Springel} V., 2005, \mnras, 364, 1105

\bibitem[{{Springel}(2010)}]{Springel_2010a}
---, 2010, \mnras, 401, 791

\bibitem[{{Springel} \& {Hernquist}(2003)}]{Springel_Hernquist_2003}
{Springel} V., {Hernquist} L., 2003, \mnras, 339, 289

\bibitem[{{Springel} {et~al}\mbox{.}(2005){Springel}, {White}, {Jenkins},
  {Frenk}, {Yoshida}, {Gao}, {Navarro}, {Thacker}, {Croton}, {Helly},
  {Peacock}, {Cole}, {Thomas}, {Couchman}, {Evrard}, {Colberg}, \&
  {Pearce}}]{Springel_et_al_2005}
{Springel} V. {et~al.}, 2005, Nature, 435, 629

\bibitem[{{Springel} {et~al}\mbox{.}(2001){Springel}, {White}, {Tormen}, \&
  {Kauffmann}}]{Springel_et_al_2001b}
{Springel} V., {White} S.~D.~M., {Tormen} G., {Kauffmann} G., 2001, \mnras,
  328, 726

\bibitem[{{Stinson} {et~al}\mbox{.}(2006){Stinson}, {Seth}, {Katz}, {Wadsley},
  {Governato}, \& {Quinn}}]{Stinson_et_al_2006}
{Stinson} G., {Seth} A., {Katz} N., {Wadsley} J., {Governato} F., {Quinn} T.,
  2006, \mnras, 373, 1074

\bibitem[{{Tornatore} {et~al}\mbox{.}(2007){Tornatore}, {Borgani}, {Dolag}, \&
  {Matteucci}}]{Tornatore_et_al_2007}
{Tornatore} L., {Borgani} S., {Dolag} K., {Matteucci} F., 2007, \mnras, 382,
  1050

\bibitem[{{Trayford} {et~al}\mbox{.}(2016){Trayford}, {Theuns}, {Bower},
  {Crain}, {Lagos}, {Schaller}, \& {Schaye}}]{Trayford_et_al_2016}
{Trayford} J.~W., {Theuns} T., {Bower} R.~G., {Crain} R.~A., {Lagos} C.~d.~P.,
  {Schaller} M., {Schaye} J., 2016, \mnras, 460, 3925

\bibitem[{{Trayford} {et~al}\mbox{.}(2015){Trayford}, {Theuns}, {Bower},
  {Schaye}, {Furlong}, {Schaller}, {Frenk}, {Crain}, {Dalla Vecchia}, \&
  {McCarthy}}]{Trayford_et_al_2015}
{Trayford} J.~W. {et~al.}, 2015, \mnras, 452, 2879

\bibitem[{{Tremonti} {et~al}\mbox{.}(2004){Tremonti}, {Heckman}, {Kauffmann},
  {Brinchmann}, {Charlot}, {White}, {Seibert}, {Peng}, {Schlegel}, {Uomoto},
  {Fukugita}, \& {Brinkmann}}]{Tremonti_et_al_2004}
{Tremonti} C.~A. {et~al.}, 2004, \apj, 613, 898

\bibitem[{{Valentino} {et~al}\mbox{.}(2015){Valentino}, {Daddi}, {Strazzullo},
  {Gobat}, {Onodera}, {Bournaud}, {Juneau}, {Renzini}, {Arimoto}, {Carollo}, \&
  {Zanella}}]{Valentino_et_al_2015}
{Valentino} F. {et~al.}, 2015, \apj, 801, 132

\bibitem[{{van Dokkum} {et~al}\mbox{.}(2010){van Dokkum}, {Whitaker},
  {Brammer}, {Franx}, {Kriek}, {Labb{\'e}}, {Marchesini}, {Quadri}, {Bezanson},
  {Illingworth}, {Muzzin}, {Rudnick}, {Tal}, \& {Wake}}]{van_Dokkum_et_al_2010}
{van Dokkum} P.~G. {et~al.}, 2010, \apj, 709, 1018

\bibitem[{{Vila-Costas} \& {Edmunds}(1992)}]{Vila-Costas_Edmunds_1992}
{Vila-Costas} M.~B., {Edmunds} M.~G., 1992, \mnras, 259, 121

\bibitem[{{Vogelsberger} {et~al}\mbox{.}(2013){Vogelsberger}, {Genel},
  {Sijacki}, {Torrey}, {Springel}, \& {Hernquist}}]{Vogelsberger_et_al_2013}
{Vogelsberger} M., {Genel} S., {Sijacki} D., {Torrey} P., {Springel} V.,
  {Hernquist} L., 2013, \mnras, 436, 3031

\bibitem[{{Vogelsberger} {et~al}\mbox{.}(2014){Vogelsberger}, {Genel},
  {Springel}, {Torrey}, {Sijacki}, {Xu}, {Snyder}, {Nelson}, \&
  {Hernquist}}]{Vogelsberger_et_al_2014}
{Vogelsberger} M. {et~al.}, 2014, \mnras, 444, 1518

\bibitem[{{von der Linden} {et~al}\mbox{.}(2010){von der Linden}, {Wild},
  {Kauffmann}, {White}, \& {Weinmann}}]{von_der_Linden_et_al_2010}
{von der Linden} A., {Wild} V., {Kauffmann} G., {White} S.~D.~M., {Weinmann}
  S., 2010, \mnras, 404, 1231

\bibitem[{{Wang} {et~al}\mbox{.}(2008){Wang}, {De Lucia}, {Kitzbichler}, \&
  {White}}]{Wang_et_al_2008}
{Wang} J., {De Lucia} G., {Kitzbichler} M.~G., {White} S.~D.~M., 2008, \mnras,
  384, 1301

\bibitem[{{Weisz} {et~al}\mbox{.}(2014){Weisz}, {Dolphin}, {Skillman},
  {Holtzman}, {Gilbert}, {Dalcanton}, \& {Williams}}]{Weisz_et_al_2014}
{Weisz} D.~R., {Dolphin} A.~E., {Skillman} E.~D., {Holtzman} J., {Gilbert}
  K.~M., {Dalcanton} J.~J., {Williams} B.~F., 2014, \apj, 789, 147

\bibitem[{{Weisz} {et~al}\mbox{.}(2015){Weisz}, {Dolphin}, {Skillman},
  {Holtzman}, {Gilbert}, {Dalcanton}, \& {Williams}}]{Weisz_et_al_2015}
---, 2015, \apj, 804, 136

\bibitem[{{Wetzel}, {Tinker} \& {Conroy}(2012){Wetzel}, {Tinker}, \&
  {Conroy}}]{Wetzel_et_al_2012}
{Wetzel} A.~R., {Tinker} J.~L., {Conroy} C., 2012, \mnras, 424, 232

\bibitem[{{Wetzel} {et~al}\mbox{.}(2013){Wetzel}, {Tinker}, {Conroy}, \& {van
  den Bosch}}]{Wetzel_et_al_2013}
{Wetzel} A.~R., {Tinker} J.~L., {Conroy} C., {van den Bosch} F.~C., 2013,
  \mnras, 432, 336

\bibitem[{{White} \& {Rees}(1978)}]{White_Rees_1978}
{White} S.~D.~M., {Rees} M.~J., 1978, \mnras, 183, 341

\bibitem[{{Wiersma}, {Schaye} \& {Smith}(2009){Wiersma}, {Schaye}, \&
  {Smith}}]{Wiersma_et_al_2009a}
{Wiersma} R.~P.~C., {Schaye} J., {Smith} B.~D., 2009, \mnras, 393, 99

\bibitem[{{Wiersma} {et~al}\mbox{.}(2009){Wiersma}, {Schaye}, {Theuns}, {Dalla
  Vecchia}, \& {Tornatore}}]{Wiersma_et_al_2009b}
{Wiersma} R.~P.~C., {Schaye} J., {Theuns} T., {Dalla Vecchia} C., {Tornatore}
  L., 2009, \mnras, 399, 574

\bibitem[{{Wuyts} {et~al}\mbox{.}(2016){Wuyts}, {Wisnioski}, {Fossati},
  {F{\"o}rster Schreiber}, {Genzel}, {Davies}, {Mendel}, {Naab},
  {R{\"o}ttgers}, {Wilman}, {Wuyts}, {Bandara}, {Beifiori}, {Belli}, {Bender},
  {Brammer}, {Burkert}, {Chan}, {Galametz}, {Kulkarni}, {Lang}, {Lutz},
  {Momcheva}, {Nelson}, {Rosario}, {Saglia}, {Seitz}, {Tacconi}, {Tadaki},
  {{\"U}bler}, \& {van Dokkum}}]{Wuyts_et_al_2016}
{Wuyts} E. {et~al.}, 2016, ArXiv e-prints

\bibitem[{{Yang} {et~al}\mbox{.}(2005){Yang}, {Mo}, {van den Bosch}, \&
  {Jing}}]{Yang_et_al_2005}
{Yang} X., {Mo} H.~J., {van den Bosch} F.~C., {Jing} Y.~P., 2005, \mnras, 356,
  1293

\bibitem[{{Yang} {et~al}\mbox{.}(2007){Yang}, {Mo}, {van den Bosch},
  {Pasquali}, {Li}, \& {Barden}}]{Yang_et_al_2007}
{Yang} X., {Mo} H.~J., {van den Bosch} F.~C., {Pasquali} A., {Li} C., {Barden}
  M., 2007, \apj, 671, 153

\bibitem[{{Zahid} {et~al}\mbox{.}(2014){Zahid}, {Dima}, {Kudritzki}, {Kewley},
  {Geller}, {Hwang}, {Silverman}, \& {Kashino}}]{Zahid_et_al_2014}
{Zahid} H.~J., {Dima} G.~I., {Kudritzki} R.-P., {Kewley} L.~J., {Geller} M.~J.,
  {Hwang} H.~S., {Silverman} J.~D., {Kashino} D., 2014, \apj, 791, 130

\bibitem[{{Zaritsky}, {Kennicutt} \& {Huchra}(1994){Zaritsky}, {Kennicutt}, \&
  {Huchra}}]{Zaritsky_et_al_1994}
{Zaritsky} D., {Kennicutt}, Jr. R.~C., {Huchra} J.~P., 1994, \apj, 420, 87

\bibitem[{{Zentner}, {Hearin} \& {van den Bosch}(2014){Zentner}, {Hearin}, \&
  {van den Bosch}}]{Zentner_et_al_2014}
{Zentner} A.~R., {Hearin} A.~P., {van den Bosch} F.~C., 2014, \mnras, 443, 3044

\bibitem[{{Zhang} {et~al}\mbox{.}(2009){Zhang}, {Li}, {Kauffmann}, {Zou},
  {Catinella}, {Shen}, {Guo}, \& {Chang}}]{Zhang_et_al_2009}
{Zhang} W., {Li} C., {Kauffmann} G., {Zou} H., {Catinella} B., {Shen} S., {Guo}
  Q., {Chang} R., 2009, \mnras, 397, 1243

\end{thebibliography}

\end{document}